\documentclass[]{pasj02}
\usepackage[switch,mathlines]{lineno}
\usepackage{natbib}
\usepackage{graphicx}
\usepackage{amsmath}
\usepackage{multirow}
\usepackage{color}
\usepackage{comment}
\usepackage{url} 
\DeclareUnicodeCharacter{2243}{\ensuremath{\simeq}}
\DeclareUnicodeCharacter{2500}{--}

\jyear{2026}
\Received{}
\Accepted{}

\newcommand{\ion}[2]{#1\,\textsc{#2}}

\newcommand{\Msun}{{\mathrm M_{\odot}}}
\newcommand{\Zsun}{{\mathrm Z_{\odot}}}
\newcommand{\PopIII}{\mbox{Pop I\hspace{-1pt}I\hspace{-1pt}I}}

\newcommand{\Mhalo}{{M_{\text{halo}}}}
\newcommand{\Luv}{{L_{\text{UV}}}}
\newcommand{\Muv}{{M_{\text{UV}}}}

\newcommand{\Kroupa}{\texttt{Model:Kroupa}}
\newcommand{\KroupaDustTrinca}{\texttt{Model:Kroupa(D:Trinca)}}
\newcommand{\zZzZ}{\texttt{Model:zZ-IMF(Z:zZ)}}
\newcommand{\zZzZDustTrinca}{\texttt{Model:zZ-IMF(Z:zZ; D:Trinca)}}
\newcommand{\zZKroupa}{\texttt{Model:zZ-IMF(Z:Kroupa)}}
\newcommand{\zZKroupaDustTrinca}{\texttt{Model:zZ-IMF(Z:Kroupa; D:Trinca)}}

\newcommand{\zZIMF}{$z$-$Z$ dependent IMF}

\setcitestyle{round}

\begin{document}

\title{
On the effect of a redshift- and metallicity-dependent top-heavy IMF on high-redshift UV luminosity functions
%ling with redshift- and metallicity-dependent top-heavy IMF and the UV luminosity function in the high-redshift Universe
}

\author{
 Keita \textsc{Fukushima},\altaffilmark{1,2}\altemailmark\email{fukushima@snu.ac.kr}\orcid{0000-0002-5045-6052}
 Shuya \textsc{Kitagawa},\altaffilmark{3}
 Akio K. \textsc{Inoue},\altaffilmark{2,3}\orcid{0000-0002-7779-8677}
 Tomoaki \textsc{Ishiyama},\altaffilmark{4}\orcid{0000-0002-5316-9171}
 Sunmyon \textsc{Chon},\altaffilmark{5, 6, 7}\orcid{0000-0002-2912-3923}
 and Kazuyuki \textsc{Omukai}\altaffilmark{7}\orcid{0000-0001-5922-180X}
}

\altaffiltext{1}{Institute for Data Innovation in Science, Seoul National University, Seoul 08826, Republic of Korea}
\altaffiltext{2}{Waseda Research Institute for Science and Engineering, Faculty of Science and Engineering, Waseda University, 3-4-1 Okubo, Shinjuku, Tokyo 169-8555, Japan}
\altaffiltext{3}{Department of Pure and Applied Physics, Graduate School of Advanced Science and Engineering, Faculty of Science and Engineering, Waseda University, 3-4-1 Okubo, Shinjuku, Tokyo 169-8555, Japan}
\altaffiltext{4}{Digital Transformation Enhancement Council, Chiba University, 1-33, Yayoi-cho, Inage-ku, Chiba 263-8522, Japan}
\altaffiltext{5}{Dipartimento di Fisica, Universit\`{a} di Roma ‘La Sapienza’, P.le Aldo Moro 2, I-00185 Roma, Italy}
\altaffiltext{6}{Max-Planck-Institut für Astrophysik, Karl-Schwarzschild-Str. 1, D-85741 Garching, Germany}
\altaffiltext{7}{Astronomical Institute, Graduate School of Science, Tohoku University, Aoba, Sendai 980-8578, Japan}


\maketitle

\begin{abstract}

JWST observations have revealed a higher-than-expected abundance of bright galaxies at $z \gtrsim 10$. Several scenarios have been proposed to explain this excess, one of which invokes a top-heavy stellar initial mass function (IMF) in the early Universe. In this work, we construct spectral energy distribution (SED) templates based on a redshift- and metallicity-dependent IMF (\zZIMF) motivated by the star-cluster simulations of \citet[MNRAS, 514, 4639]{Chon2022MNRAS} and the redshift-dependent IMF model of \citet[ApJ, 890, 19]{Steinhardt2020ApJ}. We implement these templates in a semi-analytic galaxy model built on a large-box $N$-body simulation to investigate the impact of a top-heavy IMF on the ultraviolet luminosity function (UVLF) of high-redshift galaxies.
The adopted \zZIMF\ enhances the UV luminosity of young stellar populations, particularly at low metallicity, and increases the scatter in the UV luminosity-to-star formation rate conversion factor relative to the Kroupa IMF. Without dust attenuation, both top-heavy IMF models predict a substantially enhanced bright end of the UVLF. However, when dust attenuation is included and calibrated at $z=7$, IMF-dependent metal enrichment and the resulting dust attenuation partly offset the intrinsic UV enhancement. 
The self-consistent model broadly reproduces the observed UVLF at $z\sim10$, although discrepancies remain at $z\gtrsim12$. 
For the selected high-redshift UVLF constraints, the best-fitting dust optical depths are 51\% and 21\% of the fiducial values at $z=9.7$ and $13.04$, respectively, indicating a preference for weaker effective dust attenuation. 
The highest-redshift comparison does not provide a robust constraint. Our results highlight the importance of accounting for both enhanced stellar UV emission and IMF-dependent metal enrichment and dust attenuation when assessing the impact of a top-heavy IMF on high-redshift galaxies.

\end{abstract}

%\pagewiselinenumbers

\section{Introduction}

The James Webb Space Telescope (JWST) has revealed a large number of bright galaxies at $z \gtrsim 10$ \citep[e.g.,][]{Castellano2022ApJ, Finkelstein2022ApJ, Donnan2023MNRAS_a, Naidu2022_a_arXiv}, including MoM-z14 at $z_\mathrm{spec}=14.44\pm0.02$, the most distant spectroscopically confirmed galaxy to date \citep{Naidu2026_MoMz14}.
Before JWST, only a handful of galaxies had been
spectroscopically confirmed at $z > 8$  \citep[e.g.,][]{Zitrin2015ApJ,Hashimoto18Nature,Tamura2019ApJ},
too few to statistically
characterize the early galaxy population. 
JWST has dramatically expanded
the spectroscopic census \citep{Stark_2025_reviewJWST}, enabling detailed studies of early galaxies, including their stellar populations, star formation histories, and black hole activity. 
At the same time, the unexpectedly high abundance of bright galaxies poses a challenge to many existing models of galaxy formation and evolution.

JWST measurements of the UV luminosity function (UVLF) have consistently found high number densities of UV-bright galaxies at $z\sim10$--12, exceeding many pre-JWST theoretical predictions \citep[e.g.,][]{Harikane2023a_ApJS, Harikane2024ApJ_UVLF, Harikane2024arXiv_UVLF, Bunker2023A&A, Arrabal2023Natur, Fujimoto2024ApJ_uncover, Carniani2024Natur, Finkelstein2024ApJ,Whitler2025ApJ,Franco2025arXiv250804791F}.
At higher redshifts, however, the observational picture remains uncertain. 
While the JADES measurements suggest that the bright-end excess relative to many theoretical models persists at $z\gtrsim12$ \citep{Whitler2025ApJ}, wide-area measurements from PANORAMIC and other JWST fields suggest a more rapid decline in the number density of UV-bright galaxies beyond $z\sim13$ \citep{Weibel2026ApJ, McLeod2026arXiv260416666M}.
Gravitational-lensing surveys further extend UVLF constraints to fainter magnitudes over $z\sim9$--15 \citep{Chemerynska2026MNRAS}.
Thus, although the unexpectedly high abundance of UV-bright galaxies is consistently observed at $z\sim10$--12, the evolution of the UVLF beyond $z\sim12$ remains uncertain.

The physical origin of this bright-end excess remains unclear.
Several astrophysical explanations have been proposed, including enhanced star formation efficiency (SFE) at $z\gtrsim10$ due to suppressed feedback in the first galaxies \citep{Inayoshi2022ApJ, Dekel2023MNRAS, Qin2023MNRAS, Finkelstein2024ApJ, Jeong2025ApJ, Chen2026MNRAS_FeedbackFree}; reduced dust attenuation resulting from radiation-driven dust removal in galaxies with high Eddington ratios \citep{Fiore2023ApJ_Dust, Ferrara2023MNRAS, Ferrara2024A&A_Dust_UVLF_outf}; and large scatter in the relation between UV magnitude ($\Muv$) and halo mass ($\Mhalo$) arising from stochastic star formation histories \citep{Mason2023MNRAS, Shen2023MNRAS, Sun2023ApJ, Mirocha2023MNRAS, Gelli2024ApJ, Kravtsov2024arXiv, Yung2024MNRAS, Munoz2026MNRAS, Sarkar2026MNRAS}. 
Other possibilities include additional UV emission from $\PopIII$ stars or active galactic nuclei (AGNs) \citep{Harikane2023a_ApJS, Hegde2024JCAP}, and an enhanced $\Luv$/SFR ratio resulting from a top-heavy initial mass function (IMF) at high redshift \citep{Shen2023MNRAS, Trinca2024MNRAS, Ventura2024MNRAS, Cueto2024AandA, Hutter2025AandA, Jeong2025ApJ, Durrant2026}. 
From a cosmological perspective, proposed explanations include enhanced power on $\sim1$ Mpc scales \citep{Padmanabhan2023ApJ, Parashari2023MNRAS};
a blue-tilted primordial power spectrum with positive running \citep{Hirano2024MNRAS, Kobayashi2026PRD};
a modified cosmic time--redshift relation \citep{Melia2023MNRAS, Melia2024A&A}; accelerated formation of galaxies and galaxy clusters in modified Newtonian dynamics \citep{McGaugh2024ApJ}; and a negative cosmological constant \citep{Menci2024JCAP}.
Furthermore, the consistent detection of this bright-end excess at $z\sim10$--12 in wide-area surveys spanning multiple independent fields suggests that cosmic variance alone cannot explain it \citep{McLeod2024MNRAS, Weibel2026ApJ, Weibel2026ApJ_PANORAMICII}.

Recent cosmological hydrodynamical simulations have reached different conclusions about whether standard galaxy-formation physics can account for the observed bright-end excess.
The FIREbox$^{\rm HR}$ simulation reproduces the observed UVLFs and their evolution over $z\sim6$--14 without requiring a strongly redshift-dependent SFE \citep{Feldmann2025MNRAS}.
In contrast, \citet{Lu2026arXiv260506782L} found that the dust-attenuated UVLFs predicted by COLIBRE systematically fall below observational estimates at the bright end, with the discrepancy increasing toward higher redshifts.
Without dust attenuation, COLIBRE produces sufficiently bright galaxies at $z\sim7$--12, but its brightest galaxies remain underluminous at $z=15$, suggesting that an additional boost to their intrinsic UV emission may be required.

%\citet{Mauerhofer2025A&A} constructed models with a top-heavy IMF and high star formation efficiency using both hydrodynamical simulations and semi-analytic models, and demonstrated that both models can reproduce the observed excess of bright galaxies at $z=11$.
%The introduction of a top-heavy IMF in high-$z$ galaxies has also been used to explain the presence of ionizing sources with $T_{\rm eff} \gtrsim 80,000$ K and an electron temperature of $T_e \approx 34,000$ K in the $z = 8.271$ galaxy observed by JWST \citep{Cullen2025arXiv}.
%Recent high-resolution star cluster formation simulations \citep{Chon2022MNRAS} have investigated the IMF across a metallicity range of $10^{-4} < Z < 10^{-1}$ and a redshift range of $0 < z < 20$. 

Among the astrophysical mechanisms for boosting intrinsic UV emission, IMF variations are physically motivated by the thermodynamics and chemistry of primordial and metal-poor gas, which can affect the characteristic stellar mass.
Star-cluster-scale simulations predict that the IMF becomes increasingly top-heavy at lower metallicities and higher redshifts \citep{Chon2021MNRAS, Chon2022MNRAS, Chon2024MNRAS}.
By incorporating quantities derived from the SPHINX20 radiation-hydrodynamical simulation into the DELPHI semi-analytic model, \citet{Mauerhofer2025A&A} found that such a metallicity- and redshift-dependent IMF can reproduce the observed abundance of bright galaxies at $z\gtrsim11$.
Observationally, \citet{Cullen2025arXiv} inferred a very hard ionizing spectrum for the $z=8.27$ galaxy EXCELS-63107 and suggested that an excess of stars more massive than $50\,\Msun$, possibly associated with a top-heavy IMF, could explain the spectrum. 
%We hereafter refer to the redshift- and metallicity-dependent IMF adopted in this work as the \zZIMF.

IMF variations, however, affect the observed UVLF not only by changing the intrinsic UV luminosity of stellar populations but also by modifying metal enrichment and, consequently, dust attenuation.
Their net effect on the observed UVLF therefore depends on the evolution of metals and dust, as well as the adopted dust attenuation prescription.
Recently, \citet{Durrant2026} investigated these effects using a cosmological hydrodynamical simulation with a density-dependent top-heavy IMF.
They found that the enhanced far-UV emission is partly offset by stronger dust attenuation arising from increased dust surface densities.
Their IMF prescription depends primarily on the density of star-forming gas, rather than explicitly on redshift and metallicity.
It therefore remains unclear how the competing effects of enhanced UV emission and dust attenuation shape the UVLF when the IMF depends on redshift and metallicity.

In this study, we ask whether a redshift- and metallicity-dependent top-heavy IMF can explain the observed bright-end excess in the high-redshift UVLF when its effects on both stellar UV emission and chemical enrichment are taken into account. 
We hereafter refer to the IMF adopted in this work as the \zZIMF.
We construct spectral energy distribution (SED) templates for this IMF and incorporate them into a semi-analytic galaxy model based on the Shin-Uchuu simulation \citep{Ishiyama2021}. 
To assess the role of IMF-dependent chemical enrichment, we compare a diagnostic control model in which the \zZIMF\ affects only the stellar SEDs with a self-consistent model in which it also modifies metal production and the resulting dust attenuation.
This comparison allows us to test whether the enhancement in intrinsic UV emission remains sufficient to explain the observed bright-end excess after accounting for the associated metal enrichment and dust attenuation.

Section~\ref{sec:Model_IMF_SED_Semiana} describes our model and comparison strategy. Section~\ref{sec:results} presents the intrinsic and dust-attenuated UVLFs together with galaxy and SED diagnostics that help interpret the results. Section~\ref{sec:discussion} discusses whether a variable IMF can explain the bright-galaxy excess, the implications for dust evolution, and the main limitations of our model. Section~\ref{sec:conclusion} summarizes our conclusions.
The Appendices provide details of the IMF
parameterization, simple stellar population (SSP) SED templates and diagnostics, stellar mass and metal ejection, and the semi-analytic galaxy model implementation.
They also present the $z=7$ calibration of the alternative dust attenuation model and additional UVLF comparisons with varying dust attenuation strengths.

\section{Model}\label{sec:Model_IMF_SED_Semiana}

To predict the observable UVLF, we incorporate SSP spectra based on the \zZIMF\ into a semi-analytic galaxy model and apply dust attenuation.
The following subsections describe these model components and our strategy for assessing the direct effect of the IMF on stellar UV emission and its indirect effect through metal enrichment and dust attenuation.
Technical derivations and validation tests are provided in the Appendices.

\subsection{Redshift- and metallicity-dependent IMF}\label{subsec:IMF_model}

Our \zZIMF\ is motivated by recent high-resolution star-cluster formation simulations, which show that the thermodynamic and chemical states of low-metallicity gas vary with metallicity and redshift, affecting the characteristic stellar mass and the fraction of massive stars formed \citep{Chon2021MNRAS,Chon2022MNRAS,Chon2024MNRAS}.

To construct this IMF, we combine two complementary prescriptions: the redshift-dependent IMF proposed by \citet{Steinhardt2020ApJ}, which neglects explicit metallicity dependence, and the $z$--$Z$ trends reported by \citet{Chon2022MNRAS} from star-cluster formation simulations. The resulting IMF varies continuously with both $z$ and $Z$. It is calibrated to the IMF prescriptions of \citet{Chon2022MNRAS} at low metallicities and \citet{Steinhardt2020ApJ} at solar metallicity.

The Steinhardt IMF is based on a \citet{Kroupa2001MNRAS}-type functional form, but allows the characteristic mass scale to increase with redshift \citep{Schneider2010MNRAS}. 
Specifically, the break mass is scaled by $(T_{\mathrm g}/T_{0})^{2}$, where, $T_{\mathrm g}$ denotes the gas temperature in star-forming molecular clouds and $T_{0}=20\,\mathrm{K}$ is a typical value for the Milky Way. 
This mass scale corresponds to the minimum mass for fragmentation and collapse. 
In the model, $T_{\mathrm g}$ is regulated by stellar radiation, cosmic rays, and the CMB, and generally increases toward higher redshift.
The redshift dependence of $T_{\mathrm g}$ is computed following the original formulation of \citet{Steinhardt2020ApJ} (see Appendix~\ref{sec:appendix_steinhardtIMF}). 

\citet{Chon2022MNRAS} performed star-cluster formation simulations across a grid of metallicities ($\log(Z/\Zsun)=-1,-2,-3,-4$) and redshifts ($z=0,5,10,20$), finding that the IMF becomes markedly more top-heavy toward lower metallicity and higher redshift.
Their simulations characterize how the low-mass peak and the high-mass tail of the IMF vary with metallicity and redshift.
We use these results to determine the metallicity and redshift dependence of our \zZIMF.

\begin{figure*}
    \begin{center}
    \includegraphics[width=1.0\textwidth]{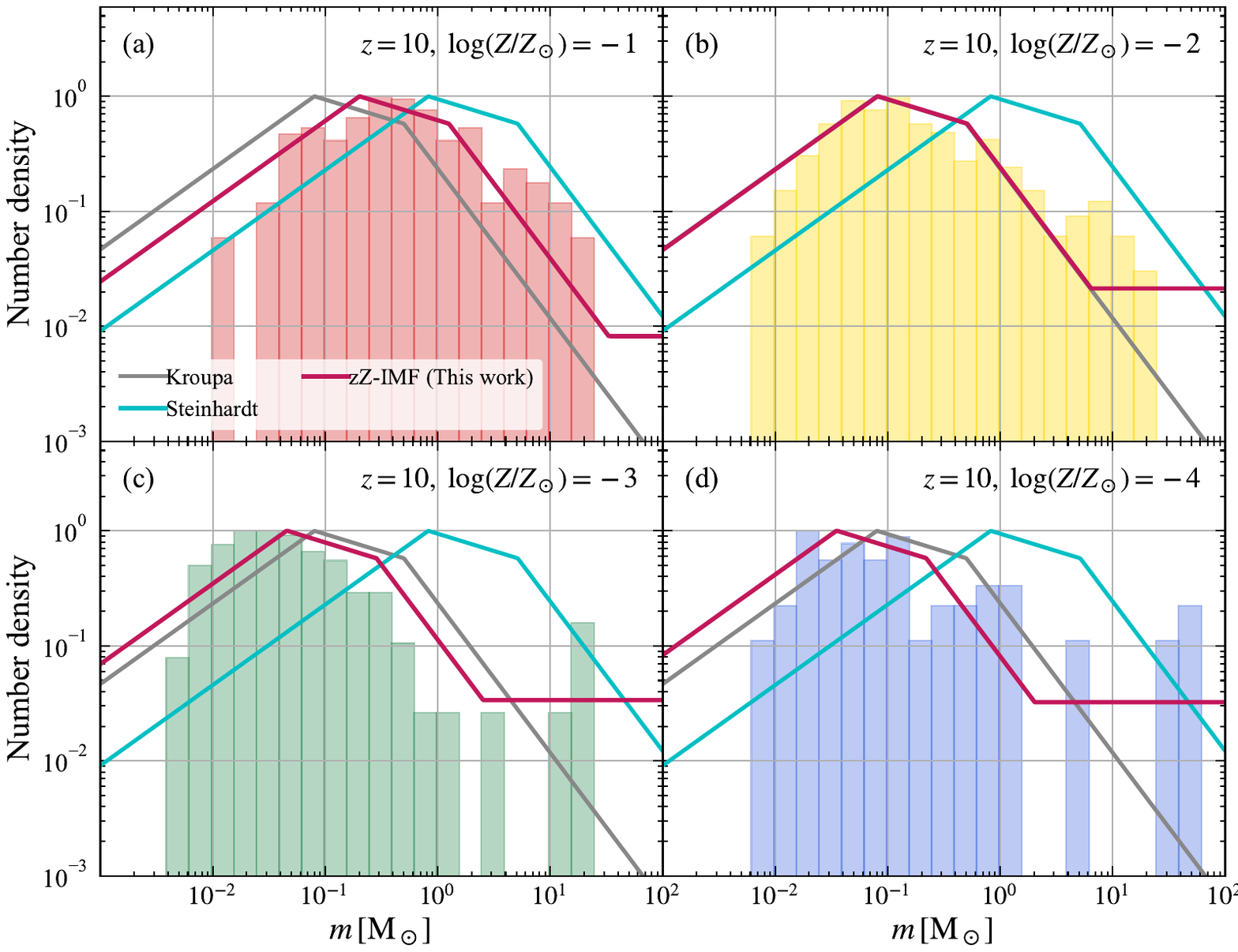}
    \end{center}
    \caption{Comparison of our
    \zZIMF\ model (red solid lines) with other IMFs at $z = 10$. The panels correspond to $\log(Z/\Zsun) = -1$ (panel~a), $-2$ (panel~b), $-3$ (panel~c), and $-4$ (panel~d). 
    The gray and light-blue lines represent the IMFs of \citet{Kroupa2001MNRAS} and \citet{Steinhardt2020ApJ}, respectively, while the histograms show the simulation results of \citet{Chon2022MNRAS}.}
    \label{fig:IMF_result_z10}
\end{figure*}

The simulations of \citet{Chon2022MNRAS} account for the redshift dependence through CMB heating but do not include radiative feedback from the forming stars.
Subsequent radiation-hydrodynamic simulations by \citet{Chon2024MNRAS} investigated the effects of stellar radiative feedback on the IMF at different metallicities, although the redshift dependence was not explored.
They found that radiative heating of dust suppresses the formation of low-mass stars, while photoionization limits the growth of massive stars.
Although these effects modify the detailed shape of the IMF, the mass fraction in massive stars remains relatively insensitive to radiative feedback.
We therefore adopt the results of \citet{Chon2022MNRAS} to construct an IMF that depends on both redshift and metallicity,
while retaining the two-component structure consisting of a low-mass Kroupa-like component and a top-heavy, log-flat component.

We parameterize our IMF as a continuous function of redshift and metallicity to reproduce the IMFs obtained in the simulations of \citet{Chon2022MNRAS}, using their results at $z=0, 5, 10$, and 20.
Figure~\ref{fig:IMF_result_z10} shows a representative comparison at $z=10$ for $\log(Z/\Zsun)=-1,-2,-3$, and $-4$.
Our model (red curves) reproduces both the low-mass peak and the high-mass regime of the simulated IMFs (histograms), with an approximately log-flat distribution at high masses.
The fitting procedure, interpolation across the $(z,\,Z)$ plane, and validation against the simulation results are described in Appendix~\ref{app:imf_technical}.
The resulting IMF prescriptions determine the stellar UV output and metal production used in the galaxy model.

\subsection{Stellar population and SED model}\label{subsec:SED_model}

We construct intrinsic SED templates for SSPs based on the \zZIMF\ described in Section~\ref{subsec:IMF_model}, using the FSPS code \citep{Conroy2009ApJ,Conroy2010ApJ}.
The templates are generated over grids of stellar age, metallicity, and redshift using the MIST isochrones \citep{Choi2016ApJ, Dotter2016ApJS} and the MILES stellar library \citep{SnchezBlazquez2006MNRAS, FalcanBarroso2011AA}.
For comparison, we also construct SED templates using the standard Kroupa IMF.
Representative full SED templates are provided in Appendix~\ref{sec:appendix_SED_temp}, while Appendix~\ref{app:sed_diagnostics} presents the age, metallicity, and redshift dependence of the SSP UV luminosity, UV slope, ionizing-photon output, and Balmer/4000-\AA\ breaks. 

\subsection{Semi-analytic galaxy model and metal evolution}\label{subsec:semiana_model}

We base our analysis on the Shin-Uchuu simulation~\footnote{\url{https://www.skiesanduniverses.org/Simulations/Uchuu/}} \citep{Ishiyama2021}, a high-resolution cosmological $N$-body simulation designed to probe the formation and evolution of dark matter halos across cosmic time. The simulation adopts a box size of $140 \, h^{-1} \, \mathrm{Mpc}$ on a side and contains $6400^3$ particles, corresponding to a dark matter particle mass of $m_p = 8.97 \times 10^5 \, h^{-1} \, \Msun$ and a gravitational softening of $0.4 \, h^{-1} \, \mathrm{kpc}$. The cosmological parameters are consistent with the \citet{Planck2020A&A} results. Halos and subhalos were identified using the ROCKSTAR finder~\footnote{\url{https://bitbucket.org/gfcstanford/rockstar/}} \citep{Behroozi2013ApJ}, and their merger trees were constructed using the consistent trees merger tree code~\footnote{\url{https://bitbucket.org/pbehroozi/consistent-trees/}} \citep{Behroozi2013bApJ}.

To construct galaxy merger histories, we utilize the halo merger trees from the simulation, spanning the redshift range from $z = 19.95$ to $z = 7.19$.
We select halos with masses exceeding $10^{10} \Msun$
at $z = 7.19$ and use only their merger trees
in our semi-analytic model.
This selection is motivated by computational constraints
and our focus on the bright-end excess in the UVLF.

We evolve galaxies along these merger trees with a deliberately simple semi-analytic model. 
The SFR is tied to the halo accretion rate,
while the gas mass and metallicity evolve
through cosmological gas accretion, star formation,
metal production, and galactic outflows.
The governing equations are given in Appendix~\ref{app:semianalytic_details}.
The galaxy SED is constructed by convolving the star formation history with IMF-dependent SSP templates, accounting for the age and metallicity of each stellar population.
For the \zZIMF\ model, the returned mass fraction and metal yields vary with redshift and metallicity; for the Kroupa model, they are fixed (see Appendix~\ref{app:appendix_Yield}). 

The merger-tree snapshot interval is used as the model timestep, without substepping. 
Consequently, if this interval becomes comparable to or longer than the timescale of individual star-formation episodes, the predicted burstiness and the UV-luminosity scatter can depend on the temporal sampling. 
We return to this limitation in Section~\ref{subsec:discussion_limitations}.

\subsection{Dust attenuation}\label{subsec:dust_model}

\subsubsection{Fiducial attenuation model}\label{subsubsec:dust_calibration}

To compare our predictions with the observed
UVLF, we model dust attenuation over
1450--1550\,\AA.
We adopt the dust attenuation prescription of \citet{Shimizu14} and calibrate its parameters using the observed UVLF at $z = 7$ \citep{Bouwens2021AJ, Harikane2022ApJS}.
This prescription assumes a ``sandwich'' configuration, where a uniformly mixed slab of stars and dust is embedded between two stellar layers \citep{Xu&Buat1995A&A}. 
The UV escape fraction $f_{\mathrm{UV}}$ is modeled as a function of the dust optical depth $\tau_{\mathrm{d}}$ and the geometry parameter $\delta$, which characterizes the spatial distribution of dust and stars:
\begin{equation}
f_{\mathrm{UV}} = \frac{1 - \delta}{2}(1 + e^{-\tau_{\mathrm{d}}}) + \frac{\delta}{\tau_{\mathrm{d}}}(1 - e^{-\tau_{\mathrm{d}}}).
\label{eq:dust_shimizu14}
\end{equation}

The optical depth is given by
$\tau_{\mathrm{d}} = 3\Sigma_{\mathrm{d}}/4a_{\mathrm{d}}s$,
where $\Sigma_{\mathrm{d}}$ is the dust surface density, defined as
$\Sigma_{\mathrm{d}} = e_{\tau} m_{Z}/\pi r^2_{\mathrm{half}}$. 
Here, $m_{Z}$ is the gas-phase metal mass,
$r_{\mathrm{half}}$ is the dark-matter halo half-mass radius,
$a_{\mathrm{d}}$ represents the characteristic radius of dust grains, and $s$ denotes their material density.
We adopt $a_{\mathrm{d}} = 0.1~\mu\mathrm{m}$ and $s = 2.5~\mathrm{g~cm^{-3}}$, consistent with the dust formation models in supernova ejecta 
\citep{Todini2001MNRAS, Nozawa2003ApJ}. 
The normalization parameter $e_{\tau}$ is defined as $e_{\tau} = e_{M_{\mathrm{d}}}/e_{\mathrm{rd}}^{2}$, where $e_{M_{\mathrm{d}}}$ denotes the ratio of dust mass to metal mass in the gas component, and $e_{\mathrm{rd}}$ represents the ratio of the effective dust distribution radius $r_{\mathrm{d}}$ to $r_{\mathrm{half}}$.
Uncertainties in $a_{\mathrm{d}}$ and $s$ can be effectively absorbed into the fitted normalization parameter $e_{\tau}$.
The metallicity evolution, which depends on the adopted IMF (see Appendix~\ref{subsubsec:semiana_model_metal}), is also reflected in the computation of $\Sigma_{\mathrm{d}}$.

In equation~(\ref{eq:dust_shimizu14}), the parameter $\delta$ ranges from 0 to 1. 
A value of $\delta = 1$ corresponds to a uniformly mixed slab in which stars and dust are homogeneously distributed, whereas $\delta = 0$ represents a geometry in which an infinitely thin dust sheet lies at the mid-plane, with stellar layers located above and below it. 
Consequently, in the limit of $\tau_{\mathrm{d}} \to \infty$, the UV escape fraction $f_{\mathrm{UV}}$ saturates at $(1 - \delta)/2$.

The parameters $e_{\tau}$ and $\delta$ are determined by fitting our model UVLF to the observed UVLF at $z = 7$ \citep{Bouwens2021AJ, Harikane2022ApJS}.  
These two parameters are fixed across all redshifts, while $\Sigma_{\mathrm{d}}$ varies with redshift through its dependence on metallicity, thereby modulating the amount of dust attenuation.

After determining the UV attenuation at 1500\,\AA, we apply the attenuation law of \citet{Calzetti2000} to describe its wavelength dependence.  

To quantify the reduction in effective attenuation required by the high-redshift UVLFs, we rescale the fiducial optical depth as $\tau_{\rm d}=R_\tau\tau_{\rm d}^{\rm fid}$ while keeping $\delta$ and the other model parameters fixed. Thus, $R_\tau=1$ recovers the fiducial model and $R_\tau=0$ gives the dust-free limit, independently of $\delta$. For each observational comparison, we use the simulation snapshot nearest to the effective observational redshift and restrict the fit to $M_{\rm UV}<-19$. The full comparisons are presented in Appendix~\ref{app:UVLF_dust_comparison}.

\subsubsection{Alternative attenuation model}\label{subsubsec:Trinca_Dust_model}

We also test an alternative dust attenuation model to examine whether our UVLF results depend on the adopted dust attenuation prescription.
In this model, the observed UV luminosity is written as
\begin{equation}
L_{\mathrm{UV,obs}} = L_{\mathrm{UV}} \exp(-\Sigma_{\mathrm{gas}} \, D \, k_{\mathrm{UV}}),
\label{eq:dust_Trinca}
\end{equation}
where $\Sigma_{\mathrm{gas}}$ is the gas surface density, $D$ is the dust-to-gas mass ratio, and $k_{\mathrm{UV}}$ is the UV extinction coefficient per unit dust mass. Following \citet{Trinca2024MNRAS}, we define
$\Sigma_{\mathrm{gas}} = M_{\mathrm{gas}}/\pi r_{\mathrm{gas}}^{2}$,
where $M_{\mathrm{gas}}$ is the gas mass in our semi-analytic model and $r_{\mathrm{gas}}$ is the characteristic radius of the gas distribution. 
Unlike our fiducial model, this prescription calculates the dust optical depth from the gas surface density and dust-to-gas ratio, assuming a simple screen geometry.

We estimate the dust-to-gas ratio from the gas metallicity using the empirical metallicity--dust relation of equations (8) and (9) of \citet{Galliano2021A&A}. 
We adopt $k_{\mathrm{UV}} = 19 \, \mathrm{pc}^{2}\,\mathrm{M}_{\odot}^{-1}$, based on the SMC-like extinction curve in \citet{Mancini2016MNRAS}. To isolate the impact of this attenuation law, we treat $r_{\mathrm{gas}}$ as the only free parameter and calibrate it at $z=7$. The best-fitting value is then fixed at all other redshifts.

\subsection{Model variants and comparison strategy}\label{subsec:model_variants}

Table~\ref{tab:model_comparison} summarizes the model variants. The star formation history is identical in all cases, so differences can be attributed to the IMF-dependent SED, enrichment, and dust treatment.

The \Kroupa\ is the baseline and uses the Kroupa IMF for both the SED and metal evolution. The \zZKroupa\ is an intentionally non-self-consistent diagnostic control: it applies the \zZIMF\ only to the SED while retaining the Kroupa-based metal history. 
This allows us to assess the direct effect of the \zZIMF\ on intrinsic UV luminosity separately from its effect on metal enrichment.
The \zZzZ\ instead applies the \zZIMF\ to both the SED and metal evolution. It is the self-consistent model and includes the indirect consequences of enhanced metal production and the resulting dust attenuation. 
Comparing \zZKroupa\ and \zZzZ\ therefore allows us to assess how IMF-dependent enrichment and the associated dust attenuation affect the intrinsic UV enhancement.
The corresponding three variants using the alternative \citet{Trinca2024MNRAS}-based attenuation law provide a robustness test of this interpretation.
Here, self-consistency refers specifically to the IMF-dependent stellar emission and metal enrichment, while the star formation and feedback prescriptions remain fixed.

\begin{table*}
    \centering
    \caption{Summary of the model variants and their treatments
    of star formation history, stellar SEDs, metal evolution,
    and dust attenuation.}
    \label{tab:model_comparison}
    \begin{tabular}{ccccc}
        \hline
        \textbf{Model} &
        \textbf{SF history} &
        \textbf{SED} &
        \textbf{Metal evolution} &
        \textbf{Dust attenuation} \\
        \hline
        \Kroupa &
        \multirow{6}{*}{Same} &
        Kroupa IMF &
        \multirow{2}{*}{Kroupa IMF} &
        \multirow{3}{*}{\citet{Shimizu14} at $z=7$} \\
        \cline{3-3}
        \zZKroupa & &
        \multirow{2}{*}{\zZIMF} & & \\
        \cline{4-4}
        \zZzZ & & &
        \zZIMF & \\
        \cline{1-1} \cline{3-5}
        \KroupaDustTrinca & &
        Kroupa IMF &
        \multirow{2}{*}{Kroupa IMF} &
        \multirow{3}{*}{\citet{Trinca2024MNRAS} at $z=7$} \\
        \cline{3-3}
        \zZKroupaDustTrinca & &
        \multirow{2}{*}{\zZIMF} & & \\
        \cline{4-4}
        \zZzZDustTrinca & & &
        \zZIMF & \\
        \hline
    \end{tabular}
\end{table*}

\section{Results}\label{sec:results}

We first examine how the \zZIMF, metal enrichment, and dust attenuation affect the UVLF, and then use galaxy and SED diagnostics to interpret the resulting differences.

\subsection{Intrinsic UVLF and the effect of the $z$-$Z$ dependent IMF}\label{subsec:UVLF_wo_Dust}

\begin{figure}
    \begin{center}
    \includegraphics[width=0.5\textwidth]{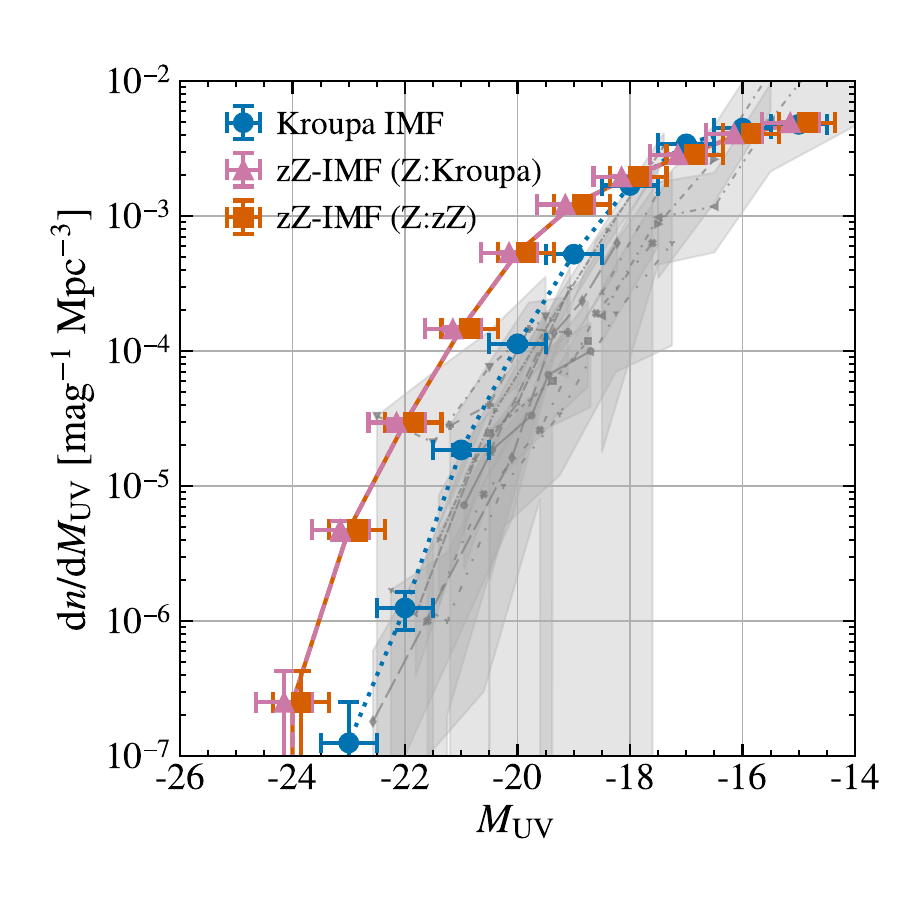}
    \end{center}
    \caption{UVLF at $z=10$ without dust attenuation.
    The orange squares, pink triangles, and blue circles represent the results of \zZzZ, \zZKroupa, and \Kroupa, respectively, with 1$\sigma$ Poisson errors. %The pink triangles and orange squares are horizontally shifted by -0.15 and +0.15 mag, respectively, for clarity; the curves represent the unshifted predictions.
    The UVLFs predicted by \zZzZ\ and \zZKroupa\ are nearly identical, causing their symbols to overlap. To make both sets of data points visible, the pink triangles and orange squares are horizontally shifted by $-0.15$ and $0.15$ mag, respectively. The curves represent the unshifted predictions.
    The two \zZIMF\ models use the same IMF prescription for their stellar SEDs but differ in their metal-enrichment histories. Their nearly identical dust-free UVLFs indicate that the differences in metal enrichment have little effect on the intrinsic UV luminosities. 
    The gray-shaded region represents the observational results \citep{Adams2024ApJ,Asada2025, Bouwens2023MNRAS_a, Castellano2023ApJ, Chemerynska2026MNRAS,Donnan2023MNRAS_a, Finkelstein2023ApJ,Harikane2024ApJ_UVLF, Oesch2018ApJ, Perez-Gonzalez2023ApJ,Weibel2026ApJ,Whitler2025ApJ}.
}
    \label{fig:UVLF_noDust}
\end{figure}

We first examine the intrinsic UVLFs predicted by our semi-analytic model with the IMF-dependent SED templates, without including dust attenuation.
No additional free parameters are introduced or recalibrated for this comparison.
Thus, the differences among the model predictions can be attributed to the adopted IMF, through its effects on the SEDs and, where applicable, the metal-enrichment history.

Figure~\ref{fig:UVLF_noDust} shows the UVLF at $z=10$ in the absence of dust attenuation. The pink, orange, and blue lines represent \zZKroupa, \zZzZ, and \Kroupa, respectively. The gray-shaded region represents the observational constraints.

In the absence of dust attenuation, the \Kroupa\ is consistent with observations at $z \sim 10$ within the error range.  
In contrast, the UVLFs predicted by both the \zZzZ\ and \zZKroupa\ are approximately 1 dex above the observations at the bright end ($M_\mathrm{UV} < -21$).
Furthermore, although the \zZKroupa\ and \zZzZ\
have different gas-phase metallicities,
their predicted UVLFs are nearly identical.
This indicates that differences in metallicity evolution have little effect on the intrinsic UV luminosities in the absence of dust attenuation.

The enhancement of the UVLFs at $M_\mathrm{UV} < -18$ in both \zZKroupa\ and \zZzZ\ relative to \Kroupa\ partly reflects the larger scatter in UV emissivity, $L_{\rm UV}/{\rm SFR}$, in the \zZIMF\ models.
This larger scatter arises from the stronger age dependence of the UV luminosity in the \zZIMF, which produces higher UV luminosities at young stellar ages but lower luminosities at older ages compared to the Kroupa IMF. Consequently, differences in stellar age distributions among galaxies lead to a broader distribution of $L_{\rm UV}/{\rm SFR}$, even though the same star formation histories are adopted across the IMF models. As shown in Figure~\ref{fig:conversion_factor}, this broadening contributes to a larger fraction of galaxies with high UV emissivity, thereby enhancing the bright end of the UVLF.

\subsection{IMF-dependent metal enrichment and dust attenuation}\label{subsec:metal_dust_effects}

We next examine how IMF-dependent metal enrichment affects galaxy properties and the resulting dust attenuation. 
We first compare the galaxy scaling relations and then present the calibration of the dust attenuation model against the observed UVLF at $z=7$.

\subsubsection{Galaxy properties and metal enrichment}\label{subsubsec:galaxy_prop_metal}

The control model (\zZKroupa) and the self-consistent \zZIMF\ model (\zZzZ) share the same IMF-dependent SEDs but differ in their metal-enrichment histories. Their comparison therefore allows us to assess the effects of IMF-dependent enrichment on the UVLF, including its impact on dust attenuation. 
The \Kroupa\ and \zZKroupa\ models share the same metal-enrichment history.
We therefore compare \Kroupa\ and \zZzZ\ to examine the differences in galaxy properties.

\begin{figure}
    \begin{center}

    \includegraphics[width=0.5\textwidth]{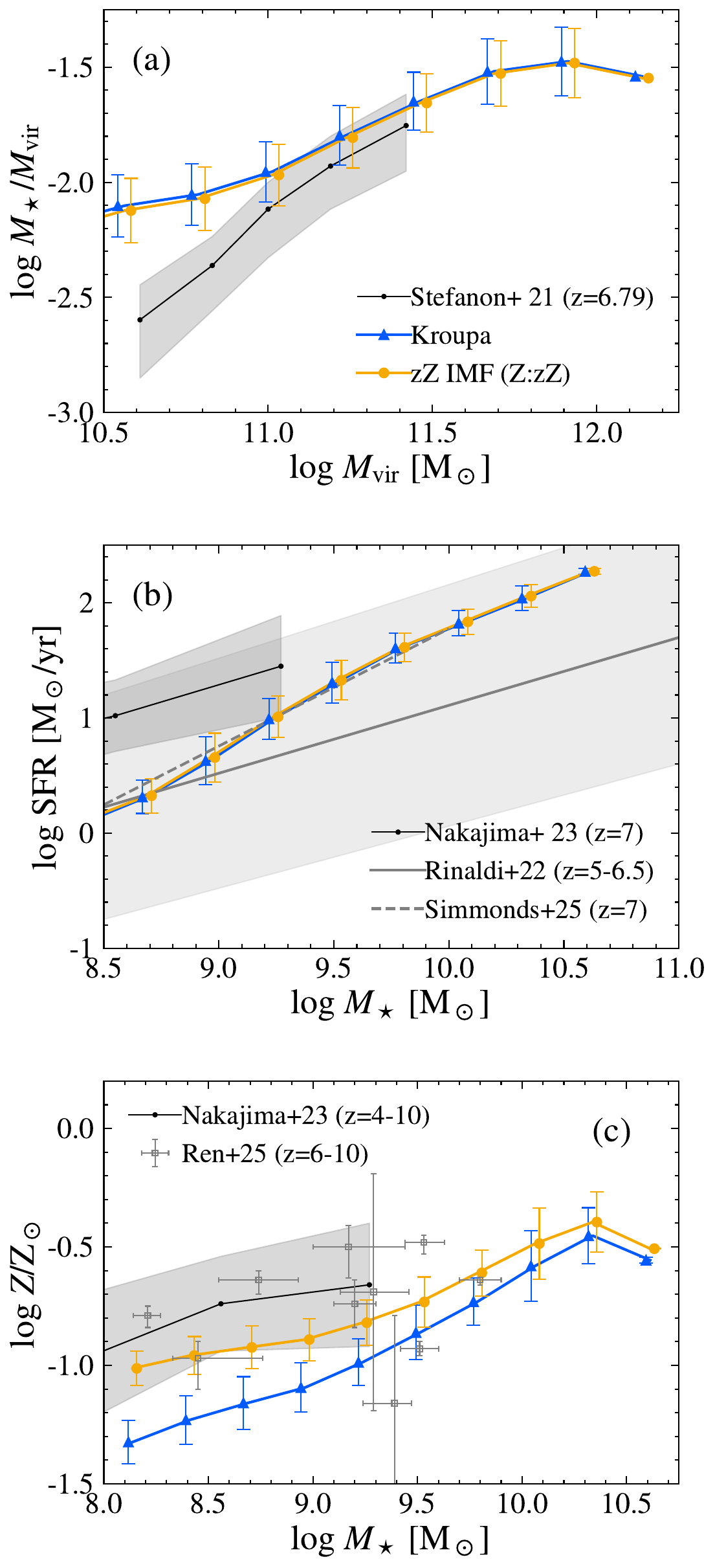}
    \end{center}
    \caption{The stellar mass-halo mass relation (panel~{a}), the star formation main sequence (panel~{b}), and the stellar mass-gas metallicity relation (panel~{c}) of our semi-analytic model at $z=7$. 
    The yellow and blue lines represent \zZzZ\ and \Kroupa, respectively, with $1\sigma$ error bars.
    %The yellow circles and blue triangles are horizontally shifted by -0.02 and +0.02 mag for panel~{a} and {b}, respectively, for clarity; the curves represent the unshifted predictions.
    In all three panels, the yellow circles and blue triangles are horizontally shifted by $-0.02$ and $+0.02$ dex, respectively, for clarity; the curves represent the unshifted predictions.
    The observational results are shown by the black and gray symbols, lines, and shaded regions: 
    \citet{Stefanon2021ApJ_SHMR} (panel~{a}), \citet{Nakajima2023ApJS_SFMS_MZR,Rinaldi2022ApJ,Simmonds2025MNRAS} (panel~{b}), and \citet{Nakajima2023ApJS_SFMS_MZR,Ren2025MNRAS} (panel~{c}).}
    \label{fig:SHMR_MS_MZR}
\end{figure}

Figure~\ref{fig:SHMR_MS_MZR} shows the stellar mass--halo mass relation (SHMR), star formation rate--stellar mass relation (SFR--$M_\star$ relation), and stellar mass--gas metallicity relation (MZR) predicted by our semi-analytic model at $z=7$, together with observational results.
The observational sample compiled by
\citet{Ren2025MNRAS}\footnote{
This sample includes galaxies from
\citet{Mawatari2025arXiv, Harshan2024ApJ,
Heintz2023ApJ, Nakajima2023ApJS_SFMS_MZR,
Harikane2025ApJ, Scholtz2025MNRAS,
Kiyota2025ApJ, Stiavelli2023ApJ,
Morishita2023ApJ_GLASS_JWST, Venturi2024A&A,
Helton2025NatAs, Marconcini2025A&A,
Witten2025arXiv, Rowland2026MNRAS,
Fisher2025MNRAS}, in addition to the
galaxies presented by Ren et al. (2025).
}
is used for comparison with the MZR.
We assume a solar metallicity of
$12+\log\left(\mathrm{O/H}\right)=8.69$
and $\Zsun = 0.0139$ \citep{Asplund2021A&A}.

In panel~(a), the SHMR is consistent with observations within the error range for $M_{\rm halo} \sim 10^{11} - 10^{11.4} M_{\odot}$.  
The \zZzZ\ shows a slightly smaller
$M_{\star}/M_{\rm halo}$ than the \Kroupa.
Despite having the same star formation history,
this difference arises because the gas return fraction $f$ in the gas evolution equation
(equation~\ref{eq:mdot_gas} in Appendix~6)
depends on the IMF.
A more top-heavy IMF leads to a larger $f$,
resulting in a greater fraction of the stellar
mass being returned to the gas.
Consequently, the stellar mass retained in galaxies is slightly lower in the \zZzZ.

In panel~(b), our results are consistent with the observations of \citet{Simmonds2025MNRAS} in the stellar-mass range $M_{\star} \lesssim 10^{10}\,\Msun$.
In contrast, our model predicts a slightly steeper
star-formation main sequence (SFMS) than those reported by
\citet{Nakajima2023ApJS_SFMS_MZR}
and \citet{Rinaldi2022ApJ}.
A similarly steep slope is also seen in the
FIRE-2 simulations, from which the SFE adopted in this study is derived
\citep{Feldmann2025MNRAS}.

In panel~(c), the MZR of the \zZzZ\ is consistent with the observations within the error range, while that of the \Kroupa\ is approximately 0.5~dex lower than the observed values of \citet{Nakajima2023ApJS_SFMS_MZR}.  
This is because a top-heavy IMF produces
more metals per unit stellar mass formed
(see Appendix~\ref{app:appendix_Yield}),
leading to more rapid chemical enrichment.
The metallicity difference between the two models decreases toward higher stellar masses.
Compared with the results of \citet{Ren2025MNRAS}, both models are consistent with the observations within the error ranges in the massive-galaxy regime ($M_\star\gtrsim10^{9.3}\,\Msun$).
In both models, the metallicity shows a slight
decrease over the stellar-mass range
$10^{10.3}$--$10^{10.6}\,\Msun$.

\subsubsection{Calibration of dust attenuation at $z=7$}\label{subsubsec:UVLF_z7_dust_calib}

\begin{figure}
    \begin{center}
    \includegraphics[width=0.5\textwidth]{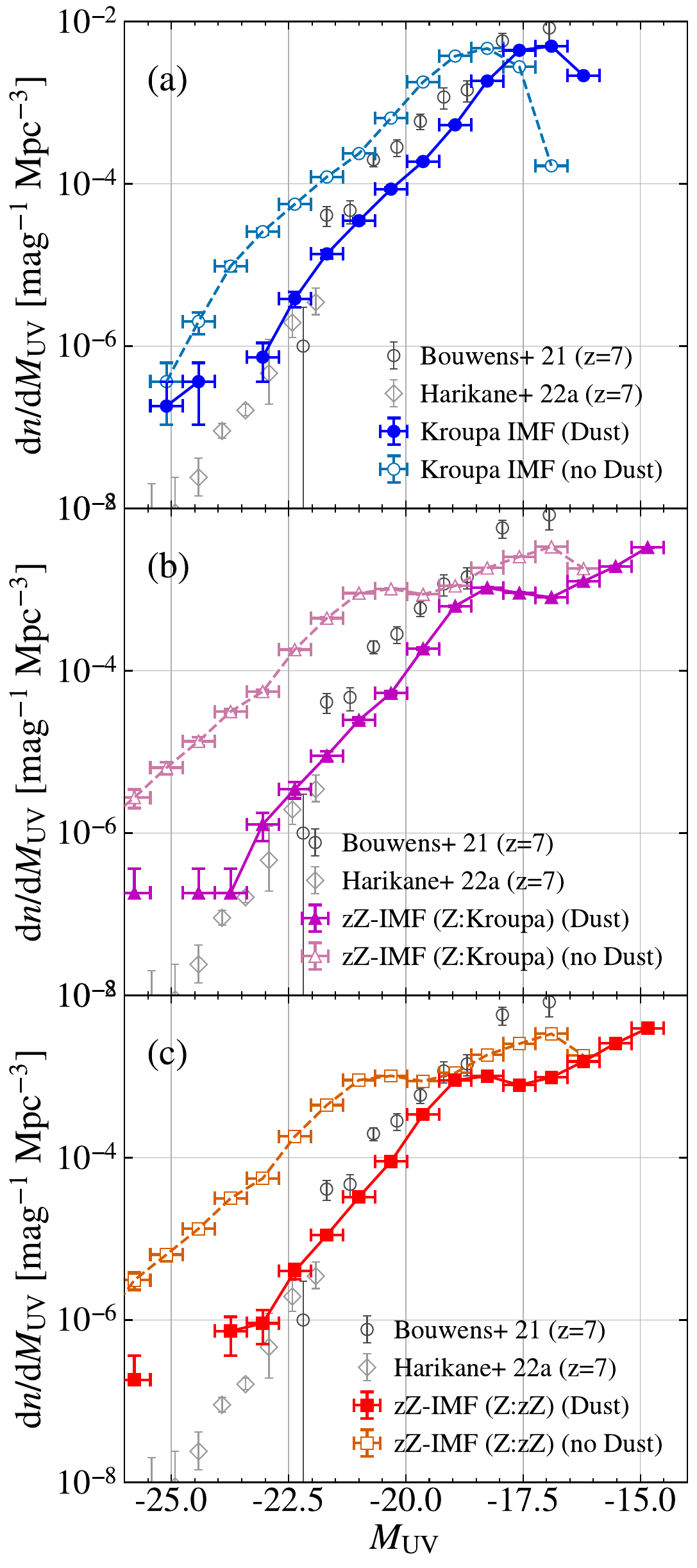}
    \end{center}
    \caption{UVLF at $z=7$ with and without dust attenuation.
    Filled and open symbols with error bars represent model predictions with and without dust attenuation, respectively.
    Observed UVLFs from \citet{Bouwens2021AJ} and \citet{Harikane2022ApJS} are shown by the yellow and light-blue solid lines, respectively.
    From top to bottom, the panels correspond to \Kroupa\ (panel~a), \zZKroupa\ (panel~b), and \zZzZ\ (panel~c). The dust attenuation parameters are calibrated over $-24 \le M_{\rm UV} \le -17$.
    }
    \label{fig:UVLF_dust_calib}
\end{figure}

Figure~\ref{fig:UVLF_dust_calib} shows the UVLFs
with and without dust attenuation at $z=7$,
together with observations. From top to bottom, the panels correspond to \Kroupa\ (panel~a),
\zZKroupa\ (panel~b), and \zZzZ\ (panel~c), respectively.
The best-fit values of $(\delta, e_\tau)$
are $(0.674, 2340)$, $(0.919, 7439)$,
and $(0.951, 2002)$ for the \Kroupa,
\zZKroupa, and \zZzZ, respectively.
For galaxies with intrinsic $M_{\rm UV}<-22$ at $z=7$, we compute the median dust optical depth, $\tau_{\rm d,\, med}$, and use it to estimate the corresponding UV escape fraction, $f_{\rm UV}$, from equation~\ref{eq:dust_shimizu14}, and UV attenuation, $A_{\rm UV}$. 
The values of
$(\tau_{\rm d,med}, f_{\rm UV}, A_{\rm UV}/{\rm mag})$
are $(25.07, 0.19, 1.80)$, $(37.07, 0.0650, 2.97)$,
and $(16.83, 0.0808, 2.73)$
for the \Kroupa, \zZKroupa, and \zZzZ, respectively.
Thus, reproducing the observed UVLF at $z=7$ requires stronger effective dust attenuation in the two \zZIMF\ models than in the \Kroupa.

This requirement can be understood in terms of
the bright-end enhancement of the UVLFs
predicted by the \zZIMF\ models.
In our UVLF modeling, star formation is assumed to occur in a single instantaneous burst during each simulation snapshot interval, resulting in a broad distribution of intrinsic UV luminosities.
This broad distribution, combined with the enhanced UV emissivity of the top-heavy IMF, increases the abundance of UV-bright galaxies, which must be offset by stronger dust attenuation to reproduce the observed UVLF.

\subsection{Dust-attenuated UVLFs across redshift
}\label{subsec:UVLF_w_Dust}

\subsubsection{Fiducial dust attenuation model}\label{subsubsec:UVLF_dust_fiducial}

\begin{figure}
    \begin{center}    
    \includegraphics[width=0.48\textwidth]{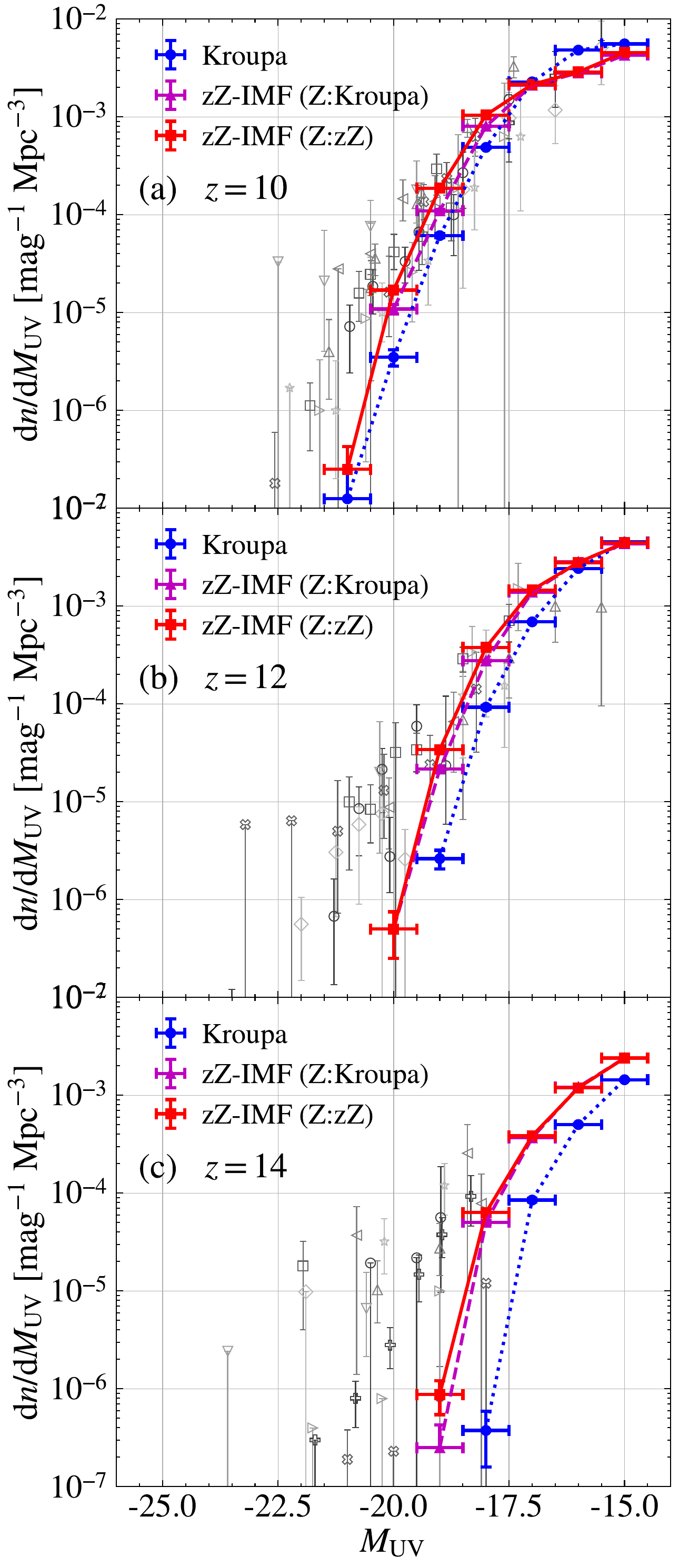}
    \end{center}
    \caption{UVLFs with dust attenuation at $z = 10$ (panel~a), $z = 12$ (panel~b), and $z = 14$ (panel~c). 
    The red squares, purple triangles, and blue circles represent the predictions of \zZzZ, \zZKroupa, and \Kroupa, respectively, with 1$\sigma$ Poisson error bars.
    %The corresponding predictions without dust attenuation are shown by the light-gray symbols.
    The gray-shaded regions represent the observational results \citep{Adams2024ApJ,Asada2025, Bouwens2023MNRAS_a, Castellano2023ApJ, Chemerynska2026MNRAS, Donnan2023MNRAS_a,  Finkelstein2022ApJ, Finkelstein2023ApJ, Franco2025arXiv250804791F, Harikane2024ApJ_UVLF, Oesch2018ApJ, Perez-Gonzalez2023ApJ, Harikane2022ApJ_b, Harikane2023a_ApJS, McLeod2026arXiv260416666M,Robertson2024ApJ, Weibel2026ApJ,Whitler2025ApJ}.
}
    \label{fig:UVLF_result_with_Dust}
\end{figure}

Figure~\ref{fig:UVLF_result_with_Dust}
shows the dust-attenuated UVLFs predicted by
our three models at $z=10$ (panel~a),
$z=12$ (panel~b), and
$z=14$ (panel~c),
together with the observational constraints.
The red squares, purple triangles, and blue circles
represent \zZzZ, \zZKroupa, and \Kroupa,
respectively.

At $z=10$, both \zZzZ\ and \zZKroupa\ are broadly consistent with the observations within their uncertainties, although their predicted number densities lie slightly below the central observational estimates at $M_{\rm UV}\simeq-21$.
At $z=12$, both \zZzZ\ and \zZKroupa\ are consistent with the observational estimates at $M_{\rm UV}\simeq-19$, but underpredict the number density at $M_{\rm UV}\simeq-20$ by approximately 1 dex.
At $z=14$, both models agree with the $z=13.5$ UVLF estimates of
\citet{McLeod2026arXiv260416666M}
at $M_{\rm UV}\simeq-18$, but their predicted
number densities at $M_{\rm UV}\simeq-19$
are more than 1 dex below the observational estimates.
Nevertheless, although the redshifts are not directly matched, the model predictions lie relatively close to the lower number-density estimates at $z\simeq15.8$ reported by
\citet{McLeod2026arXiv260416666M}.

Similar to Figure~\ref{fig:UVLF_noDust}, both $zZ$-IMF models exhibit higher $dn/dM_{\rm UV}$ than \Kroupa.  
However, in contrast to Figure~\ref{fig:UVLF_noDust}, the \zZzZ\ predicts number densities more than 0.2 dex higher than \zZKroupa\ at $M_\mathrm{UV} < -20$ for $z \sim 10$ and $z \sim 12$, and at $M_\mathrm{UV} < -19$ for $z \sim 14$.  
Although \zZzZ\ predicts higher gas-phase metallicities than \zZKroupa\ (see Figure~\ref{fig:SHMR_MS_MZR}), the dust attenuation parameters are calibrated separately for each model at $z=7$. In particular, the best-fit value of $e_\tau$ is larger in \zZKroupa\ than in \zZzZ, compensating for its lower metal content. Consequently, the difference in the dust-attenuated UVLFs reflects the combined effects of metal enrichment and the calibrated dust attenuation parameters, rather than metallicity alone.

Thus, while adopting a top-heavy IMF enhances the bright end of the intrinsic UVLF, the accompanying dust attenuation partly offsets this enhancement. Neither model fully reproduces the observed UVLF, particularly at $z \gtrsim 12$.

Because the Shin-Uchuu box is approximately $200\,\mathrm{Mpc}$ on a side, number densities below $\sim10^{-6}\,\mathrm{Mpc}^{-3}\,\mathrm{mag}^{-1}$ are sampled by very few objects. Predictions in this regime, especially at the brightest magnitudes, should therefore be interpreted cautiously even when statistical error bars are shown.

\subsubsection{Robustness to the attenuation prescription}\label{subsubsec:UVLF_dust_trinca}

\begin{figure}
    \begin{center}
    \includegraphics[width=0.5\textwidth]{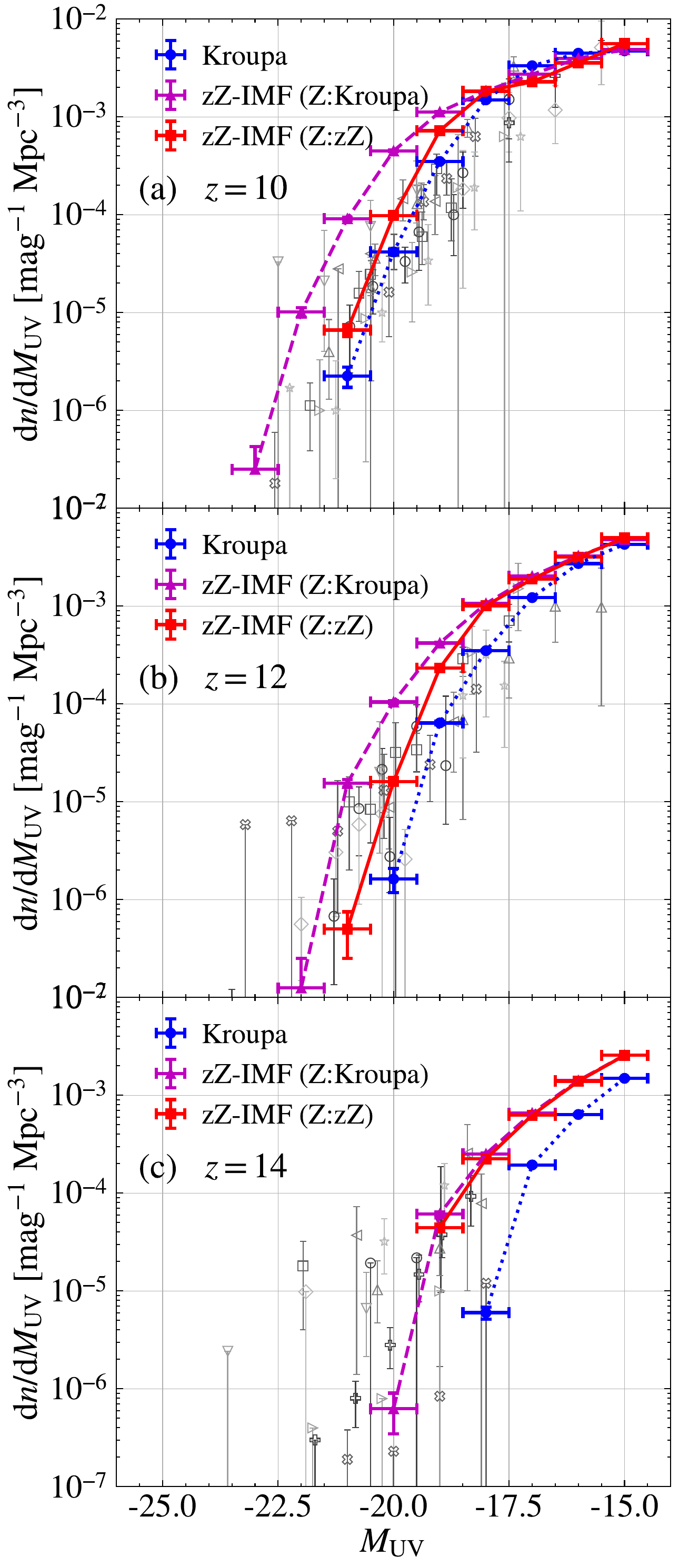}
    \end{center}
    \caption{Same as Figure~\ref{fig:UVLF_result_with_Dust}, but using the dust attenuation model of \citet{Trinca2024MNRAS}.}
    \label{fig:UVLF_result_with_Dust_Trinca}
\end{figure}

Figure~\ref{fig:UVLF_result_with_Dust_Trinca} shows the same comparison as in Figure~\ref{fig:UVLF_result_with_Dust}, but using the dust attenuation model of \citet{Trinca2024MNRAS}. 

In this model, the difference between the two \zZIMF\ models can be understood more directly in terms of metallicity-dependent dust content. Specifically, the \zZKroupaDustTrinca\ model predicts lower metallicities than the \zZzZDustTrinca\ model. This results in weaker dust attenuation and hence a higher observed UVLF.
By contrast, the physically self-consistent \zZzZDustTrinca\ model predicts a UVLF relatively close to that of the Kroupa result, without a large enhancement at the bright end.
The enhancement of the intrinsic UVLF expected from a top-heavy IMF can be substantially reduced by dust attenuation when metal enrichment is treated self-consistently.

The similarity of the results obtained with the two dust attenuation prescriptions supports the robustness of this qualitative conclusion.
The corresponding $z=7$ calibration is shown in Appendix~\ref{app:UVLF_dust_comparison}.

\subsection{Required effective dust attenuation at high redshift}\label{subsec:dust_required}

We next vary the dust optical-depth scaling factor $R_\tau$ for the self-consistent \zZzZ\ while leaving its galaxy properties and intrinsic UV luminosities unchanged. 
The best-fitting values of $R_\tau$ are 0.51 and 0.21 for the \citet{Weibel2026ApJ} sample at $z=9.7$ and the \citet{Franco2025arXiv250804791F} sample at $z=13.04$, respectively, corresponding to reductions in the fiducial dust optical depth of 49\% and 79\%.
The inferred values are sensitive to the adopted observational sample and magnitude range. For example, restricting the fit to the \citet{Weibel2026ApJ} sample at $z=9.7$ to $M_{\rm UV}\leq-20$ changes the best-fitting value from 0.51 to 0.35. Moreover, alternative observational samples at similar redshifts yield substantially different values of $R_\tau$. Therefore, the inferred attenuation strengths should be regarded as model- and sample-dependent estimates rather than precise constraints on dust evolution. Further details of the fitting procedure are provided in Appendix~\ref{app:UVLF_dust_comparison}.

Comparing the \zZzZ\ prediction at
$z=14.3$ with the observations at $z=13.5$
\citep{McLeod2026arXiv260416666M}
yields a formal minimum near $R_\tau=0.04$.
We do not regard this value as a robust constraint because too few bright magnitude bins are adequately sampled by the model, and the predicted number densities remain below the observed values even close to the dust-free limit. The full UVLF comparisons and the observational data used in each panel are shown in Figure~\ref{fig:UVLF_all_redshifts_zZzZ_dust_comparison} in Appendix~\ref{app:UVLF_dust_comparison}. Because $R_\tau$ is an effective rescaling of the attenuation, it should not be interpreted directly as a measurement of the dust mass or dust-to-metal ratio.

\begin{figure}
    \begin{center}
    \includegraphics[width=0.5\textwidth]{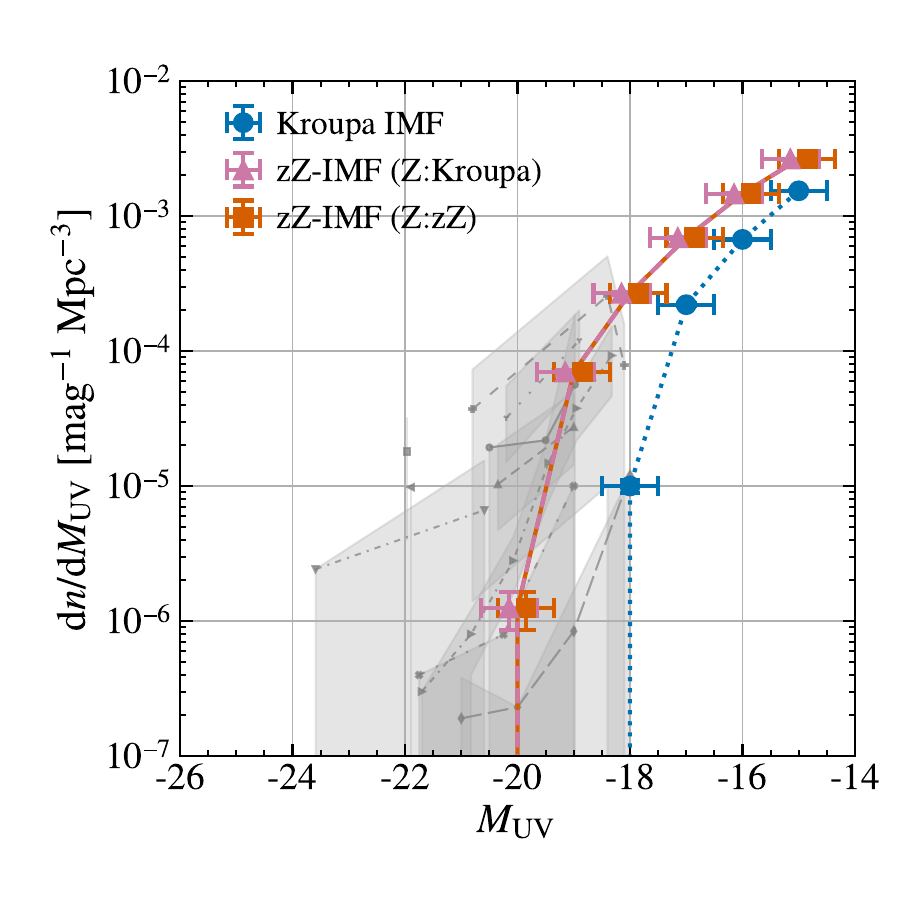}
    \end{center}
\caption{UVLF at $z\sim14$ without dust attenuation, compared with the observational data shown in the $z\sim14$ panel of Figure~\ref{fig:UVLF_result_with_Dust}. The colors and symbols follow the same conventions as in Figure~\ref{fig:UVLF_result_with_Dust}. The predictions of \zZKroupa\ and \zZzZ\ nearly overlap.
    %The orange squares and line represent the predictions of \zZzZ, while the blue circles and line represent those of  \zZKroupa.
}

\label{fig:UVLF_z14_wo_dust}
\end{figure}

Figure~\ref{fig:UVLF_z14_wo_dust} presents the UVLFs predicted by our models at $z=14$ without dust attenuation.
At $M_\mathrm{UV}\simeq-20$, the dust-free number densities predicted by \zZKroupa\ and \zZzZ\ approach the range of recent observational constraints at $z\gtrsim13$, but do not reproduce all measurements.
The JADES analysis of \citet{Whitler2025ApJ} reports comparatively high number densities over $z\sim12$--16.
Bearing in mind the differences in effective redshift, our $z=14$ predictions lie between the wider-area results of \citet{McLeod2026arXiv260416666M}: the dust-free \zZzZ\ prediction lies slightly below their $z=13.5$ measurement, whereas the dust-free \zZKroupa\ and \zZzZ\ predictions both lie above their tentative $z=15.8$ upper limit at $M_\mathrm{UV}\simeq-20$.
Both model predictions also remain consistent with the less stringent $z\simeq17$ upper limit from PANORAMIC \citep{Weibel2026ApJ}.
Thus, reducing dust attenuation alone cannot fully reconcile our model predictions with all the observational constraints at $z\gtrsim13$.

\subsection{Additional diagnostics of UV emission and dust attenuation}\label{subsec:supporting_properties}
Here, we examine three additional diagnostics: the UV luminosity-to-SFR conversion factor and its scatter, the IRX--$\beta$ relation, and the $M_{\rm UV}$--$\beta$ relation.
These diagnostics help us understand the effects of the IMF, dust attenuation, and nebular emission on the UV properties of galaxies.
More extensive SSP diagnostics are presented in Appendix~\ref{app:sed_diagnostics}.

\subsubsection{UV luminosity-to-SFR conversion factor}
\label{subsubsec:uv_sfr_conversion}

\begin{figure}
    \begin{center}
    \includegraphics[width=0.5\textwidth]{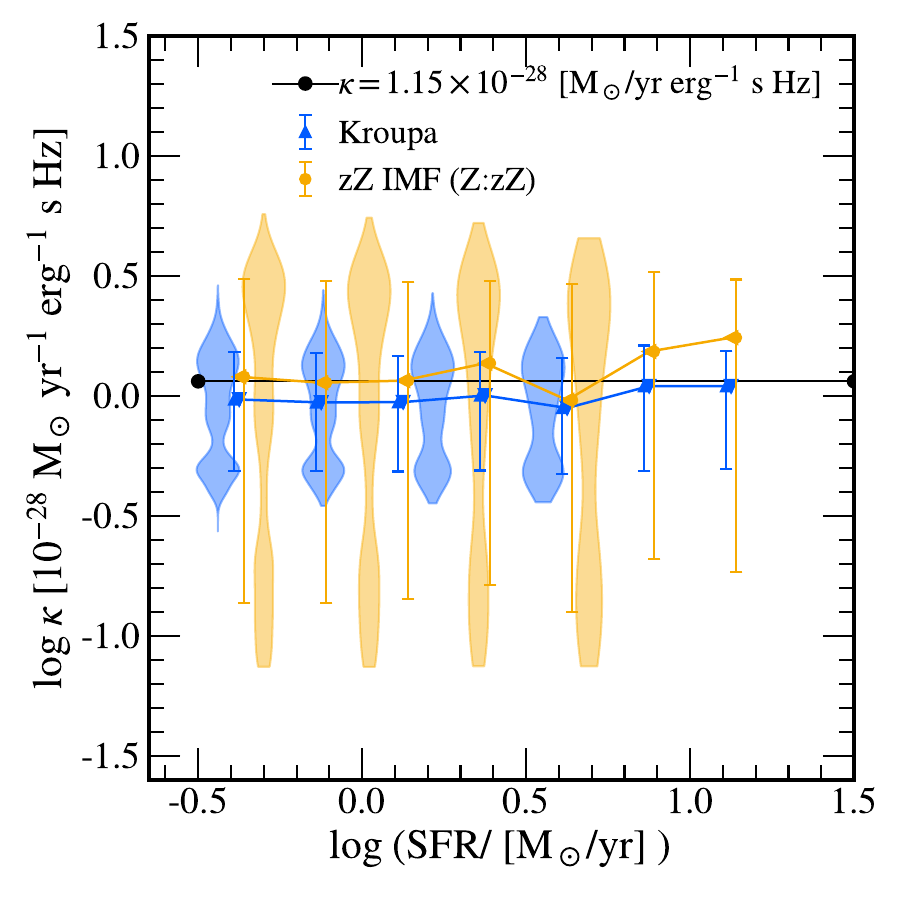}
    \end{center}
    \caption{Relationship between the SFR and the UV luminosity-to-SFR conversion factor $\kappa=\mathrm{SFR}/L_\mathrm{UV}$ at $z=10$, where $L_\mathrm{UV}$ is the intrinsic specific luminosity at 1500 \AA.
    The orange and blue lines represent the median values for \zZzZ\ and \Kroupa, respectively, while the violin plots show the distributions of $\kappa$.
    For clarity, the median markers for \zZzZ\ and \Kroupa\ are horizontally shifted by $+0.015$ and $-0.015$ dex, respectively, from the centers of the 0.25-dex bins, while the curves remain unshifted.
    The violin plots are computed in four broader SFR bins and positioned at the median $\log\mathrm{SFR}$ of the galaxies in each bin, with horizontal shifts of $+0.07$ and $-0.07$ dex for \zZzZ\ and \Kroupa, respectively.
    The highest-SFR bin spans $0.5\leq\log(\mathrm{SFR}/[\mathrm{M_\odot\,yr^{-1}}])\leq1.5$.
    The commonly adopted value, $\kappa=1.15\times10^{-28}\,\mathrm{M_\odot\,yr^{-1}\,erg^{-1}\,s\,Hz}$ \citep[e.g.,][]{Madau14,Harikane2022ApJS}, is shown by the black solid line.
    The \zZzZ\ model exhibits a larger scatter in $\kappa$ than the \Kroupa\ model.
     }

    \label{fig:conversion_factor}
\end{figure}

Figure~\ref{fig:conversion_factor} shows the relationship between the SFR and the UV-to-SFR conversion factor $\kappa=\mathrm{SFR}/L_\mathrm{UV}$ at $z=10$, where $L_\mathrm{UV}$ is the intrinsic specific luminosity at 1500 \AA. 

The median $\kappa$ values in both models are approximately consistent with the value adopted by \citet{Harikane2022ApJS}, although \zZzZ\ exhibits a larger dispersion than \Kroupa.
The larger dispersion in the \zZzZ\ arises because young stellar populations with a top-heavy IMF are brighter in the UV, whereas older stellar populations become fainter (see Figure~\ref{fig:SSP_SED_result}
in Appendix~\ref{sec:appendix_SED_temp}).

The larger dispersion in $\kappa$ for \zZzZ\ compared to \Kroupa\ arises despite adopting the same star formation histories in both models, suggesting that differences in the IMF-dependent stellar emission amplify the scatter in UV luminosity at a given SFR. This enhanced scatter may contribute to the dispersion in the $M_\mathrm{UV}$--$M_\mathrm{halo}$ relation, which has been proposed as a possible explanation for the bright-end excess in the UVLF \citep[e.g.,][]{Mason2023MNRAS}.  

In our semi-analytic model, the SFR depends on the dark matter mass accretion rate, which is determined by $M_\mathrm{halo}$ and redshift (see Appendix~\ref{subsubsec:semiana_model_stellarmass}).  
As a result, the SFR–$M_\mathrm{halo}$ relation does not exhibit significant dispersion.  
However, the intrinsic scatter in $\kappa$ at a given SFR introduces scatter in the $M_\mathrm{UV}$--$M_\mathrm{halo}$ relation, affecting the shape of the UVLF.
\citet{Shen2023MNRAS} showed that the observed UVLFs can be reproduced if the scatter in the $M_{\rm UV}$–$M_{\rm halo}$ relation, $\sigma_{\rm UV}$, is set to 0.75, 1.5, 2.0, and 2.5 at $z = 9$, 10, 12, and 16, respectively.  
In the \zZIMF\ model, $\sigma_{\mathrm UV}$ at $z = 10$ is approximately 0.6, indicating that this scatter alone is insufficient to fully reproduce the observed UVLF at $z=10$.

Therefore, in our UVLF analysis, the top-heavy IMF affects the UVLF through changes in the distribution of $L_\mathrm{UV}$ per unit SFR, particularly its increased scatter, which broadens the UV luminosity distribution at fixed SFR.

\subsubsection{IRX--$\beta$ relation}
\label{subsubsec:irx_beta}

\begin{figure*}
    \begin{center}
    \includegraphics[width=0.95\textwidth]{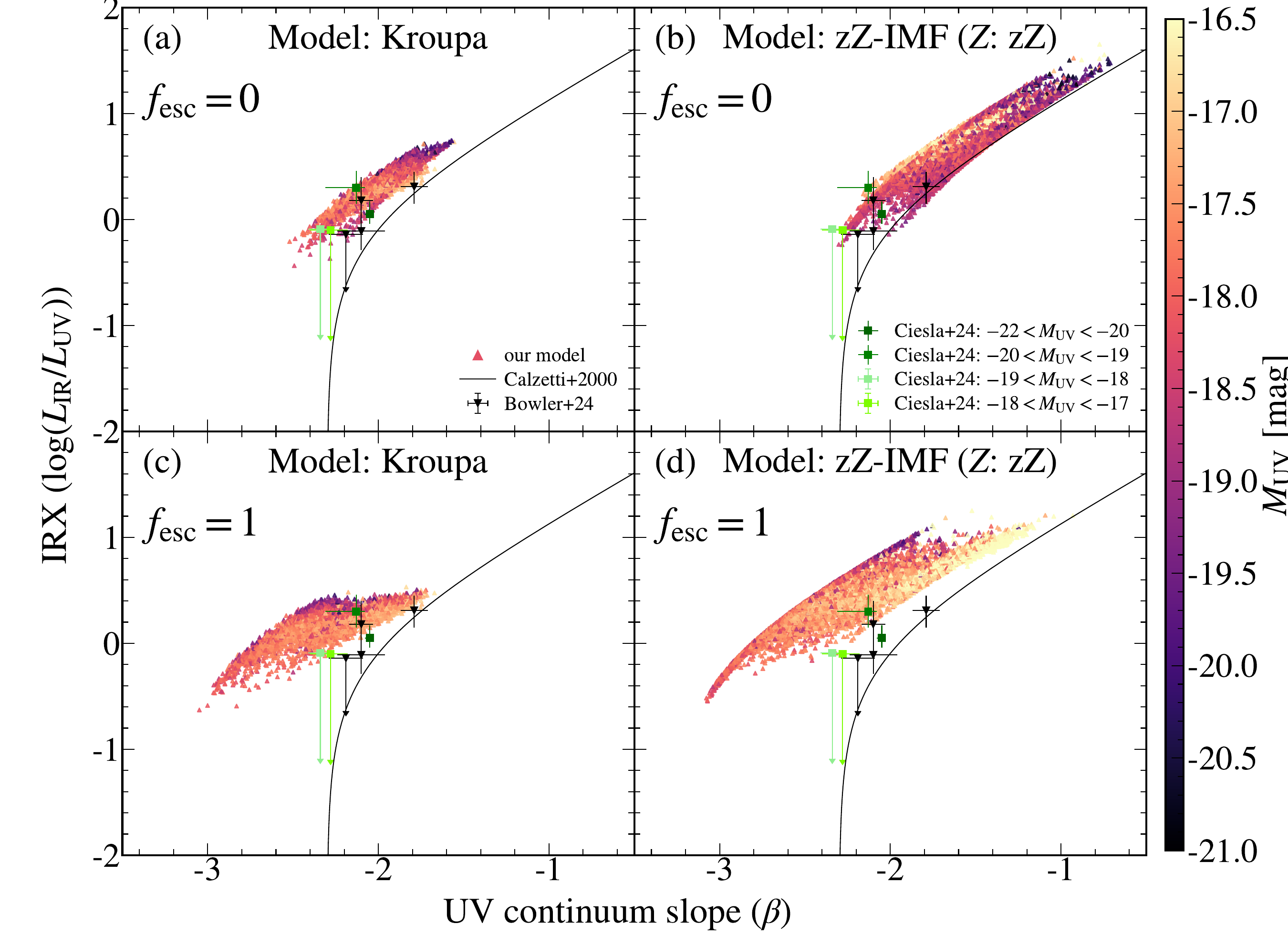}
    \end{center}
    \caption{IRX–$\beta$ relation at $z = 10$ with dust attenuation.  
    The left and right columns show the predictions of \Kroupa\ and \zZzZ, respectively.
    The upper and lower rows correspond to $f_{\mathrm esc}=0$ and $f_{\mathrm esc}=1$, respectively.
    The black line represents the reference relation obtained using the attenuation law of \citet{Calzetti2000}, assuming an intrinsic UV slope of $\beta = -2.3$.  
    Black downward-pointing triangles with error bars represent the observations of \citet{Bowler2024}, while green squares with error bars show those of \citet{Ciesla2024arXiv}.  
    The model galaxies are color-coded by $M_\mathrm{UV}$, as indicated by the color bars.
    }
    \label{fig:IRX-beta}
\end{figure*}

To investigate the effects of dust attenuation
and nebular emission on the UV properties of galaxies,
we examine the IRX--$\beta$ relation.
We compute the UV continuum slope $\beta$
from the dust-attenuated UV SED.
IRX is computed from the ratio of the infrared luminosity to the dust-attenuated UV luminosity. 
Because we do not explicitly model the full infrared dust-emission SED, we estimate the infrared luminosity from the stellar luminosity absorbed by dust, obtained by integrating the difference between the intrinsic and dust-attenuated SEDs over wavelength.
This absorbed luminosity is assumed to be re-emitted in the infrared.

Figure~\ref{fig:IRX-beta} shows that the IRX--$\beta$ distribution depends strongly on both the escape fraction and the adopted IMF. The left panels present the \Kroupa\ results, while the right panels show the \zZzZ\ results. The upper panels correspond to $f_\mathrm{esc}=0$, and the lower panels correspond to $f_\mathrm{esc}=1$. 
The model galaxies are color-coded by $M_\mathrm{UV}$.

For comparison, we include the observations
of \citet{Ciesla2024arXiv} and \citet{Bowler2024},
along with the reference IRX--$\beta$ relation
derived using the attenuation law of
\citet{Calzetti2000}, assuming an intrinsic
UV slope of $\beta=-2.3$.

The escape fraction has a strong effect on the UV slope. Compared with the $f_\mathrm{esc}=1$ case, the $f_\mathrm{esc}=0$ models have systematically redder $\beta$ because nebular emission contributes more strongly when ionizing photons do not escape. 
This effect is more pronounced for UV-fainter galaxies
(i.e., those with higher $M_\mathrm{UV}$),
suggesting a larger relative contribution of
nebular emission to their UV continuum.
By contrast, the $f_\mathrm{esc}=1$ models extend to much bluer $\beta$ values. This shows that the escape fraction controls not only the mean UV slope but also the blue edge of the IRX--$\beta$ distribution.

The IMF dependence is strongest in the $f_\mathrm{esc}=0$ case.
Under this condition, \Kroupa\ extends to lower $\beta$ values and produces more galaxies with $\beta<-2.5$ than \zZzZ.
This is because the top-heavy IMF in the \zZzZ\ produces more ionizing photons, enhancing the relative contribution of nebular emission and shifting $\beta$ to redder values.
In the $f_\mathrm{esc}=1$ case, this IMF dependence becomes much weaker. This indicates that the difference between the two IMF models in the IRX--$\beta$ plane is driven mainly by nebular emission rather than by the stellar UV continuum alone.

The observed trend is broadly more consistent with the $f_\mathrm{esc}=0$ models. The data from \citet{Ciesla2024arXiv} show that $\beta$ increases with $M_\mathrm{UV}$, and this behavior is broadly reproduced when nebular emission is included. 
This suggests that nebular emission may play
an important role in explaining the observed trend.
At the same time, the simulated galaxies do not lie on a single tight IRX--$\beta$ relation. Instead, they show a broad distribution even at fixed $M_\mathrm{UV}$. This implies that the location of galaxies in the IRX--$\beta$ plane is affected not only by dust attenuation but also by intrinsic variations in stellar populations and nebular emission.

\subsubsection{$M_{\rm UV}$--$\beta$ relation}
\label{subsubsec:muv_beta}

\begin{figure*}
    \begin{center}
    \includegraphics[width=\textwidth]{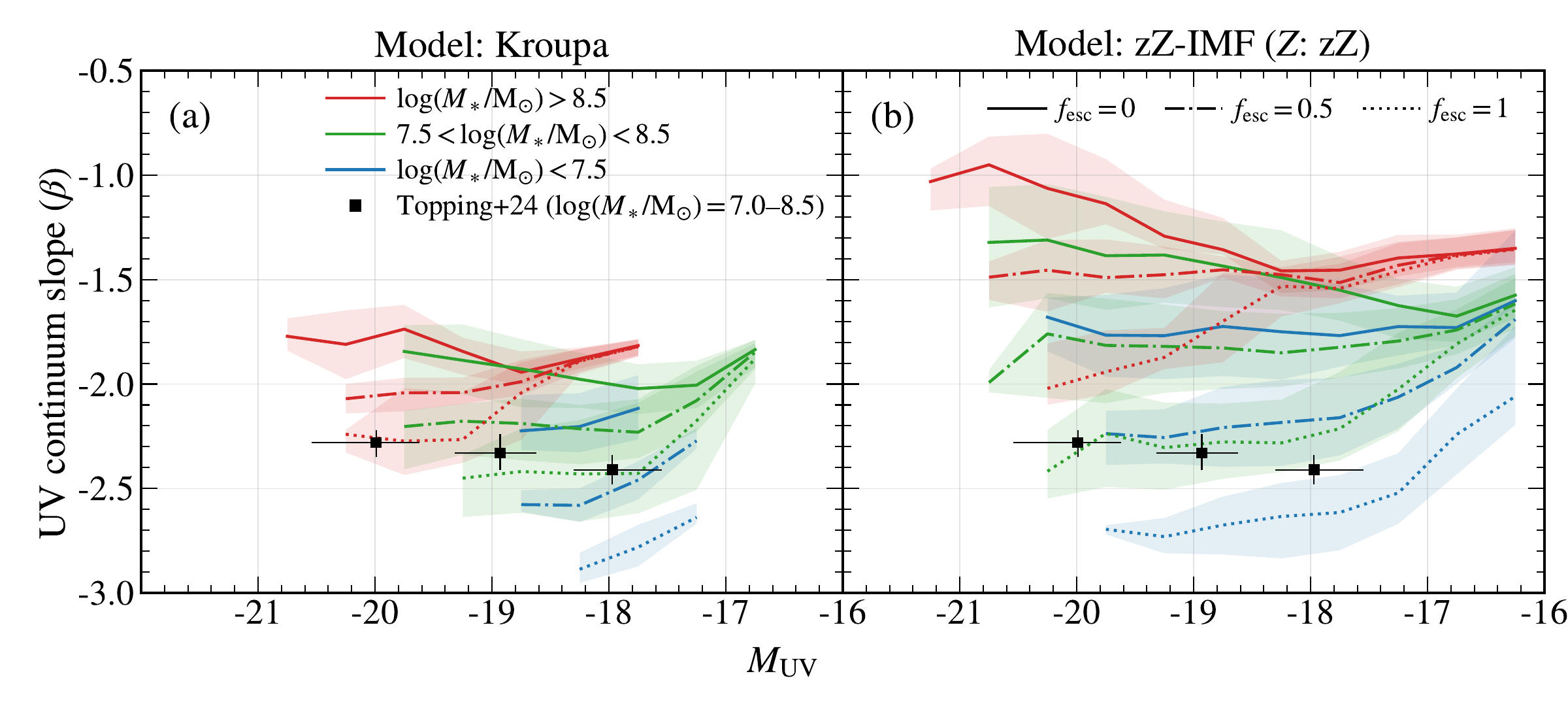}
    \end{center}
    \caption{
    Relation between $M_{\mathrm{UV}}$ and $\beta$ at $z=10$. The line colors indicate stellar-mass bins: red for $\log(M_\star/\Msun) > 8.5$, green for $7.5 < \log(M_\star/\Msun) < 8.5$, and blue for $\log(M_\star/\Msun) < 7.5$. The line styles denote the escape fraction $f_{\mathrm{esc}}$: solid for $f_{\mathrm{esc}}=0$, dash-dotted for $f_{\mathrm{esc}}=0.5$, and dotted for $f_{\mathrm{esc}}=1$. 
    The panel~a and panel~b show the results of \Kroupa\ and \zZzZ, respectively.
    The black points with error bars represent the observations of \citet{Topping2024MNRAS}.
    }

    \label{fig:Muv-beta}
\end{figure*}

To further examine the dependence of the UV slope on galaxy properties, we investigate the $M_{\rm UV}$--$\beta$ relation.
Figure~\ref{fig:Muv-beta} shows the predicted relations at $z=10$ for \Kroupa\ (panel~a) and \zZzZ\ (panel~b), compared with the observations of \citet{Topping2024MNRAS}.
The results are divided into stellar-mass bins and shown for different escape fractions.

When comparing galaxies at fixed $M_{\rm UV}$, more massive systems exhibit systematically redder UV continuum slopes ($\beta$). This trend is naturally explained if dust attenuation is more effective in massive galaxies.

Turning to the escape fraction of ionizing photons, $f_{\rm esc}$, we find that smaller $f_{\rm esc}$ corresponds to redder $\beta$.
This is because a lower $f_{\rm esc}$ means that a larger fraction of ionizing photons is absorbed by the interstellar gas, enhancing nebular emission, particularly the nebular continuum, and thereby making the UV continuum redder.

These qualitative trends are common to both \Kroupa\ and \zZzZ. However, the mass-dependent reddening, i.e., larger $\beta$ at higher mass, is more pronounced in \zZzZ.

Observationally, \citet{Topping2024MNRAS} found that $\beta$ becomes redder toward higher UV luminosities, corresponding to a negative slope in the $M_{\rm UV}$--$\beta$ relation. In our models, this observed slope is reproduced when nebular emission is significant, i.e., small $f_{\rm esc}$, whereas models with $f_{\rm esc}=1$ predict the opposite trend. The stellar masses of the robustly identified extremely blue galaxies ($\beta<-2.8$) in \citet{Topping2024MNRAS} are mostly in the range $\log(M_\ast/M_\odot)=7.0$--$8.5$, with a median of $\log(M_\ast/M_\odot)=7.5\pm0.2$. Therefore, the most direct comparison with their sample is provided by our low- and intermediate-mass bins. In these bins, \zZzZ\ tends to predict redder $\beta$ values than the observed relation, suggesting that nebular continuum emission may be too strong in this model when $f_{\rm esc}$ is small.

\section{Discussion}\label{sec:discussion}

\subsection{Can a variable IMF explain the overabundance of UV-bright galaxies at high-redshift?}\label{subsec:discussion_variable_imf}

Our results indicate that the top-heavy \zZIMF\ increases the intrinsic UV luminosity relative to the Kroupa IMF. When the same IMF is also used to compute metal enrichment, however, the more rapid buildup of metals strengthens dust attenuation and limits the resulting enhancement of the observed UVLF. 
This qualitative trend holds for both dust attenuation prescriptions considered here: the fiducial \citet{Shimizu14}-like model and the alternative \citet{Trinca2024MNRAS}-based model.

While a non-negligible amount of dust attenuation is required at $z\sim7$, the high-redshift comparisons favor substantially weaker effective attenuation, with $R_\tau=0.51$ and $0.21$ at $z_{\rm obs}=9.7$ and $13.04$, respectively.
The formal value near $R_\tau=0.04$ at $z_{\rm obs}=13.5$ is not a robust constraint, and the model remains below the highest reported number densities even near the dust-free limit. Efficient dust removal or suppressed dust attenuation, as proposed by \citet{Ferrara2023MNRAS, Ferrara2024A&A_Dust_UVLF_outf}, may help reconcile the models with observations, but cannot by itself explain all observational estimates.

\citet{Fiore2023ApJ_Dust} demonstrated that the Eddington ratio $\lambda_\text{E}$ of galaxies at $z > 6.5$ increases with redshift and that there is a strong anti-correlation between $\lambda_\text{E}$ and the dust UV optical depth. This is attributed to radiation-driven dusty outflows, which expel dust from galaxies.
Such outflows have been suggested to produce a bluer UV continuum slope and enhance Ly$\alpha$ transmission through velocity offsets \citep{Ferrara2024A&A_Dust_UVLF_outf}.

This partial offset of the intrinsic UV enhancement by dust attenuation is qualitatively consistent with the recent cosmological hydrodynamical simulations presented by \citet{Durrant2026}.
Their density-dependent top-heavy IMF enhances the intrinsic far-UV luminosities, but also increases metal enrichment and dust surface densities, thereby partly offsetting the UV enhancement.
This qualitative agreement emerges despite substantial differences in the IMF prescriptions and galaxy-formation frameworks: their IMF prescription depends on the density of star-forming gas and is implemented in a hydrodynamical simulation, whereas ours depends explicitly on redshift and metallicity and is implemented in a semi-analytic framework.

%This partial offset of the intrinsic UV enhancement is also found in recent studies adopting different IMF prescriptions and galaxy-formation frameworks. In the cosmological hydrodynamical simulations of \citet{Durrant2026}, a density-dependent top-heavy IMF enhances the intrinsic far-UV luminosity but also increases metal enrichment and dust surface densities, thereby partly offsetting the UV enhancement. Similarly, \citet{Ferrara2026arXivUVLF} found that the intrinsic UV brightening produced by a metallicity-dependent top-heavy IMF is largely counteracted by stronger supernova feedback and dust attenuation. Using the \texttt{DELPHI} semi-analytic model, they showed that the resulting UVLF differs from that predicted with a Salpeter IMF by less than 0.5 dex at $z\simeq11$--17. Unlike their model, our halo-accretion-based SFR prescription does not explicitly account for the suppression of star formation by IMF-dependent supernova feedback.

Thus, the ability of the \zZIMF\ model to explain the observed overabundance of UV-bright galaxies cannot be assessed from its intrinsic stellar UV emissivity alone; the associated metal enrichment and dust attenuation must also be taken into account. A fully self-consistent treatment of IMF-dependent stellar feedback will be important for future investigations.

Thus, the ability of the \zZIMF\ model to explain the observed overabundance of UV-bright galaxies cannot be assessed from its intrinsic stellar UV emissivity alone; the associated metal enrichment and dust attenuation must also be taken into account.

\subsection{Implications for dust evolution at high redshift}\label{subsec:discussion_dust}

The strong influence of dust is also evident in the fiducial COLIBRE simulations of \citet{Lu2026arXiv260506782L}, whose dust-attenuated UVLFs fall increasingly below observations at the bright end toward higher redshift.
Neglecting dust attenuation largely alleviates this deficit at $z\lesssim12$, but not at $z=15$.
\citet{Jeong2025ApJ} similarly highlighted the competing roles of metal enrichment and dust removal: a top-heavy IMF accelerates metal enrichment, whereas intense starbursts driven by enhanced SFE can evacuate dust from star-forming regions.
Observationally, \citet{Franco2025arXiv250804791F} found little evolution in the median UV continuum slope at $z>8$, which they interpreted as suggesting either low effective dust attenuation or a spatial separation between dust and star-forming regions.

The decrease in $R_\tau$ from 0.51 at $z_{\rm obs}=9.7$ to 0.21 at $z_{\rm obs}=13.04$ is consistent with weaker effective attenuation toward earlier times. However, because these values are derived from different observational samples and $R_\tau$ is degenerate with dust geometry, spatial distribution, and dust-to-metal ratio, we do not infer a specific functional form for its redshift evolution.

Our dust-free predictions agree more closely with the available observational constraints than the dust-attenuated predictions.
However, differences between the model and observational redshifts, together with the predominance of upper limits at the highest redshifts, limit the precision of this comparison.
This suggests that reduced effective dust attenuation may play an important role at $z\sim14$. However, even in the dust-free limit, our model does not reproduce all the reported bright-end UVLF measurements at the highest redshifts (Fig.~\ref{fig:UVLF_z14_wo_dust}).

This indicates that reducing dust attenuation alone may be insufficient and that additional mechanisms enhancing the intrinsic UV luminosities, such as increased SFE or bursty star formation histories, may also be required. These findings highlight the importance of understanding both the processes governing dust production, spatial distribution, and removal, and those regulating star formation in the earliest galaxies.

\subsection{Comparison with previous studies and model limitations}\label{subsec:discussion_limitations}

Other mechanisms may also enhance the bright end of the UVLF.
One possibility is an increase in the mean SFE.
\citet{Donnan2025MNRAS_UVLF} reproduced the observed UVLF at $z=12.5$ using an SFE--$M_{\rm halo}$ relation in which the SFE at $\log_{10}(M_{\rm halo}/\Msun)=11.7$ is 0.6 dex higher than that inferred from the \citet{Harikane2022ApJS} relation adopted in our model.
Using a high-resolution cosmological zoom-in simulation of a massive galaxy at $z\sim10$, \citet{Chen2026MNRAS_FeedbackFree} identified feedback-free starbursts in dense stellar clusters and found a global SFE of $\simeq0.2$--0.3.
Most clusters formed their stars in bursts lasting $\lesssim3\,\mathrm{Myr}$.
These results suggest that high SFEs and short-duration starbursts may boost UV luminosities without invoking IMF variations.

Variations in the star formation history can also enhance the bright end of the UVLF by increasing the scatter in UV luminosity at fixed halo mass, without requiring a systematic increase in the mean SFE.
By jointly modeling the UVLF, galaxy clustering, and H$\alpha$-to-UV ratios, \citet{Munoz2026MNRAS} found evidence for stronger burstiness in lower-mass halos.
Similarly, \citet{Sarkar2026MNRAS} found that shorter characteristic star formation timescales can reproduce the slow evolution of the UVLF at $z>10$. 
They also found that, within their model, neither dust-free conditions nor a top-heavy IMF alone reproduces the  UVLF measurements at $z\sim14$.
These mechanisms are complementary to the IMF-driven effect identified in our model:
burstiness introduces scatter through temporal variations in the SFR, whereas the top-heavy IMF introduces scatter through variations in UV luminosity per unit SFR.
Both effects could broaden the $M_{\rm UV}$--$M_{\rm halo}$ relation and enhance the bright end of the UVLF.
Low-luminosity AGNs may also enhance the UV luminosities of high-redshift galaxies, although accounting for their contribution would require modeling black hole formation and evolution.

A limitation of our semi-analytic model is its temporal resolution.
We use the merger-tree snapshot intervals as timesteps without internal substepping.
At $z=7$--$14$, these intervals range from 18.5 to 42.0 Myr, with a median of 27.9 Myr.
The interval following the $z=14.3$ snapshot is 17.6 Myr.
These timescales exceed the $\lesssim 3$ Myr starbursts found by \citet{Chen2026MNRAS_FeedbackFree} and are comparable to the lifetimes of core-collapse supernova progenitors ($\sim 3$--40 Myr). 
Although formation times are assigned to stellar populations within each snapshot interval when constructing the SEDs, the model does not resolve individual short starbursts or the delay between star formation and supernova feedback. 
This limitation may affect the predicted duty cycle of UV-bright phases and the scatter in UV luminosity.
Tests with shorter timesteps or internal substepping
are needed to quantify these effects.

Our IMF prescription also has limitations.
It combines the redshift-dependent model of \citet{Steinhardt2020ApJ} with the metallicity and redshift trends found by \citet{Chon2022MNRAS}.
Although this provides a continuous prescription for use in SSP-based SED templates, the model relies on extrapolation at $\log(Z/\Zsun)>0$, $\log(Z/\Zsun)<-4$, and $z>20$.
Predictions in these regimes should therefore be interpreted cautiously.

Furthermore, stellar radiative feedback is not explicitly included in the simulations used to construct our IMF.
Radiation-hydrodynamic simulations by \citet{Chon2024MNRAS} show that dust heating suppresses low-mass fragmentation, while photoionization inhibits accretion onto massive stars.
These effects modify both ends of the IMF, with the high-mass distribution becoming more Salpeter-like at higher metallicity, around $\log(Z/\Zsun)\simeq-2$ to $-1$.
At lower metallicity, circumstellar disks can shield ionizing radiation and allow massive stars to grow to $\sim100\,\Msun$.
Our prescription may therefore overestimate the degree of top-heaviness at higher metallicity and the abundance of low-mass stars relative to models including radiative feedback.

IMF variations may also be accompanied by changes in the SFE.
\citet{Chon2024MNRAS} found that the SFE decreases toward lower metallicity, consistent with previous studies
\citep[e.g.,][]{He2019MNRAS,FukushimaHajime2020MNRAS_StarClusterFormation_RadFB_Z,Guszejnov2022MNRAS}. 
Possible causes include stronger stellar feedback, higher temperatures and pressures in \ion{H}{ii} regions, and reduced shielding against ionizing radiation.
In this study, we keep the SFE prescription fixed to isolate the effects of the IMF on UV emission
and chemical enrichment.
A model in which the IMF and SFE evolve together remains a subject for future work.

Our \zZIMF\ is also not intended to describe extreme star-formation regimes that produce supermassive stars (SMSs).
Such stars may form when H$_2$ cooling is suppressed by strong external radiation fields \citep[e.g.,][]{Bromm2003ApJ, Shang2010MNRAS, Sugimura2014MNRAS, Kimura2025ApJ, Chon2026Nature}, or by the CMB at extremely high redshifts \citep{Ito2024PASJ}, or by internal FUV radiation from newly formed stars \citep{Kiyuna2026arXiv}.
Simulations by \citet{Chon2020MNRAS, Chon2025MNRAS} suggest that low-metallicity clouds exposed to strong FUV radiation may form a central SMS of $\sim10^4\,\Msun$ at $\log(Z/\Zsun)\lesssim-3$.
At $\log(Z/\Zsun)=-2$, enhanced fragmentation instead leads to star clusters containing very massive stars (VMSs) of $\sim10^3\,\Msun$.
Such populations may contain distinct low-mass and SMS or VMS components.
Our IMF, which has an upper stellar mass of $\sim100\,\Msun$ and uses SSP-based templates, cannot represent these regimes.

Future cosmological hydrodynamic simulations
will be useful for testing these effects
self-consistently.
The COLIBRE simulations of \citet{Durrant2026} highlight the importance of self-consistently modeling UV emission, stellar feedback, metal enrichment, and dust production with a density-dependent IMF.
The Gadget4-Osaka code \citep{Oku2024} already implements a different redshift- and metallicity-dependent top-heavy IMF together with SN and AGN feedback.
Implementing our \zZIMF\ prescription in this framework would allow us to investigate its effects on UV emission, metal enrichment, dust attenuation, and feedback in a cosmological hydrodynamic simulation.

\section{Conclusions}\label{sec:conclusion}

We investigated whether a redshift- and metallicity-dependent top-heavy IMF (\zZIMF) can explain the bright-end excess in the high-redshift UVLF, accounting for its effects on stellar UV emission, metal enrichment, and dust attenuation.
To distinguish these effects, we compared a model in which the IMF affects only the stellar SED with one in which its effects on chemical evolution are also included self-consistently.
Our main findings are as follows:

\begin{enumerate}

\item In the absence of dust attenuation,
adopting a redshift- and metallicity-dependent
top-heavy IMF enhances the bright end of the
UVLF relative to a standard Kroupa IMF,
regardless of the treatment of metal enrichment.
This enhancement reflects both increased UV
emission and larger scatter in UV luminosity.
However, matching the observed UVLF at
$z\sim7$ requires substantial dust attenuation,
particularly for the top-heavy IMF.

\item The reduction of the bright-end UVLF enhancement by dust attenuation is not specific to a single attenuation model. Both the fiducial Shimizu-like model and the alternative \citet{Trinca2024MNRAS}-based model show that, when metal enrichment is treated consistently with the IMF, the resulting stronger dust attenuation partially offsets the bright-end enhancement produced by the top-heavy IMF.

\item When the effects of the top-heavy IMF on metal enrichment are treated self-consistently,
the model broadly reproduces the observed UVLF
at $z\sim10$.
However, better agreement with observations at higher redshifts requires weaker effective dust attenuation, with optical depths reduced to approximately 50\% and 20\% of the fiducial values at $z_{\rm obs}=9.7$ and $13.04$, respectively.
At $z_{\rm obs}=13.5$, even nearly dust-free predictions fall below the highest reported number densities.
Thus, a variable IMF and reduced dust attenuation alone cannot reproduce all the reported bright-end UVLF measurements at the highest redshifts.
 
\item The redshift- and metallicity-dependent top-heavy IMF increases the scatter in UV luminosity per unit SFR, while leaving its median value close to that for a standard Kroupa IMF.
This larger scatter arises because the UV output of stellar populations depends more strongly on age and metallicity.
It contributes to the enhancement of the bright end of the UVLF.

Previous studies have invoked scatter in the SFR or variations in the star formation history at fixed halo mass as a possible origin of the broadened $M_{\rm UV}$--$M_{\rm halo}$ relation \citep[e.g.,][]{Mason2023MNRAS,Shen2023MNRAS}. Our results suggest that the \zZIMF\ can further broaden this relation by increasing the scatter in the UV luminosity-to-SFR conversion factor.

\item For the SSP SED templates, the \zZIMF\ produces systematic metallicity- and redshift-dependent variations in the UV luminosity, UV continuum slope $\beta$, and ionizing photon output, in addition to the standard age evolution of stellar populations. 
In the young, UV-bright phase, lower-metallicity SSPs are brighter in the UV and have stronger ionizing outputs because the \zZIMF\ becomes more top-heavy. 
Quantitatively, the enhancement of $L_{\lambda,\mathrm{UV}}$ relative to a standard Kroupa IMF reaches a factor of $\sim 5$--6 across the explored metallicity range, while the redshift dependence is weaker, with variations of a factor of $\sim 3$--3.5.
Similar trends are found for other stellar
population diagnostics presented in
Appendix~\ref{app:sed_diagnostics}.
\end{enumerate}

While the \zZIMF\ framework improves the modeling of high-redshift galaxy SEDs and provides insights into key diagnostics of stellar populations, several avenues remain for future work. These include incorporating metallicity-dependent star formation efficiency, accounting for black hole and AGN contributions, and testing the \zZIMF\ in cosmological hydrodynamic simulations with explicit feedback physics. The SED templates generated in this study are publicly available to facilitate further work in this area.

\begin{ack}
We thank Raffaella Schneider and Alessandro Trinca
for valuable comments and helpful discussions.
Computational analyses were performed using the
analysis servers at the Center for Computational
Astrophysics, National Astronomical Observatory of Japan.
This work was supported by JSPS KAKENHI Grant Numbers
23H00131 (AI), 22H00149, 26H02060 (KO), and 26H02062 (TI)
and by the Global-LAMP Program of the National Research
Foundation of Korea (NRF), funded by the Ministry of
Education (Grant No. RS-2023-00301976; KF).
TI has been supported by IAAR Research Support Program in Chiba University Japan.

We thank Instituto de Astrofisica de Andalucia (IAA-CSIC), Centro de
Supercomputacion de Galicia (CESGA) and the Spanish academic and
research network (RedIRIS) in Spain for hosting Uchuu DR1, DR2 and DR3
in the Skies \& Universes site for cosmological simulations. The Uchuu
simulations were carried out on Aterui II supercomputer at Center for
Computational Astrophysics, CfCA, of National Astronomical Observatory
of Japan, and the K computer at the RIKEN Advanced Institute for
Computational Science. The Uchuu Data Releases efforts have made use
of the skun@IAA\_RedIRIS and skun6@IAA computer facilities managed by
the IAA-CSIC in Spain (MICINN EU-Feder grant EQC2018-004366-P).
\end{ack}

\section*{Data availability}
The stellar SSP SED templates generated in this study are publicly available in HDF5 format at \url{https://github.com/KeitaFukushima/stellar-ssp-sed-template_zZIMF}. 
The methodology used to construct the templates is described in Section~\ref{subsec:SED_model}.

\appendix

\section{Modeling the Steinhardt IMF}
\label{sec:appendix_steinhardtIMF}
\begin{figure}
    \begin{center}
    \includegraphics[width=0.5\textwidth]{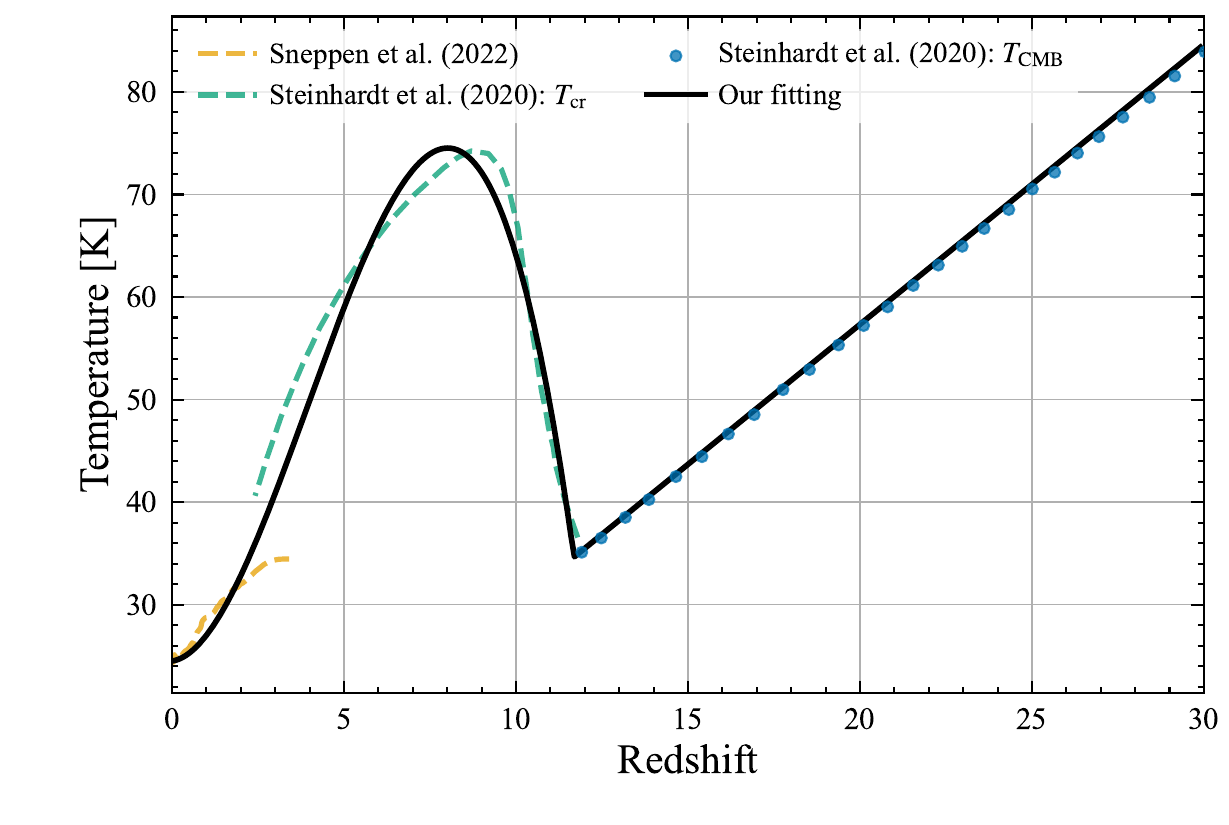}
    \end{center}
    \caption{Redshift dependence of $T_{\text{cr}}$ and $T_{\text{CMB}}$ from \citet{Steinhardt2020ApJ}, and $T_{\mathrm g}$ from \citet{Sneppen2022ApJ}. 
    The black line shows our fit to $T_{\mathrm g}$.
    The values at $z = 0, 5, 10$, and $20$ were used to construct the $z$--$Z$-dependent IMF.
}

    \label{fig:Steinhardt_IMF}
\end{figure}

The IMF proposed by \citet{Steinhardt2020ApJ} incorporates a temperature-dependent formulation.
To model the temperature--redshift relation,
we used the results shown in Figure~5 of
\citet{Steinhardt2020ApJ} for $z>3.4$
and Figure~6 of \citet{Sneppen2022ApJ}
for $z<3.4$.
We fitted the resulting temperature--redshift
relation using the \texttt{curve\_fit} function
from Python's \texttt{scipy} library,
obtaining the following expression:
\begin{equation}
\label{eq:fitting_SteinhardtIMF}
T_{\mathrm g}(z) =
\begin{cases} 
a z^3 + b z^2 + c z + 24.5 & \text{if } z < 11.7, \\
2.73 (1 + z) & \text{if } z \geq 11.7,
\end{cases}
\end{equation}
where the coefficients are
$a = -0.187$, $b = 2.215$, and $c = 0.483$.
The resulting fit is shown in Figure~\ref{fig:Steinhardt_IMF}.
The values of $m_0(z)=0.08\Msun(T_{\mathrm g}(z)/20\mathrm{K})^2$ for $Z=\Zsun$ at $z=0$, 5, 10, and 20 are $0.120$, $0.694$, $0.822$, and $0.657\,\Msun$, respectively.

\section{Construction and evaluation of the $z$--$Z$ dependent IMF}\label{app:imf_technical}

In this appendix, we describe the construction
and validation of the \zZIMF.
We first determine the IMF parameters from
the results of \citet{Chon2022MNRAS} and
\citet{Steinhardt2020ApJ}, and then construct
a continuous parameterization in redshift
and metallicity.
We also present additional IMF comparisons
at different redshifts and examine the
properties of the resulting IMF.

\subsection{Fitting the Chon and Steinhardt IMF prescriptions}\label{subsubsec:IMF_modelling}

We model the \zZIMF\ based on the IMF distributions
obtained by \citet{Chon2022MNRAS}, using the following
piecewise power-law form:
\begin{equation}\label{eq:IMF_formula}
\frac{dn}{dm} \propto 
\begin{cases} 
m^{-0.3} & \text{for } \log m_{\text{min}} < \log m < \log m_0, \\ 
m^{-1.3} & \text{for } \log m_0 < \log m < \log m_1, \\ 
m^{-2.3} & \text{for } \log m_1 < \log m < \log m_2, \\ 
m^{-1} & \text{for } \log m_2 < \log m < \log m_{\text{max}}.
\end{cases}
\end{equation}
The first three segments follow the standard Kroupa IMF at low and intermediate masses. 
To capture the systematic flattening of the IMF at the high-mass end seen in \citet{Chon2022MNRAS} (their Figure~11), we represent the top-heavy component by a log-flat behavior, implemented here as $dn/dm \propto m^{-1}$ above $m_2$. 
Throughout, we adopt $m_{\rm min}=0.003\,\Msun$ and $m_{\rm max}=100\,\Msun$.

The break masses in equation~(\ref{eq:IMF_formula}), namely $m_0$, $m_1$ and $m_2$, are chosen to reproduce the following key IMF diagnostics defined by \citet{Chon2022MNRAS}. 
The median mass, $M_{\rm median}$, is the stellar mass that divides the IMF by number. The half-mass scale, $M_{\rm half}$, is the stellar mass at which the cumulative stellar mass reaches half of the total mass integrated over the full IMF mass range. The massive-star fraction, $f_{\rm massive}$, is defined here as the mass fraction in stars with $m > 10\,\Msun$.

\citet{Chon2022MNRAS} do not provide a model at $Z=\Zsun$. 
We therefore adopt the redshift-dependent IMF of \citet{Steinhardt2020ApJ} as the reference in this regime.
This gives 20 grid points in the $(z,Z)$ plane, defined by four redshifts and five metallicities.

We do not treat $m_1$ as an independent parameter in order to reduce degeneracies while preserving the Kroupa-like structure below the top-heavy component. 
Instead, we fix it through the relation $\log m_1 = \log m_0 + 0.80$. 
The offset $0.80$ is taken from the original Kroupa IMF and represents the logarithmic width of the $m^{-1.3}$ segment. 
This assumption is also supported by \citet{Chon2022MNRAS} (their Figure~11), 
whose results suggest that the intermediate-mass behavior is governed mainly by shifts in the break masses rather than by changes in the widths of the individual segments.

%In Equation~\ref{eq:IMF_formula}, the parameter $m_1$ is not treated as an independent variable but is determined from $m_0$ through the relation $\log m_1 = \log m_0 + 0.80$. 
%The value of $0.80$ is derived from the Kroupa IMF and corresponds to the difference between $m_0$ and $m_1$ in logarithmic mass space in the original Kroupa formulation. 
%The reason for fixing the mass interval of the $m^{-1.3}$ slope to this Kroupa-based width is that, similar to the log-flat high-mass component, this behavior reflects a common trend seen in the results of Fig.~11 of \citet{Chon2022MNRAS}.

We then determine the best-fit $(m_0,m_2)$ values at 20 $(z,Z)$ grid points. 
We construct a grid of candidate IMFs by varying $\log m_0$ from $-1.8$ to $-0.2$ and $\log m_2$ from $-0.8$ to $2.0$, both in steps of $0.1$. 
For each pair $(m_0, m_2)$, we compute $M_{\rm median}$, $M_{\rm half}$, and $f_{\rm massive}$.
We identify the values that best reproduce the diagnostics of \citet{Chon2022MNRAS} at $\log(Z/\Zsun)=-1,-2,-3,-4$ for $z=0,5,10,20$.
Comparisons between our fit and the results of
\citet{Chon2022MNRAS} are shown in
Figure~\ref{fig:IMF_result_z10} and
Appendix~\ref{subsec:appendix_zZIMF}.
At $\log(Z/\Zsun)=0$, we instead fit to the IMF of \citet{Steinhardt2020ApJ} at the same redshifts. 

%The gas temperature $T$ in their prescription is modeled following their original formulation (see Appendix~\ref{sec:appendix_steinhardtIMF}). 
%We use the resulting best-fit $(m_0,m_2)$ values for the continuous $(z,Z)$ parameterization described in the next subsection.

%We constructed IMFs based on Equation~\ref{eq:IMF_formula} by varying $\log m_0$ from $-1.8$ to $-0.2$ in increments of $0.1$, and $\log m_2$ from $-0.8$ to $2.0$ in increments of $0.1$.
%For all IMFs with varying $m_0$ and $m_2$, $m_{\text{median}}$, $m_{\text{half}}$, and $f_{\text{massive}}$ were calculated. The $m_0$ and $m_2$ values that best match the results of \citet[][their Figures 11 and 12]{Chon2022MNRAS} and, for $\log(Z/\Zsun) = 0$, the results of \citet{Steinhardt2020ApJ}, were determined for 20 combinations of $z = 0, 5, 10, 20$ and $\log(Z/\Zsun) = 0, -1, -2, -3, -4$.
%For the case of $\log(Z/\Zsun)=0$, we adopted the Steinhardt IMF because the IMF presented by \citet{Chon2022MNRAS} does not include a model at $Z=\Zsun$.  
%The temperature $T$ in the Steinhardt IMF was modeled based on their original formulation; details are provided in Appendix~\ref{sec:appendix_steinhardtIMF}.

\subsection{Parameterization across redshift and metallicity}\label{subsubsec:IMF_parametrization}

We next construct a smooth interpolation function for the \zZIMF\ in the $(z,Z)$ plane, using the best-fit $(m_0,m_2)$ values obtained at the 20 discrete grid points. This function returns $(m_0,m_2)$ at arbitrary redshift and metallicity within the range covered by the reference models.

To this end, we adopt a second-order polynomial in $(1+z)$ and $\log (Z/\Zsun)$:
\begin{equation}
\label{eq:fitting_formula_mo_and_m2}
\begin{aligned}
\log m_{0}(z, Z) = \sum_{i=0}^{2}\sum_{j=0}^{2-i} c_{0,ij}(1+z)^i \left[\log(Z/\Zsun)\right]^j,\\
\log m_{2}(z, Z) = \sum_{i=0}^{2}\sum_{j=0}^{2-i} c_{2,ij}(1+z)^i \left[\log(Z/\Zsun)\right]^j.
\end{aligned}
\end{equation}
Here, $c_{0,ij}$ and $c_{2,ij}$ denote the polynomial coefficients for $m_0$ and $m_2$, respectively. 
The best-fit coefficients are
$c_{0,00} = -0.684$, $c_{0,01} = 0.541$, $c_{0,02} = 0.070$, $c_{0,10} = 0.060$, $c_{0,11} = 0.0056$, $c_{0,20} = -0.00116$, 
$c_{2,00} = 3.165$, $c_{2,01} = 1.335$, $c_{2,02} = 0.153$, $c_{2,10} = -0.067$, $c_{2,11} = -0.015$, and $c_{2,20} = 0.0009$.

\begin{figure}
    \begin{center}
    \includegraphics[width=0.5\textwidth]{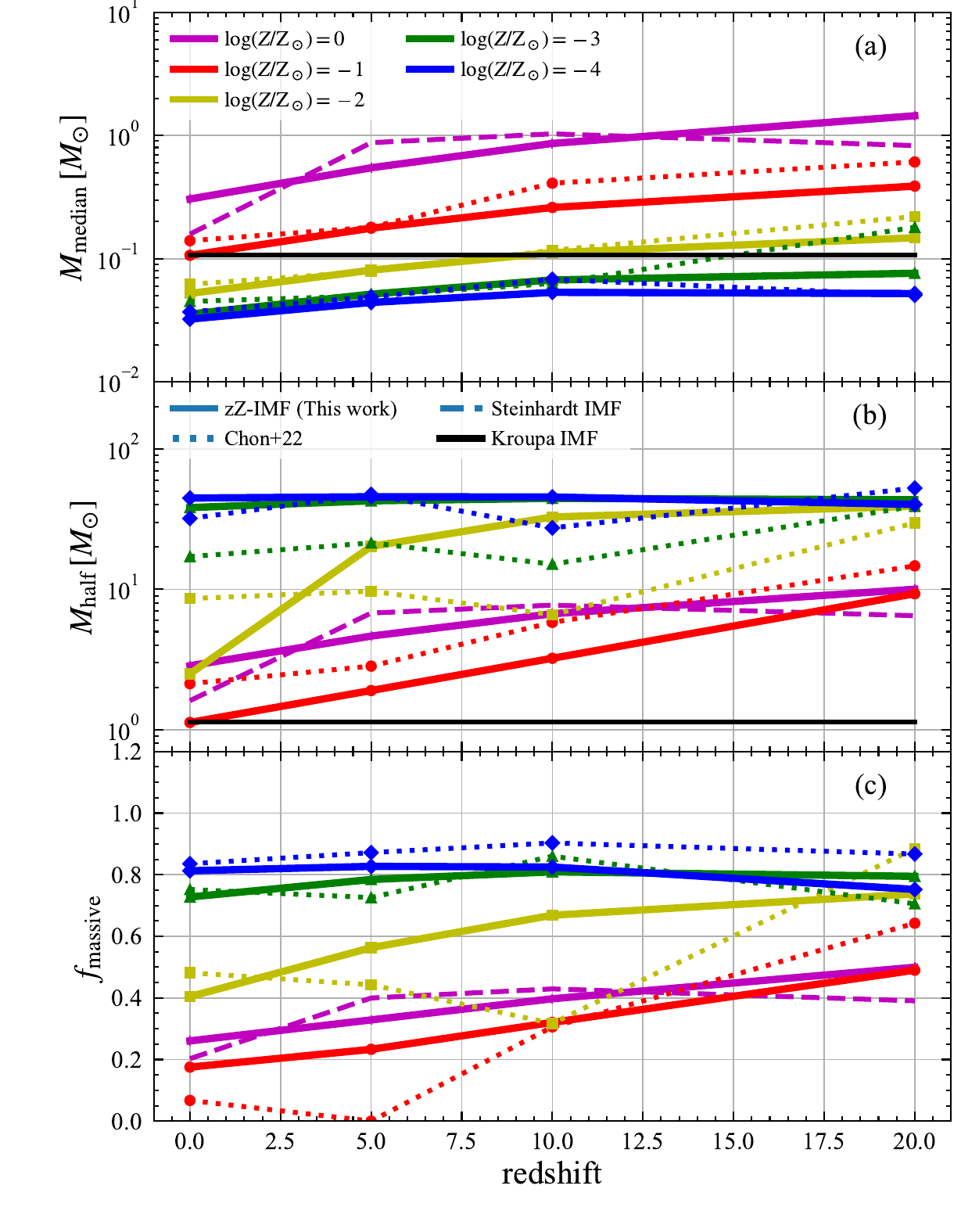}
    \end{center}
    \caption{Redshift evolution of $M_{\rm median}$ (panel~a), $M_{\rm half}$ (panel~b), and $f_{\rm massive}$ (panel~c). Here, $M_{\rm median}$ is the stellar mass that divides the IMF by number, $M_{\rm half}$ is the mass above which stars contain half of the total stellar mass, and $f_{\rm massive}$ is the mass fraction in stars with $m>10\,\Msun$.
    Our fitted results are shown by colored solid lines. The magenta, red, yellow, green, and blue lines correspond to $\log(Z/\Zsun)=0$, $-1$, $-2$, $-3$, and $-4$, respectively. The results of \citet{Chon2022MNRAS} for $\log(Z/\Zsun)\leq -1$ are indicated by colored dotted lines for each metallicity, while the Steinhardt IMF result at $\log(Z/\Zsun)=0$ is shown by a magenta dashed line.  In panels~(a) and (b), the black solid line shows the Kroupa IMF for reference. In panel~(c), $f_{\rm massive}$ is computed in the same way for both our $z$-$Z$ dependent IMF and the \citet{Chon2022MNRAS} histograms, using a threshold of $10\,M_\odot$.
    }    \label{fig:IMF_params_fitting_result}
\end{figure}

\begin{figure*}
    \begin{center}
    \includegraphics[width=0.9\textwidth]{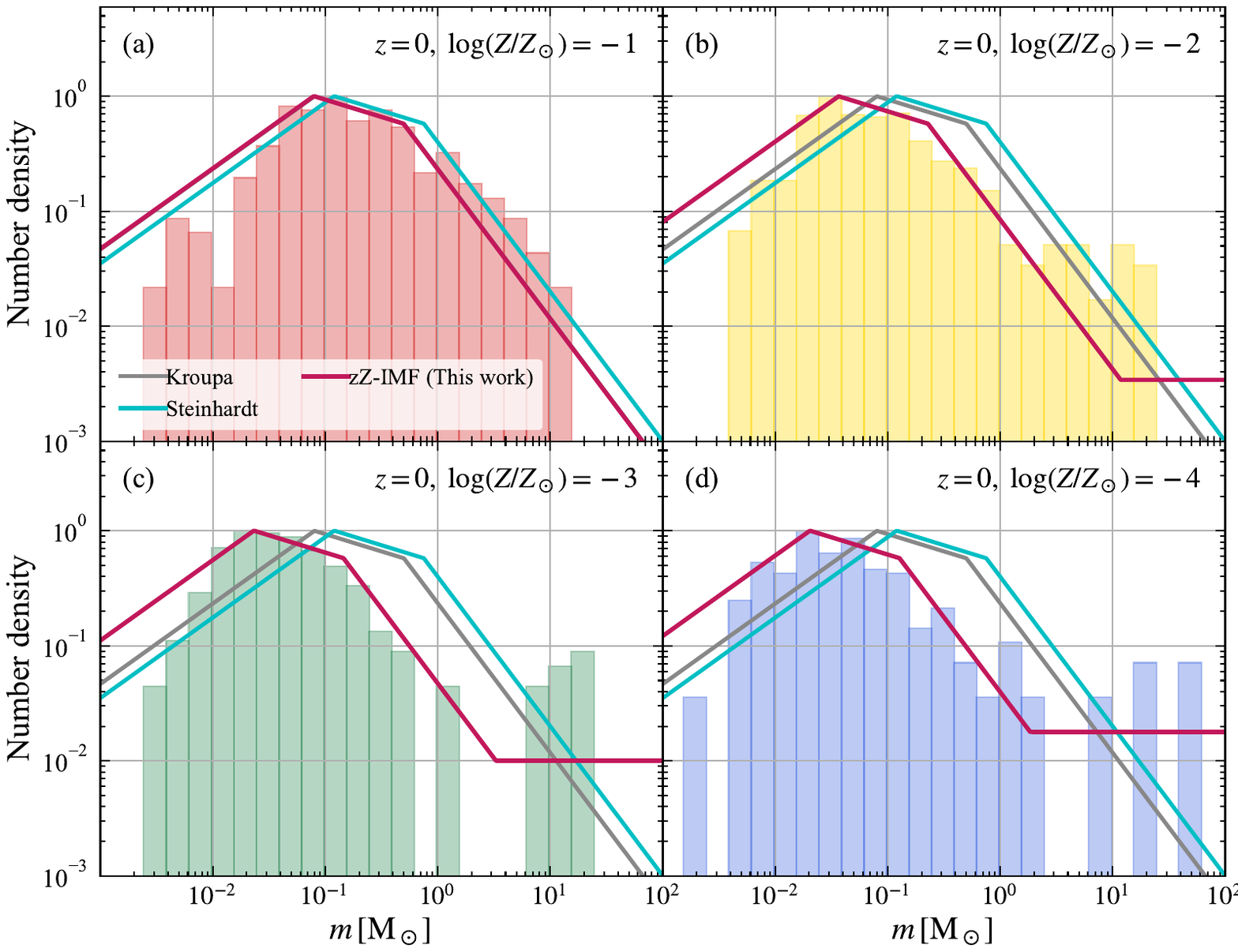}
    \end{center}
    \caption{Same as Figure~\ref{fig:IMF_result_z10}, but for $z=0$.}

    \label{fig:IMF_result_z0}
\end{figure*}

The polynomial parameterization reproduces the redshift evolution of the key IMF diagnostics across metallicity. Figure~\ref{fig:IMF_params_fitting_result} compares the parameterized \zZIMF\ (solid lines) with the original \citet{Chon2022MNRAS} trends at $\log(Z/\Zsun)\le -1$  (dotted lines) and the solar-metallicity case from \citet{Steinhardt2020ApJ} (magenta dashed line). The comparison includes three diagnostics: $M_{\rm median}$, $M_{\rm half}$, and $f_{\rm massive}$. For reference, panel~(b) additionally shows the Kroupa IMF. In panel~(c), $f_{\rm massive}$ is defined consistently for both our model and the \citet{Chon2022MNRAS} histograms as the mass fraction in stars with $m>10\,\Msun$. The close agreement indicates that the parameterization preserves the intended $z$ and $Z$ dependences of the underlying reference models. This continuous \zZIMF\ is therefore suitable for the SED and UVLF calculations that follow.

\begin{figure*}
    \centering
    \includegraphics[width=0.9\textwidth]{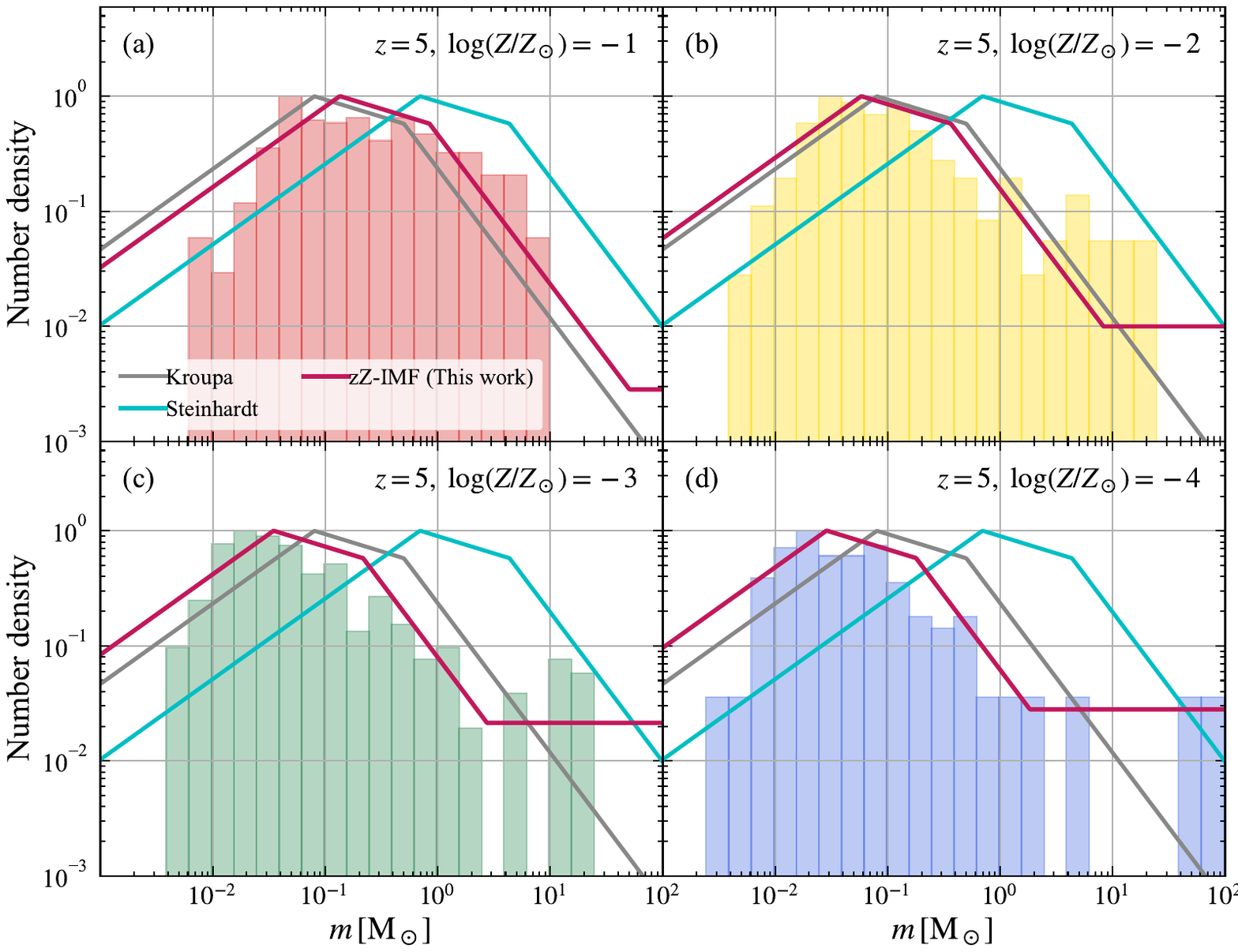}
    
    \caption{Same as figure~\ref{fig:IMF_result_z10}, but for $z=5$.}

    \label{fig:IMF_result_z5}
\end{figure*}

\begin{figure*}
    \centering
    \includegraphics[width=0.9\textwidth]{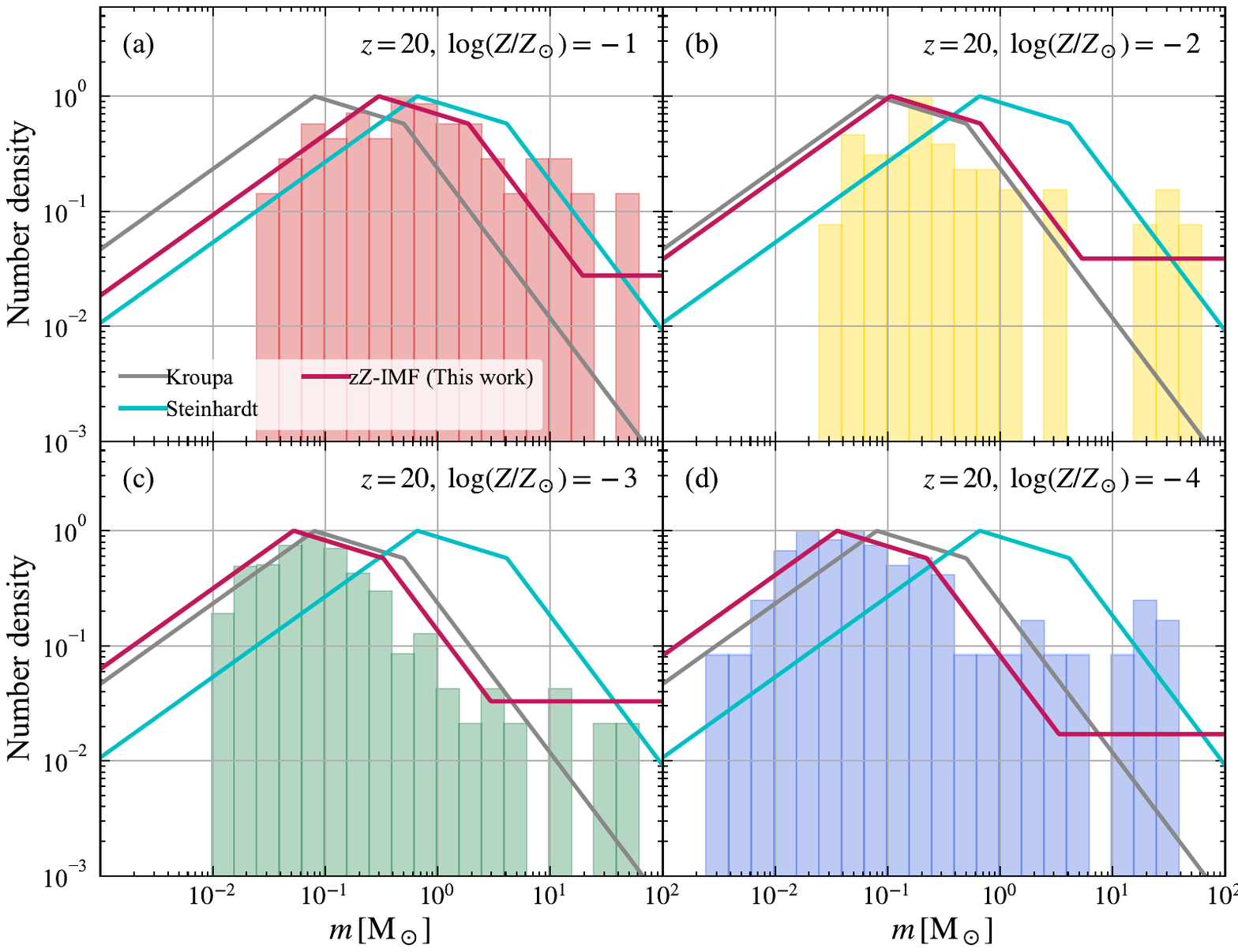}
    
    \caption{Same as figure~\ref{fig:IMF_result_z10}, but for $z=20$.}

    \label{fig:IMF_result_z20}
\end{figure*}

\subsection{IMF comparisons at other redshifts}\label{subsec:appendix_zZIMF}

Figure~\ref{fig:IMF_result_z10} presents the IMF comparison at $z=10$, while Figures~\ref{fig:IMF_result_z0}, \ref{fig:IMF_result_z5}, and \ref{fig:IMF_result_z20} show the corresponding results at $z=0$, 5, and 20, respectively. 

\subsection{Properties of the resulting IMF}\label{subsubsec:Our_IMF_model}

Figure~\ref{fig:IMF_result_z10} and Appendix~\ref{subsec:appendix_zZIMF} highlight two main differences between the Chon-simulation IMFs and the Kroupa IMF. First, the low-mass peak shifts to smaller stellar masses toward lower metallicity. Second, the simulations show an enhanced massive-star component, producing a flatter high-mass tail than the Kroupa IMF. The \zZIMF\ is constructed to reproduce these two features: it preserves the metallicity-dependent shift of the low-mass peak while also capturing the enhanced massive-star component. By contrast, the Steinhardt IMF becomes top-heavy, but its overall shape does not match the Chon-simulation IMFs, especially at low metallicity, because it underpredicts the low-mass stellar population. 
These trends are also evident in Figure~\ref{fig:IMF_params_fitting_result}, which compares our \zZIMF\ with the results of \citet{Chon2022MNRAS} and the reference IMFs in terms of  $M_{\rm median}$, $M_{\rm half}$, and $f_{\rm massive}$. 

The redshift dependence of $M_{\rm median}$ weakens markedly toward lower metallicity. 
In panel~(a), $M_{\rm median}$ decreases with decreasing metallicity, while its evolution with redshift becomes progressively smaller. 
At $\log(Z/\Zsun)\lesssim -3$, the redshift evolution is minimal, amounting to only $\sim 0.1$~dex across the full redshift range considered. 
At higher metallicity, by contrast, $M_{\rm median}$ increases systematically with redshift. 
For example, at $\log(Z/\Zsun)=0$, it rises by $\sim 0.5$~dex from $z=0$ to $z=20$ in our parameterization.

A similar metallicity-dependent trend is seen
for $M_{\rm half}$ in panel~(b).
At low metallicity, $M_{\rm half}$ varies by only $\lesssim 0.1$~dex over $z=0$--20. 
At higher metallicity, however, its evolution with redshift is much stronger. 
For example, at $\log(Z/\Zsun)=-1$, $M_{\rm half}$ increases by $\sim 0.8$~dex from $z=0$ to $z=20$. 
This behavior indicates a progressively more top-heavy mass distribution at earlier times. 
A qualitatively similar trend is also seen in the Steinhardt IMF at solar metallicity.

The massive-star fraction remains high in the lowest-metallicity regime. 
Panel~(c) shows that $f_{\rm massive}\gtrsim 0.75$ over the full redshift range for $\log(Z/\Zsun)\le -3$. 
This behavior implies a strongly top-heavy IMF in extremely metal-poor environments. 

The metallicity dependence of the redshift evolution can be understood from the underlying cooling and fragmentation physics. 
At $\log(Z/\Zsun)\sim -3$ to $-4$, efficient cooling by metals and dust allows the gas to reach low temperatures. 
This lowers the Jeans mass and fragmentation scale and therefore favors the formation of many low-mass stars \citep{Chon2022MNRAS}. 
At higher metallicities, however, the CMB-imposed temperature floor becomes increasingly important at high redshift. 
As a result, the gas evolves closer to isothermal conditions ($\gamma\simeq 1$), and fragmentation in filaments and circumstellar disks is suppressed. 
This shifts the characteristic stellar mass to higher values and strengthens the top-heavy nature of the IMF \citep{Chon2022MNRAS}. 
These effects explain why $M_{\rm median}$ and $M_{\rm half}$ show stronger redshift evolution at higher metallicity but remain nearly constant at the lowest metallicities.

\begin{figure*}
    \begin{center}
    \includegraphics[width=\textwidth]{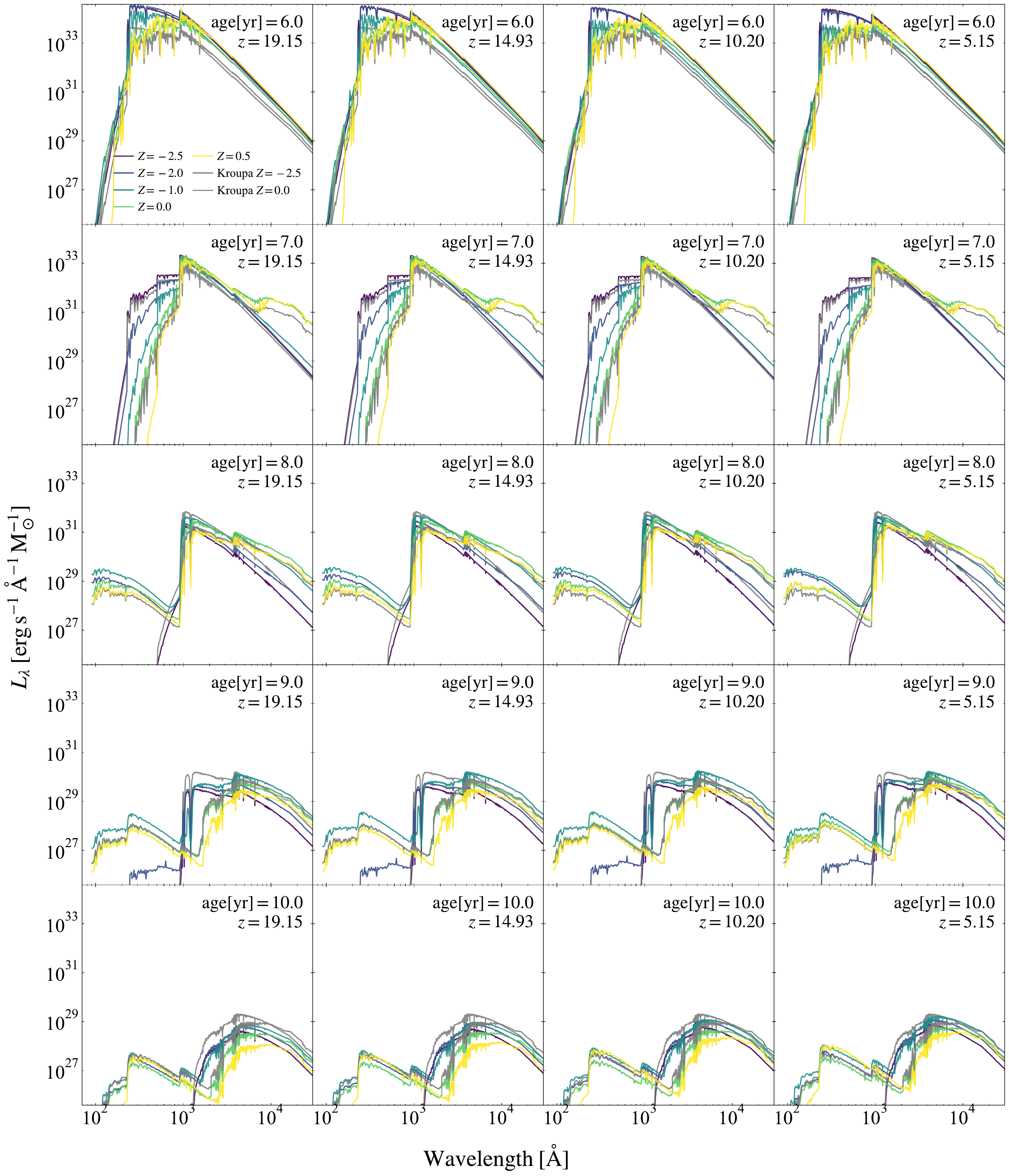}
    \end{center}
    \caption{SEDs of stellar populations modeled as SSPs using the \zZIMF.
    The columns correspond to redshifts $z = 19.15,\ 14.93,\ 10.20,$ and $5.15$, while the rows correspond to SSP ages of $10^6$, $10^7$, $10^8$, $10^9$ and $10^{10}$~yr.
    The purple, navy, cyan, yellow-green, and yellow lines correspond to metallicities of $\log(Z/\Zsun) = -2.5,\ -2.0,\ -1.0,\ 0.0,\ \text{and}\ 0.5$, respectively.
    For comparison, SEDs computed with the Kroupa IMF are also shown for $\log(Z/\Zsun) = -2.5$ (gray solid line) and $\log(Z/\Zsun) = 0.0$ (gray dashed line).}
    \label{fig:SSP_SED_result}
\end{figure*}

\begin{figure*}
   \begin{center}
    \includegraphics[width=\textwidth]{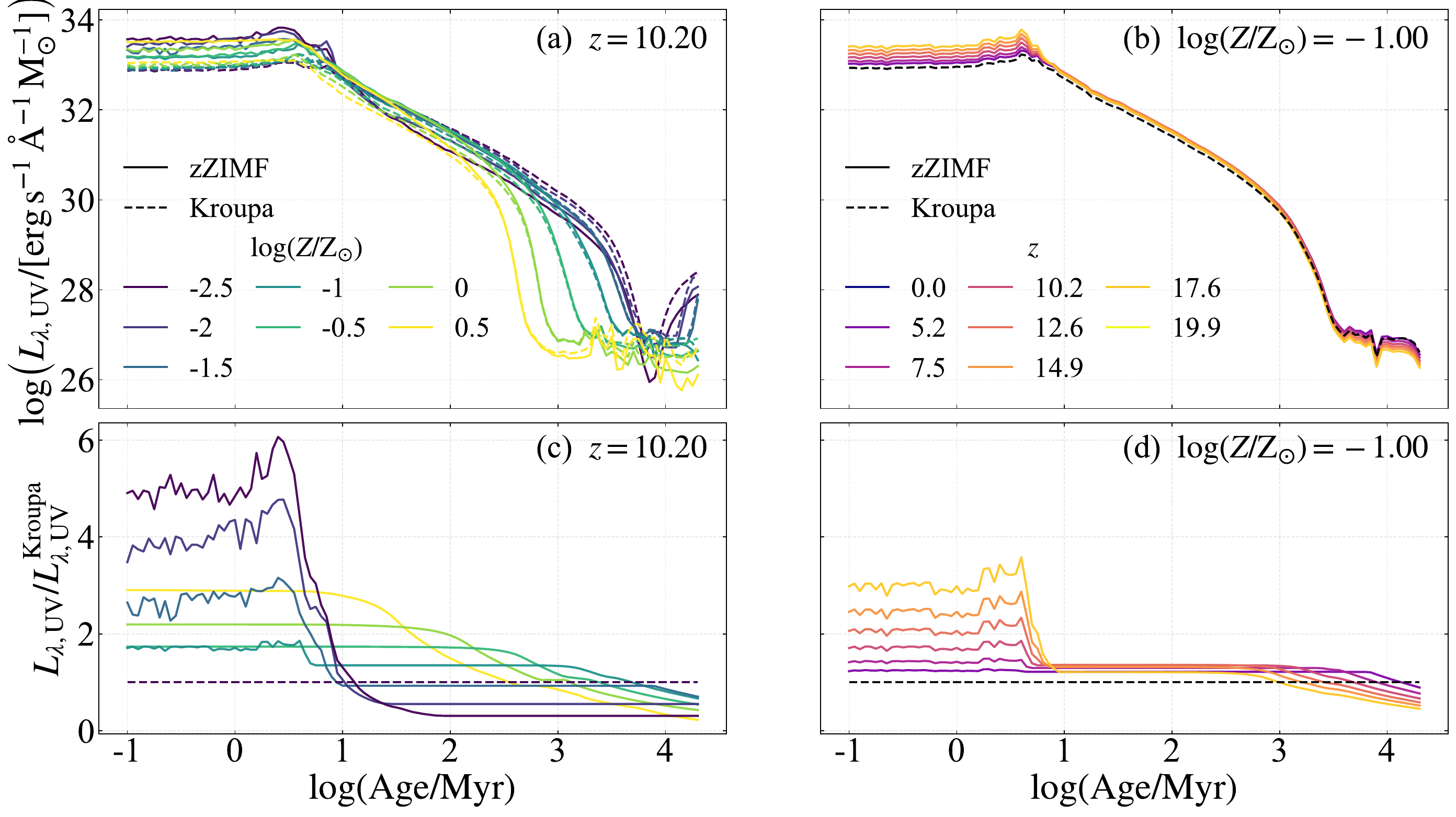}
   \end{center}
    \caption{Top panels~(a and b): Rest-frame UV luminosity at 1500\,\AA\ per unit initial stellar mass, $\log[L_{\lambda,\mathrm{UV}}/(\mathrm{erg\,s^{-1}\,\mathring{\mathrm{A}}^{-1}\,\Msun^{-1}})]$, as a function of stellar age ($\log(\mathrm{Age/Myr})$), comparing the \zZIMF\ (solid lines) with the Kroupa IMF (dashed lines). 
    Bottom panels (c and d): Ratio of $L_{\lambda,\mathrm{UV}}$ between the \zZIMF\ to that predicted by the Kroupa IMF as a function of stellar age. In the left column, metallicity is varied at fixed redshift, and both the \zZIMF\ and Kroupa SSPs are computed at the metallicities indicated by the colors. In the right column, redshift is varied at fixed metallicity; the Kroupa IMF itself has no redshift dependence.}    
    \label{fig:SSP_SED_time_evolution_Luv}
\end{figure*}

\begin{figure}
    \begin{center}
    \includegraphics[width=\columnwidth]{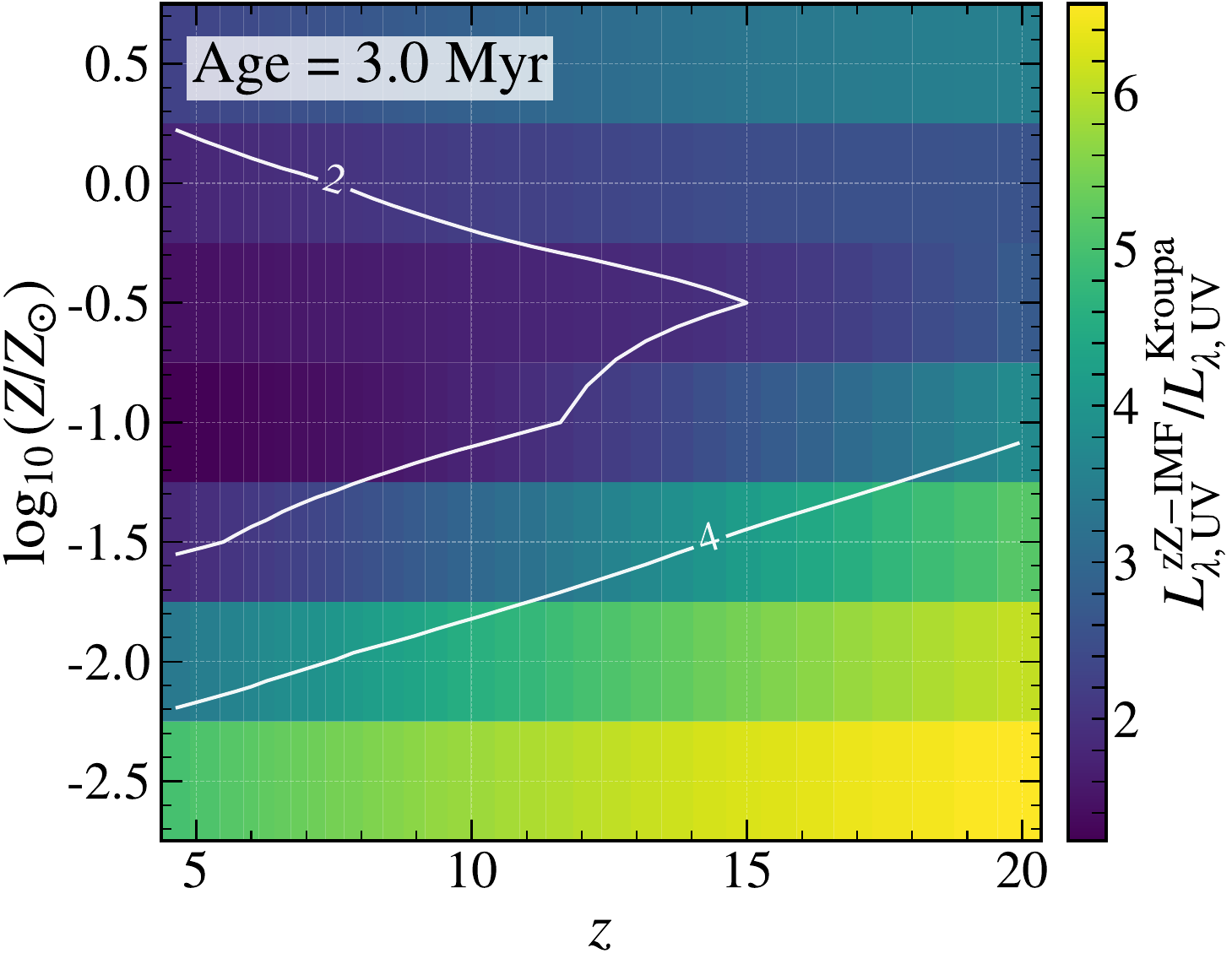}
    \end{center}
    \caption{Ratio of the rest-frame UV luminosity at 1500~\AA\ per unit initial stellar mass predicted by the \zZIMF\ to that predicted by the Kroupa IMF, as a function of redshift and metallicity.
    The SSP age is fixed at 3.0 Myr.
    The color scale indicates the UV luminosity ratio, and the white contours mark ratios of 2 and 4.}
    \label{fig:zZ_Luv_at_3Myr}
\end{figure}

\begin{figure*}
    \begin{center}
    \includegraphics[width=\textwidth]{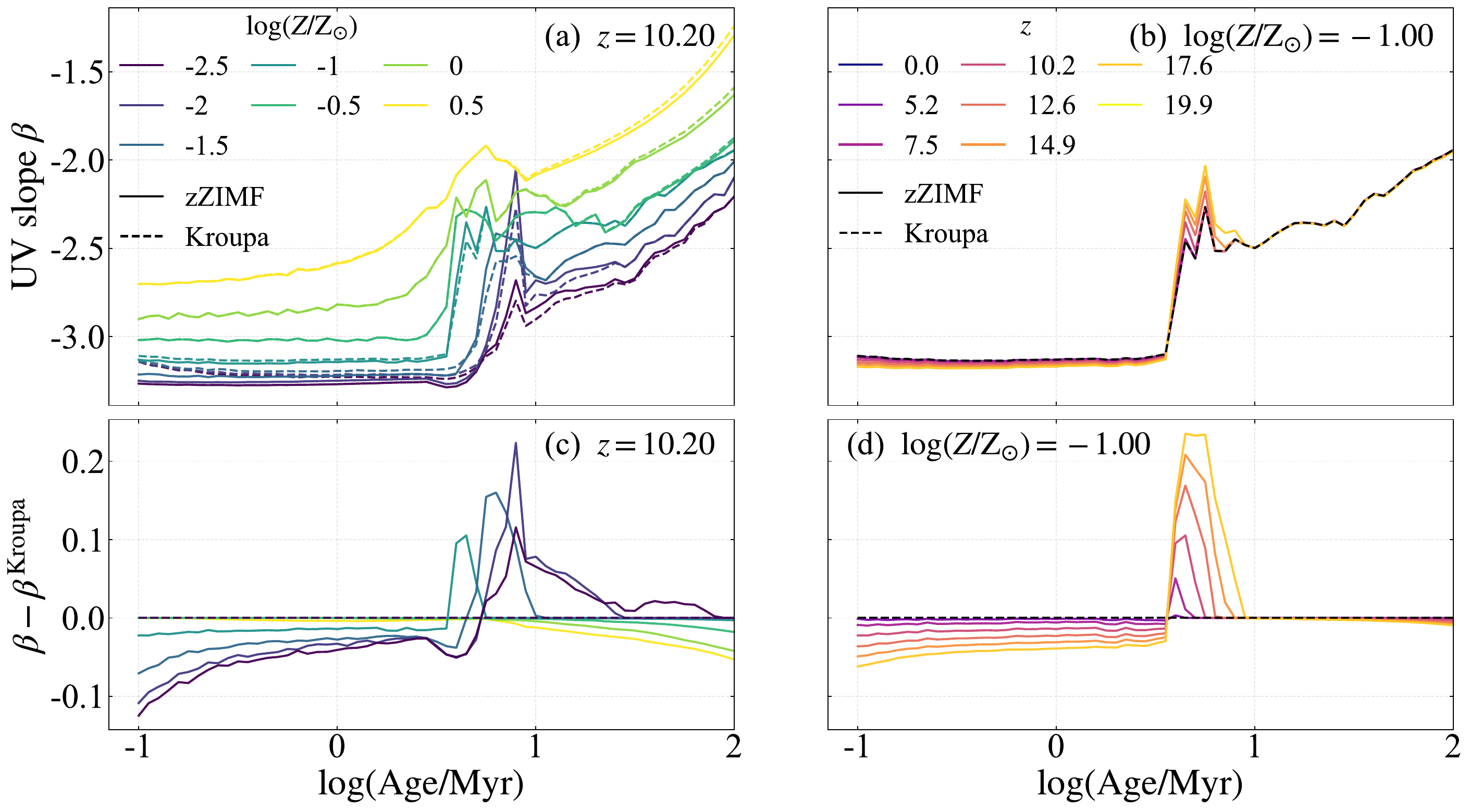}
    \end{center}
    \caption{Same as Figure~\ref{fig:SSP_SED_time_evolution_Luv}, but for the evolution of the UV continuum slope $\beta$ of the SSPs. The top panels~(a and b) show $\beta$ for the \zZIMF\
    (solid lines) and the Kroupa IMF (dashed lines),
    while the bottom panels~(c and d) show the difference
    between the two IMF models.}
    \label{fig:SSP_SED_time_evolution_beta}
\end{figure*}

The \zZIMF\ differs from the Steinhardt IMF in how it produces top-heavy behavior. 
The Steinhardt IMF has no explicit metallicity dependence. 
Its top-heaviness increases with redshift because the characteristic gas temperature $T_{\rm g}$ rises, suppressing low-mass star formation and enhancing the massive-star component. 
The \zZIMF\ also predicts a higher massive-star fraction at higher redshifts, although the strength of this trend depends on metallicity.
In addition, it further predicts that decreasing metallicity can enhance the low-mass component while also increasing the massive-star fraction. 
This simultaneous enhancement of the low- and high-mass components is a key distinction between the temperature-regulated Steinhardt IMF and the \zZIMF\ constructed based on the simulations of \citet{Chon2022MNRAS}.

\section{SED templates for the \zZIMF}
\label{sec:appendix_SED_temp}

Figure~\ref{fig:SSP_SED_result} shows the SEDs
of stellar populations modeled as SSPs using
the \zZIMF\ at different redshifts,
metallicities, and stellar ages.
For comparison, SEDs computed with the
Kroupa IMF are also shown at
$\log(Z/\Zsun)=-2.5$ and $0.0$.

At $\log(t/\mathrm{yr})=6$, the luminosity density $L_\lambda$ is higher
for $\log(Z/\Zsun)=-2$ and $-2.5$ than for
$\log(Z/\Zsun)=-1$ and $0$ at
$\lambda>200$~\AA\ across all redshifts considered,
including wavelengths longward of the Lyman limit.
This trend arises because the massive-star fraction
$f_{\rm massive}$ increases toward lower metallicity,
as shown in figure~\ref{fig:IMF_params_fitting_result},
resulting in stronger radiation from massive stars
in metal-poor stellar populations.

At $\log(t/\mathrm{yr})=7$, $L_\lambda$
decreases at wavelengths shorter than the
Lyman limit for all metallicities.
This is because massive stars with lifetimes
shorter than $10^7$~yr have already died.
In this wavelength range, $L_\lambda$ remains
higher at lower metallicities, whereas
the spectra become nearly identical across
metallicities between the Lyman limit
and approximately 4000~\AA.
At wavelengths longer than 4000~\AA,
SSPs with $\log(Z/\Zsun)=0$ or $0.5$
exhibit higher $L_\lambda$ than those
with lower metallicities.
This feature is attributed to red supergiants
(RSGs), which contribute significantly to
the emission at these ages.
A similar trend is also seen in
\citet{Leitherer1999ApJS_Starburst99}.
Because the Hayashi limit shifts toward
lower effective temperatures with increasing
metallicity \citep{Hayashi1961PASJ, Hayashi1962},
RSGs at higher metallicity can evolve to
cooler and thus redder states than those
at lower metallicity \citep{Levesque2006ApJ}.

At $\log(t/\mathrm{yr})=8$, the model with
$\log(Z/\Zsun)=-1$ shows the highest $L_\lambda$
at wavelengths shorter than the Lyman limit.
This may be attributed to the contribution
from post-AGB stars.
At longer wavelengths, the spectral slope
becomes shallower with increasing metallicity.
This trend reflects the growing contribution
from evolved stellar populations, including
red giants, AGB stars, and thermally pulsing
AGB (TP-AGB) stars, whose combined emission
shapes the integrated SSP spectrum.

At $\log(t/\mathrm{yr})=9$, the UV luminosity
of high-metallicity SSPs declines sharply.
The metallicity dependence of $L_{\rm UV}$
shown in figure~\ref{fig:SSP_SED_time_evolution_Luv}
reflects this behavior.
A likely explanation is that higher metallicity
leads to cooler main-sequence turnoff stars,
resulting in redder spectra and reduced
UV emission \citep{Vazdekis2016MNRAS}.

At $\log(t/\mathrm{yr})=10$, the model with
$\log(Z/\Zsun)=-1$ continues to exhibit
higher $L_\lambda$ at wavelengths shorter
than the Lyman limit at $z=19.15$ and $14.93$.
At $z=10.20$ and $5.15$, the model with
$\log(Z/\Zsun)=0.5$ instead shows the
highest $L_\lambda$ in this wavelength range.
At wavelengths longer than 4000~\AA,
the model with $\log(Z/\Zsun)=-1$
shows the highest $L_\lambda$
among the metallicities considered.
The enhanced emission below the Lyman limit
at these late SSP ages may be associated
with the contribution from post-AGB stars,
which becomes important after the most
massive stars have evolved off the main sequence.

\section{Detailed SED diagnostics}\label{app:sed_diagnostics}

%Figures~\ref{fig:SSP_SED_time_evolution_Luv}, \ref{fig:SSP_SED_time_evolution_beta}, \ref{fig:SSP_SED_time_evolution_Qion}, \ref{fig:SSP_SED_time_evolution_xiion}, and \ref{fig:SSP_SED_time_evolution_Dn4000}, show, respectively, the rest-frame UV luminosity at $1500$~\AA\ per unit initial stellar mass $L_{\lambda,\rm UV}$, the UV continuum slope $\beta$, the ionizing photon production rate $Q_{\rm ion}$, the ionizing photon production efficiency, $\xi_{\rm ion}=Q_{\rm ion}/L_{\nu,\mathrm{UV}}$, and the narrow 4000~\AA\ break strength $D_{\rm n}4000$ as a function of SSP age. 
%These figures compare the \zZIMF\ (solid lines) with the Kroupa IMF (dashed lines). 
%The left column shows the metallicity dependence at fixed redshift ($z=10.2$), while the right column shows the redshift dependence at fixed metallicity ($\log(Z/\Zsun)=-1$). 

We define the SED diagnostics used below. 
The UV continuum slope is defined by $f_\lambda \propto \lambda^\beta$ and is measured by fitting the model spectrum over $1300$--$2600$~\AA. 
We compute $D_{\rm n}4000$ following \citet{Hamilton1985ApJ}. 
It is defined as the ratio of the mean flux densities in the red band (4000--4100~\AA) to that in the blue band
(3850--3950~\AA).

\begin{figure*}
    \begin{center}
    \includegraphics[width=\textwidth]{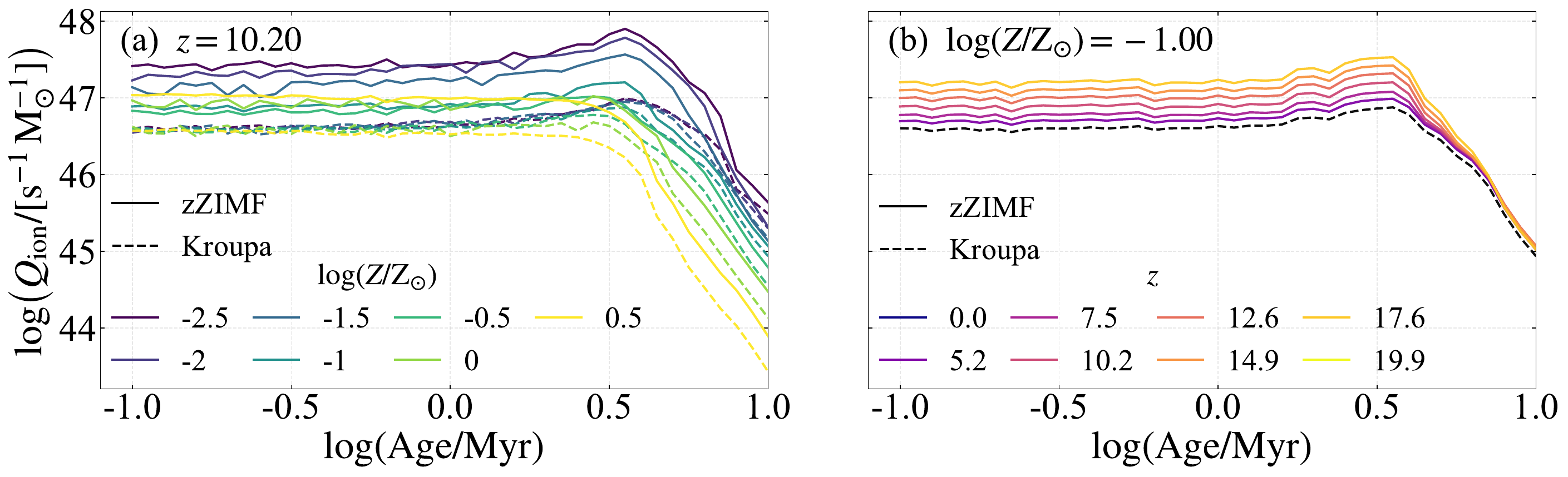}
    \end{center}
    \caption{Same as Figure~\ref{fig:SSP_SED_time_evolution_Luv}, 
    but for the evolution of the ionizing photon
    production rate per unit initial stellar mass,
    $\log[Q_{\rm ion}/(\mathrm{s^{-1}\,\Msun^{-1}})]$}
    \label{fig:SSP_SED_time_evolution_Qion}
\end{figure*}

\begin{figure*}
\begin{center}
\includegraphics[width=\textwidth]{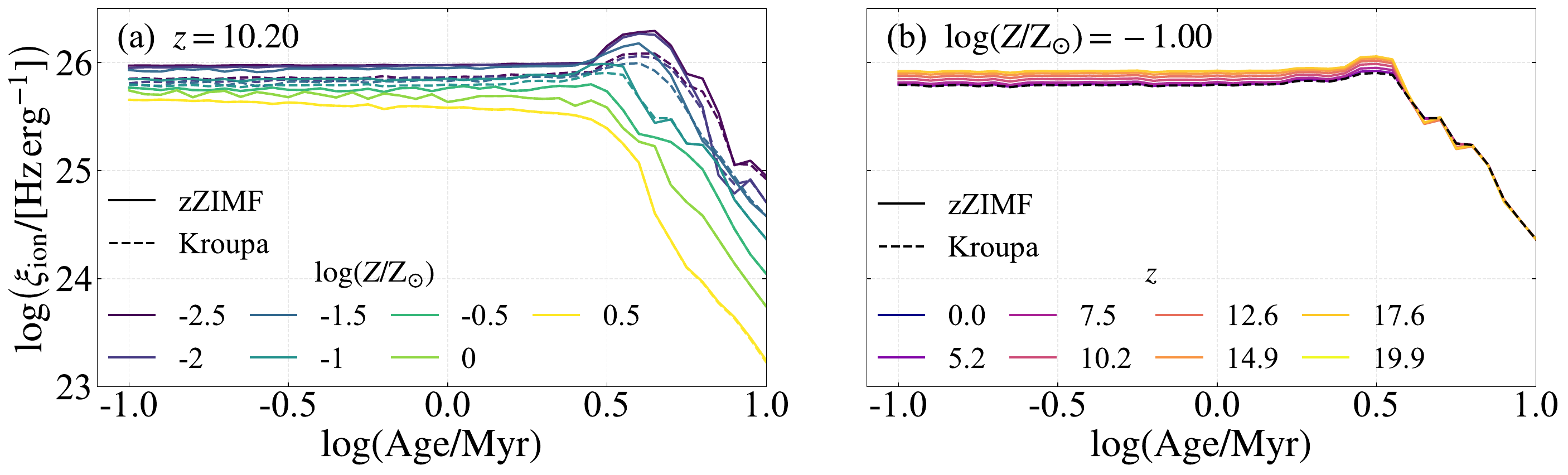}
\end{center}
\caption{Same as Figure~\ref{fig:SSP_SED_time_evolution_Luv}, but for the evolution of the ionizing photon production efficiency $\xi_{\rm ion}$ of the SSPs.}
\label{fig:SSP_SED_time_evolution_xiion}
\end{figure*}

Figure~\ref{fig:SSP_SED_time_evolution_Luv}
shows that the UV luminosity enhancement
produced by the \zZIMF\ is most pronounced
in young SSPs, and that both its strength and duration depend more strongly on metallicity than on redshift.
In the top panels, $L_{\lambda,\mathrm{UV}}$
is higher at lower metallicity and at higher
redshift for young populations, but the
variation with metallicity at fixed redshift
is clearly larger than that with redshift
at fixed metallicity.
Relative to the Kroupa IMF, the \zZIMF\ enhances the UV luminosity by a factor of $\sim 5$--$6$ across the metallicity range considered, whereas the corresponding enhancement across the redshift range is only $\sim 3$--$3.5$.

This behavior reflects the higher massive-star
fraction predicted by the \zZIMF\ at lower
metallicity and higher redshift.
In young SSPs, the UV output is dominated by massive O stars, and the UV luminosity enhancement relative to the Kroupa IMF is therefore greatest during the first few Myr.
Indeed, the bottom panels show that the ratio $L_{\lambda,\mathrm{UV}}^{\mathrm{zZIMF}}/L_{\lambda,\mathrm{UV}}^{\mathrm{Kroupa}}$ reaches its maximum near $\log(\mathrm{Age/Myr}) \sim 0.4$. This likely corresponds to the phase when very massive stars are still alive and contribute most strongly to the total UV
emission of the SSP.

By contrast, after $\sim 2$--$4$ Myr, the most massive stars begin to disappear rapidly as they reach the end of their lives, and the UV enhancement produced by the \zZIMF\ correspondingly weakens. Because the \zZIMF\ depends more strongly than the Kroupa IMF on these short-lived massive stars, its UV luminosity declines more steeply after the peak. As a result, although low-metallicity or high-redshift SSPs are the brightest at young ages, SSPs at higher metallicity
or lower redshift become brighter in
$L_{\lambda,\mathrm{UV}}$ at ages older than
$\sim 10$ Myr because their UV luminosity
declines more slowly.
In other words, the combination of \(Z\) and \(z\) that produces the highest UV luminosity changes with SSP age.

At ages $\gtrsim 100$ Myr, a further sharp
decline in the UV luminosity occurs, with
the decline beginning earlier at higher metallicity.
This is consistent with the metallicity dependence of the main-sequence turnoff temperature: lower-metallicity populations have hotter turnoff stars and maintain bluer emission for longer \citep{Vazdekis2016MNRAS}. 

To summarize the combined redshift and metallicity dependence of the UV enhancement, Figure~\ref{fig:zZ_Luv_at_3Myr} shows the ratio of the rest-frame UV luminosity predicted by the \zZIMF\ to that predicted by the Kroupa IMF at an SSP age of 3.0 Myr.
This age corresponds approximately to the peak
of the UV luminosity enhancement shown in
figure~\ref{fig:SSP_SED_time_evolution_Luv}.
The enhancement is strongest at low metallicity, particularly at $\log_{10}(Z/\Zsun)\lesssim-2$, where the \zZIMF\ produces substantially more UV emission than the Kroupa IMF.
At higher metallicities, especially at $-1\lesssim\log_{10}(Z/\Zsun)\lesssim0$, the ratio is lower, and the difference between the two IMFs becomes less pronounced.
At fixed metallicity, the ratio also tends to be higher at higher redshifts.

Figure~\ref{fig:SSP_SED_time_evolution_beta}
shows that SSPs with the \zZIMF\ generally
have bluer UV continua than those with
the Kroupa IMF, especially at young ages.
In the top panels, young SSPs at lower
metallicity and higher redshift have smaller
$\beta$, indicating bluer UV continua.
However, the difference in $\beta$ between
the two IMF models is not a constant offset
but varies substantially with SSP age.
In particular, the metallicity dependence
at fixed redshift is stronger and exhibits
more complex age-dependent behavior than
the redshift dependence at fixed metallicity.

A distinct feature appears around $\log(\mathrm{Age/Myr})\sim 0.6$--1, where the difference $\beta-\beta_{\mathrm{Kroupa}}$ temporarily shifts to larger values. 
Similar behavior has been reported in previous studies and attributed to the emergence of luminous post-main-sequence stars \citep{Zetterlund2015ApJ}. 
This interpretation is consistent with our results.

Once the SSP age exceeds $\sim 10$ Myr, the contribution of massive, hot stars to the UV continuum drops sharply, and $\beta$ correspondingly increases, indicating that the spectrum becomes redder as the stellar population ages. This age-dependent reddening should be distinguished from the differences associated with the IMF, metallicity, and redshift that characterize younger SSPs.

Figure~\ref{fig:SSP_SED_time_evolution_Qion} shows that the time evolution of the ionizing photon production rate $Q_{\rm ion}$ follows a trend similar to that of $L_{\lambda,\mathrm{UV}}$, with young SSPs exhibiting higher values at lower metallicity and at higher redshift. However, compared with $L_{\lambda,\mathrm{UV}}$, $Q_{\mathrm {ion}}$ declines more rapidly with SSP age after the first few Myr, as the most massive stars disappear.
This is because $Q_{\rm ion}$ depends more strongly on the hottest massive stars than does the UV continuum around 1500\,\AA. The metallicity dependence at fixed redshift is also more pronounced than the redshift dependence at fixed metallicity over the parameter ranges considered here. Thus, the \zZIMF\ not only enhances the UV luminosity of young stellar populations but also produces a stronger enhancement in their ionizing photon production.

Figure~\ref{fig:SSP_SED_time_evolution_xiion} shows the corresponding evolution of the ionizing photon production efficiency, $\xi_{\rm ion}=Q_{\rm ion}/L_{\nu,\mathrm{UV}}$.
At young ages, lower-metallicity SSPs have larger $\xi_{\rm ion}$, indicating that metal-poor stellar populations produce more ionizing photons per unit UV luminosity.
The \zZIMF\ enhances $\xi_{\rm ion}$ relative to the Kroupa IMF mainly at low metallicity, where the IMF becomes more top-heavy.
At fixed metallicity, the redshift dependence of $\xi_{\rm ion}$ is weaker than its metallicity dependence. 
Nevertheless, SSPs with the \zZIMF\ tend to have slightly larger $\xi_{\rm ion}$, reflecting the increasing top-heaviness of the IMF.

\begin{figure*}
    \begin{center}
    \includegraphics[width=\textwidth]{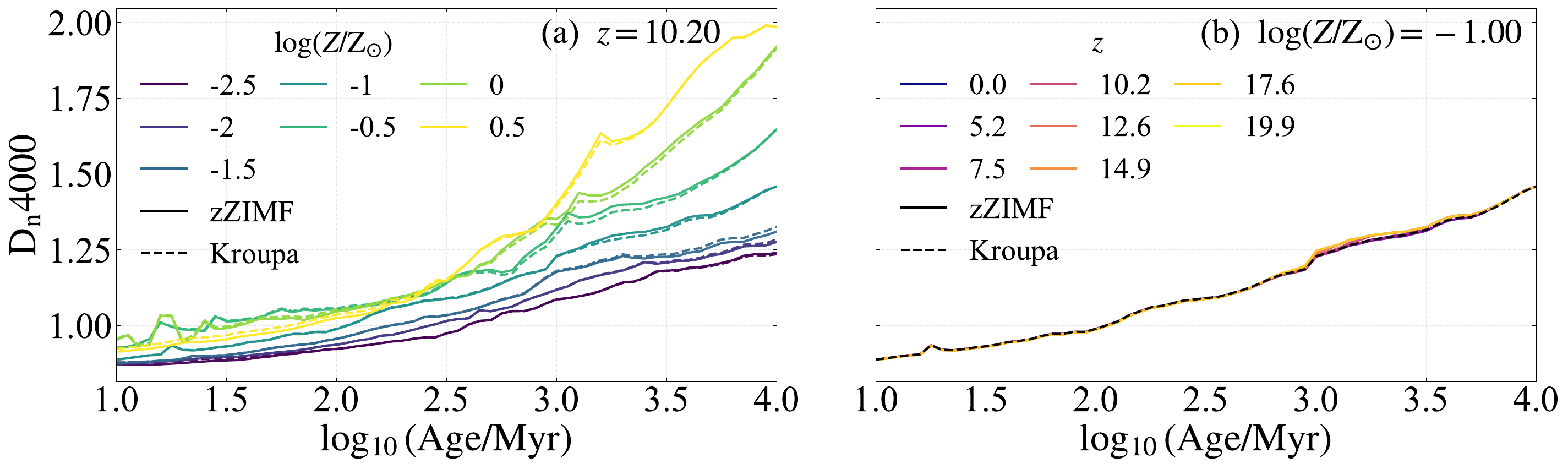}
    \end{center}
    \caption{Same as Figure~\ref{fig:SSP_SED_time_evolution_Luv}, but for the evolution of the narrow 4000~\AA\ break strength  $D_{\rm n}4000$ of the SSPs, shown in the top panels.}
    \label{fig:SSP_SED_time_evolution_Dn4000}
\end{figure*}

Figure~\ref{fig:SSP_SED_time_evolution_Dn4000} shows that $D_\mathrm{n}4000$ remains low and nearly constant in young SSPs but increases with age, particularly at higher metallicity. At young ages, the $D_\mathrm{n}4000$ values of all models cluster around $\sim 0.9$, with only weak metallicity dependence. As the population ages, $D_\mathrm{n}4000$ becomes systematically larger at higher metallicity, and the separation among the metallicity tracks becomes increasingly pronounced. This behavior is consistent with the 4000\,\AA\ break being weak in populations dominated by young massive stars but becoming increasingly sensitive to both age and metallicity in more evolved stellar populations \citep{Moresco2020ApJ}. By contrast, the redshift dependence at fixed metallicity is very weak, and the differences between the \zZIMF\ and the Kroupa IMF remain small overall.

\begin{figure}
    %\centering
    \begin{center}
    \includegraphics[width=0.5\textwidth]{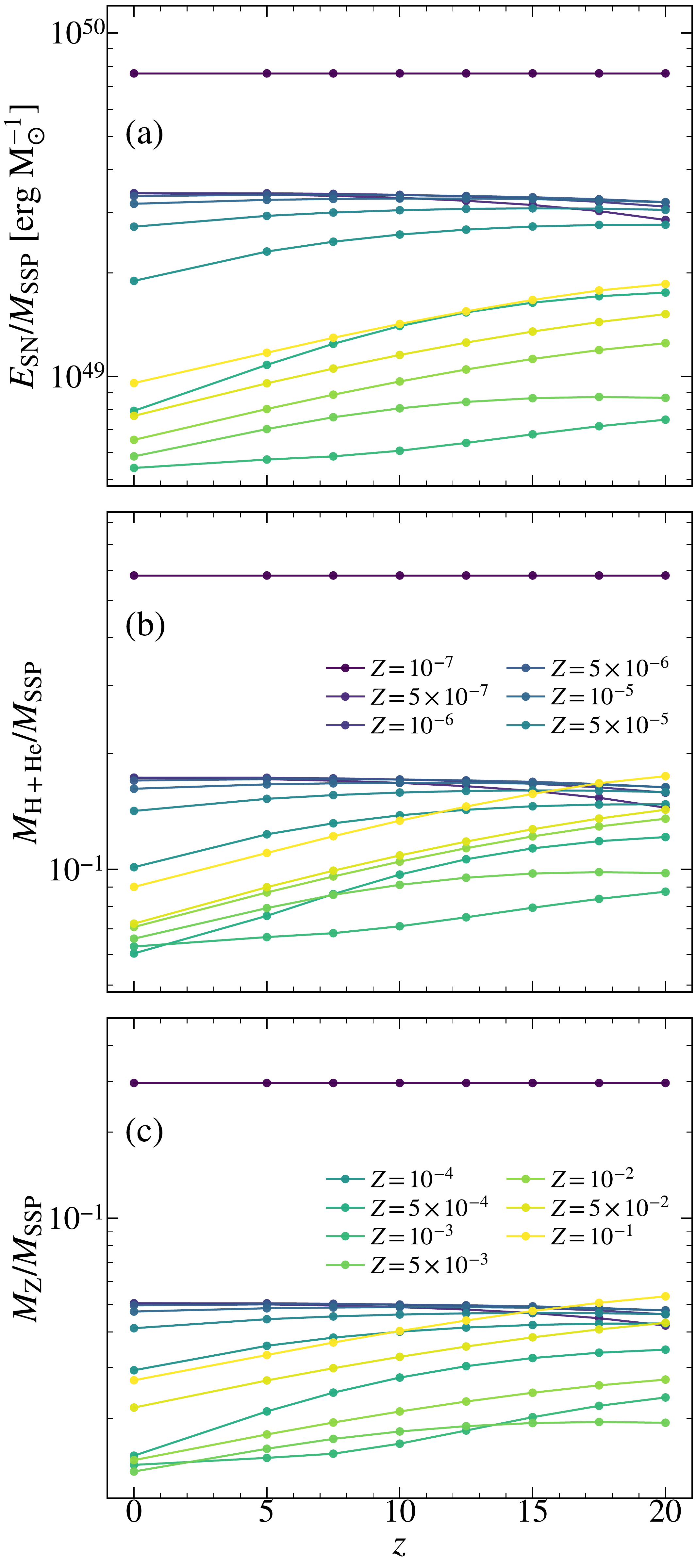}
    \end{center}
    \caption{
    Cumulative SN energy (panel~a), ejected H+He mass (panel~b), and ejected metal mass (panel~c), normalized by the initial stellar mass of the SSP, for the \zZIMF.
    The quantities are integrated up to an SSP age of $10^8$ yr.
    The horizontal axis indicates the formation redshift of the star clusters, while the colors represent their metallicities.
    }
    \label{fig:CELib}
\end{figure}

%\section{Energy and metal ejection from a top-heavy IMF}

\section{Semi-analytic model implementation and nebular emission}\label{app:semianalytic_details}

In this appendix, we describe the equations
and numerical implementation of the semi-analytic
galaxy model used in this study.
We first explain how galaxies are evolved
along the merger trees (Sec.~A.6.1)
and how their star formation rates are determined
(Sec.~A.6.2).
We then describe the treatment of gas accretion
and stellar feedback (Sec.~A.6.3),
followed by chemical enrichment, including
IMF-dependent mass and metal ejection
(Sec.~A.6.4).
Finally, we present the prescription for nebular
line and continuum emission used to construct
the galaxy SEDs (Sec.~A.6.5).

\subsection{Evolution along the merger trees}

We evolve galaxies along the merger trees while tracking star formation, gas flows, and chemical enrichment.
The time step is determined directly from the redshift intervals between successive merger-tree snapshots.
At each time step, the model accounts for halo mergers, gas inflow, star formation, metal enrichment, and gas outflow.

When advancing from timestep $i$ to $i+1$, we first account for halo mergers by summing the baryonic mass components of the progenitor halos:
\begin{equation}
m_{\{\text{gas, star, Z}\}i+1} = \sum_j \left( m_{\{\text{gas, star, Z}\}i,j} + \dot{m}_{\{\text{gas, star, Z}\}i,j} \cdot \Delta t \right),
\label{eq:merger_and_dmdt}
\end{equation}
where $m_{\{\text{gas, star, Z}\}}$ represents the gas mass, stellar mass, and gas-phase metal mass, respectively, and $\dot{m}_{\{\text{gas, star, Z}\}}$ denotes their corresponding rates of change.
The summation $\sum_j$ includes all progenitor halos $j$ at timestep $i$ that contribute to the descendant halo at timestep $i+1$.
Here, $\Delta t$ is the cosmic-time interval
between timesteps $i$ and $i+1$.

\subsection{Stellar mass and star formation}\label{subsubsec:semiana_model_stellarmass}

To model star formation along the halo assembly
histories derived from $N$-body simulations,
we adopt empirical relations that link the SFR
to the halo mass accretion rate, following
\citet{Harikane2022ApJS} and
\citet{Feldmann2025MNRAS}.
These prescriptions are motivated by the idea
that galaxy star formation is regulated by
the gas supply associated with halo growth.
Specifically, we compute the SFR using
equation~(56) of \citet{Harikane2022ApJS}
and equation~(2) of \citet{Feldmann2025MNRAS}.
For the latter, we adopt the parameters
corresponding to $\mathrm{SFR}(20\ \mathrm{Myr})$.
The halo mass accretion rate is calculated
following Appendix B of
\citet{Behroozi2015ApJ}.

Star formation is calculated for all halos
at each redshift step stored in the merger tree.
Under the SSP assumption, stellar populations
are formed with the same metallicity as the
gas from which they form.
The formation time of each stellar population
is randomly assigned within the cosmic-time
interval between the previous and current
redshift steps.
The resulting stellar ages and metallicities
are then used to calculate the SEDs.

\subsection{Gas mass and stellar feedback}\label{subsubsec:semiana_model_gas}

The gas accretion rate onto halos is calculated by multiplying the dark matter accretion rate $\dot{m}$ by the cosmic baryon fraction $f_\mathrm{b}=0.16$ \citep{Planck2020A&A}. 
The gas mass then evolves through accretion, star formation, and stellar feedback.
Its rate of change is given by:
\begin{equation}
\dot{m}_\mathrm{gas} = f_\mathrm{b} \, \dot{m} - (1-f) \, \dot{m}_\mathrm{star} - \eta \, \dot{m}_\mathrm{star} .
\label{eq:mdot_gas}
\end{equation}
The three terms on the right-hand side represent gas accretion, net gas consumption by star formation, and gas outflow driven by stellar feedback, respectively. 
Here, $f$ is the gas return fraction from dying stars, $\eta$ is the mass-loading factor, which we define following \citet{Somerville2015} as 
\begin{equation}
\label{eq:eta}
\eta \equiv \frac{\dot{m}_\mathrm{out}}{\dot{m}_\mathrm{star}} = \epsilon_{\rm SN} \left( \frac{V_0}{V_c} \right)^{\alpha_{\rm SN}}. 
\end{equation}
Here, $V_{\mathrm{c}}$ denotes the maximum circular velocity of a galaxy, which we approximate by the maximum circular velocity of its host dark matter halo, $V_{\max}$. 
The parameters $\epsilon_{\mathrm{SN}}$ and $\alpha_{\mathrm{SN}}$ characterize the normalization and velocity dependence of the mass-loading factor, respectively. Following \citet{Somerville2015}, we adopt $\epsilon_{\mathrm{SN}} = 1.5$, $\alpha_{\mathrm{SN}} = 2.2$ and the normalization velocity $V_{0} = 200~\mathrm{km~s^{-1}}$.

\begin{figure}
    \begin{center}
    \includegraphics[width=0.5\textwidth]{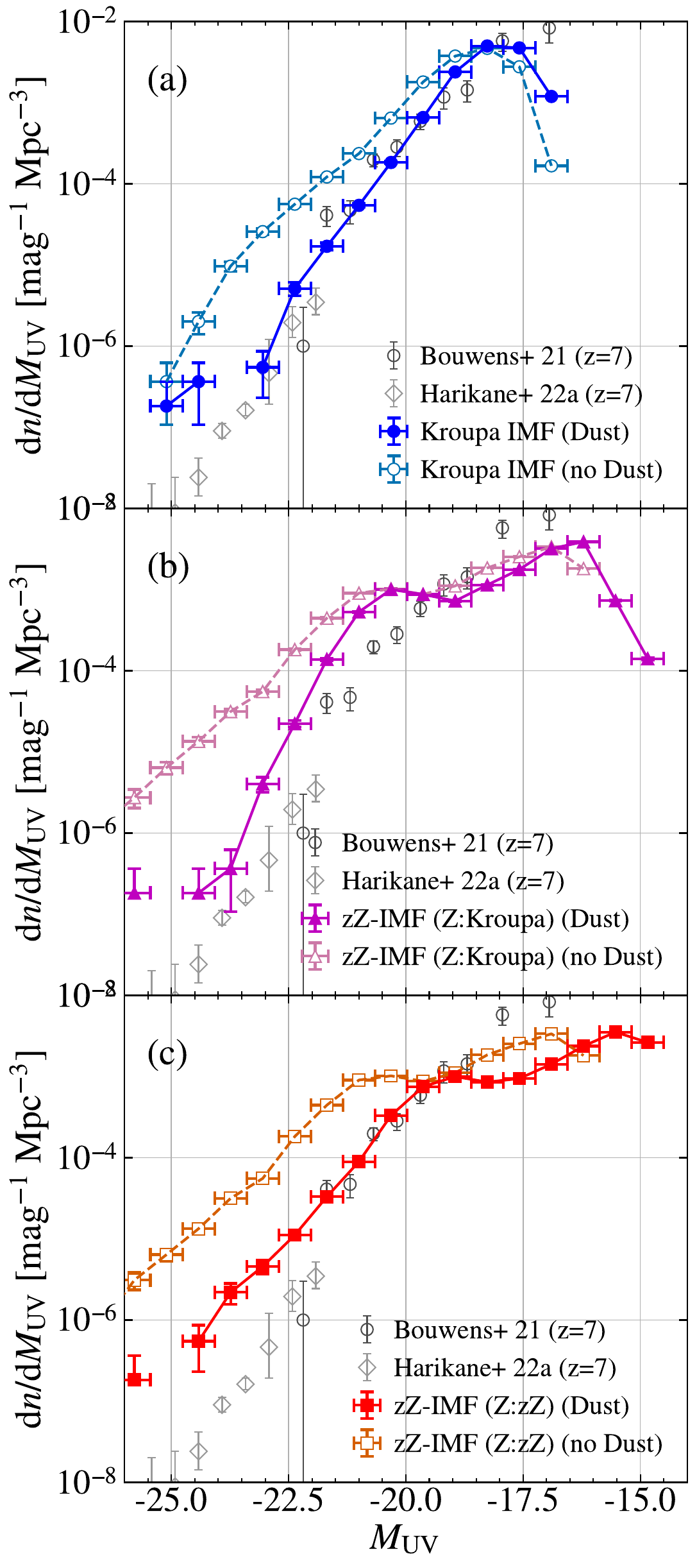}
    \end{center}
    \caption{Same as figure~\ref{fig:UVLF_dust_calib}, but using the alternative dust attenuation model of \citet{Trinca2024MNRAS}.}
    \label{fig:UVLF_dust_calib_Trinca}
\end{figure}

\begin{figure*}
  \centering
  \includegraphics[width=\textwidth]{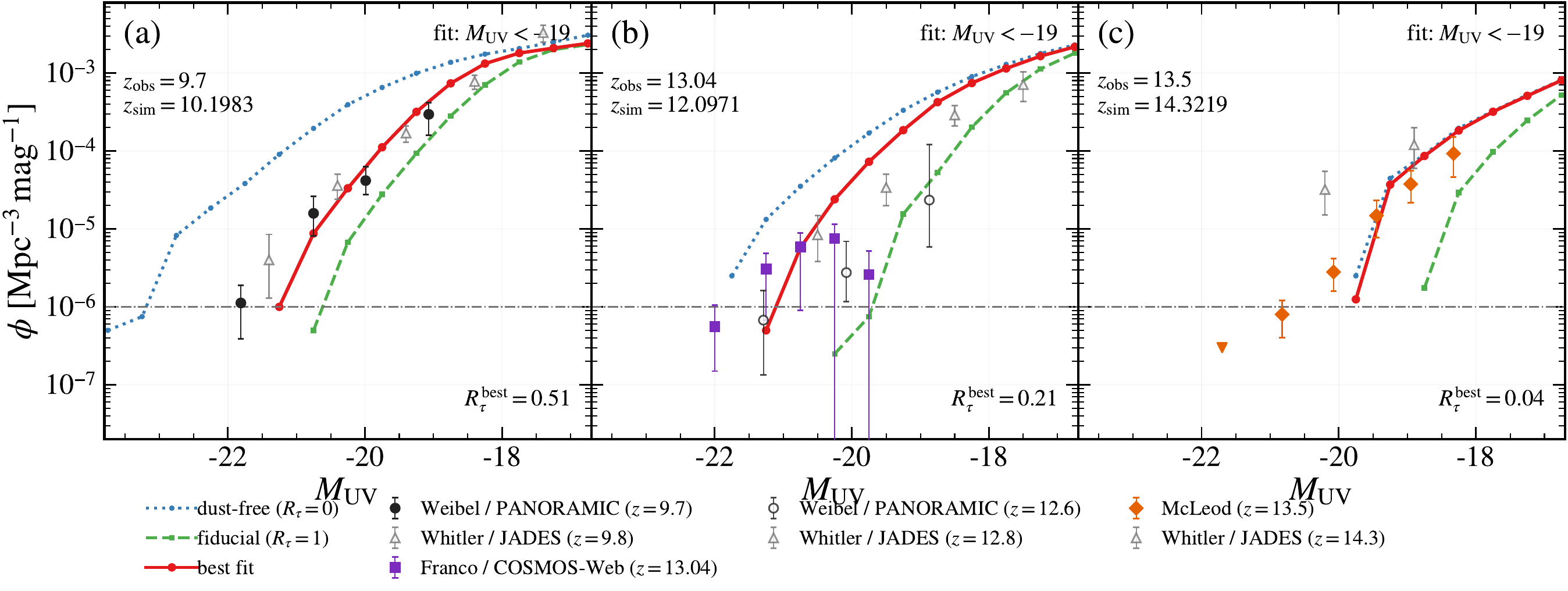}
  \caption{UVLFs predicted by the self-consistent \zZzZ\ for different effective dust optical depths. 
  The dotted and dashed curves show the dust-free ($R_\tau=0$) and fiducial ($R_\tau=1$) cases, respectively, while the solid curve in each panel shows the prediction obtained with the formal best-fitting value of $R_\tau$. 
  In the panels~(a), (b), and (c), the model is fitted to the observational UVLFs of \citet{Weibel2026ApJ} at $z=9.7$, \citet{Franco2025arXiv250804791F} at $z=13.04$, and \citet{McLeod2026arXiv260416666M} at $z=13.5$, respectively. Additional observational data from \citet{Whitler2025ApJ} at $z=9.8$, $12.8$, and $14.3$, and \citet{Weibel2026ApJ} at $z=12.6$ are shown for comparison. The observational effective redshift, adopted simulation snapshot, and fitted value of $R_\tau$ are indicated in each panel. Fits are restricted to $M_{\rm UV}<-19$.}
  \label{fig:UVLF_all_redshifts_zZzZ_dust_comparison}
\end{figure*}

\subsection{Metal enrichment}\label{subsubsec:semiana_model_metal}

Metals produced by CCSNe are assumed to be released
instantaneously following star formation
and to mix uniformly within the halo.
For the Kroupa IMF, we adopt a fixed mass return fraction $f = 0.085$, based on the value obtained for the \zZIMF\ at $z = 0$ and $Z = 0.01$. 
For the \zZIMF, $f$ varies with redshift and metallicity (see Appendix~\ref{app:appendix_Yield} for details).

The rate of change of the gas-phase metal mass during the timestep from $i$ to $i+1$
is calculated as:
\begin{equation}
\dot{m}_Z(i+1) = Z_\mathrm{yield} \, \dot{m}_\mathrm{star}  - Z_{(i)} \, (1-f) \, \dot{m}_\mathrm{star} - Z_{\mathrm{mix},(i+1)} \, \eta \, \dot{m}_\mathrm{star} ,
\label{eq:mdot_metal}
\end{equation}
where we assume that the accreted gas is
primordial and therefore contains no metals.
Consequently, there is no metal accretion term,
unlike the gas accretion term in
equation~\ref{eq:mdot_gas}.
The three terms on the right-hand side
represent metal enrichment from stars,
the incorporation of metals into newly
formed stars, and metal loss through
outflows driven by stellar feedback,
respectively.
The IMF affects the metal yield
$Z_{\rm yield}$ and the mass return fraction
$f$, thereby altering the evolution of
the gas-phase metallicity and the metallicities $Z_{(i)}$
assigned to newly formed stellar populations.

For the Kroupa IMF, the metal yield was set to $Z_\mathrm{yield} = 0.014$. 
For the $z$–$Z$ dependent IMF, $Z_\mathrm{yield}$ also varies as a function of redshift and metallicity (see Appendix~\ref{app:appendix_Yield} for details).

The metal mass incorporated into newly formed stars is calculated using the post-merger metallicity $Z_{(i)}$.
In contrast, the metal mass removed by outflows is calculated using $Z_{\mathrm{mix},(i+1)}$, because we assume that metals released by stars mix instantaneously and uniformly with the gas in the host halo.
In equation~\ref{eq:mdot_metal},
$Z_{\mathrm{mix},(i+1)}$ is first determined
from the balance between metal enrichment
and the incorporation of metals into newly
formed stars, represented by the first two terms.
The metal mass removed by galactic outflows
is then calculated using the final term.
Because the outflowing gas is assumed to
have the same metallicity as the well-mixed
ISM, the outflow reduces the gas and metal
masses in the same proportion, leaving
the gas-phase metallicity unchanged.

\subsection{Nebular emission}\label{subsubsec:Nebular_Emission}

To investigate the UV continuum slope $\beta$ of galaxies, we calculate nebular emission using an updated version of the method of \citet{Inoue2011}.
The ionizing photon production rate $Q_{\rm ion}$ is calculated as
\begin{equation}
Q_{\rm ion} = \int_0^{912\AA} \frac{L_{\lambda}}{hc/\lambda} \, d\lambda.
\label{eq:ionizing_photons_Q}
\end{equation}
This rate depends on the stellar populations of individual galaxies and the adopted IMF.  
We then calculate nebular line and continuum emission using the machine-readable tables provided by \citet{Inoue2011}, which were generated using \textsc{Cloudy} \citep{Ferland1998PASP}.
We use updated electron-temperature tables and an extended line list containing 124 emission lines from Ly$\alpha$ to 1~$\mu$m, with line intensities normalized to H$\beta$.

We calculate the H$\beta$ luminosity following
equation~(1) of \citet{Inoue2011}, assuming
$f_{\rm dust}=0$, i.e., neglecting the absorption
of ionizing photons by dust within ionized regions:
\begin{equation}
\label{eq:Ha_emission}
L_{\mathrm{H}\beta} = 
\frac{\gamma_{\mathrm{H}\beta}(T_{\mathrm{e}}) \, Q_{\rm ion} \, (1 - f_{\mathrm{esc}} - f_{\mathrm{dust}})}
{\alpha_{\mathrm{B}}(T_{\mathrm{e}}) + \alpha_{1}(T_{\mathrm{e}}) \, (f_{\mathrm{esc}} + f_{\mathrm{dust}})}.
\end{equation}
This expression follows equation (1) of \citet{Inoue2011}. 
We adopt an escape fraction $f_{\rm esc}=0$ as the fiducial value, while also exploring $f_{\rm esc}=0.5$ and $1$ to assess the effects of ionizing photon escape.
The electron temperature $T_{\mathrm{e}}$ is determined as a function of metallicity following Table~2 of \citet{Inoue2011}.
The luminosity of each emission line is then calculated using its intensity ratio relative to H$\beta$ given in the tables.

For the nebular continuum emission, we include hydrogen bound--free, free--free, and two-photon emission processes, while neglecting the helium continuum. 
Although this approximation leads to a slight underestimation of the total nebular continuum, the contribution from helium is expected to be small \citep{Inoue2010MNRAS}. 
We also neglect the continuum emission from metals.
The luminosity density of the hydrogen nebular continuum at frequency $\nu$, $L_{\nu}$, is calculated as
\begin{equation}
\label{eq:continuum_emission}
L_\nu = 
\frac{\gamma_\nu(T_{\mathrm{e}}) \, Q_{\rm ion} \, (1 - f_{\mathrm{esc}} - f_{\mathrm{dust}})}
{\alpha_{\mathrm{B}}(T_{\mathrm{e}}) + \alpha_{1}(T_{\mathrm{e}}) \, (f_{\mathrm{esc}} + f_{\mathrm{dust}})} ,
\end{equation}
where $\gamma_{\nu}$ is the nebular continuum emission coefficient. 

The coefficients $\gamma_{\mathrm{H}\beta}$ and $\gamma_{\nu}$, as well as the recombination coefficients $\alpha_{\mathrm{B}}$ and $\alpha_{1}$, are computed following the prescriptions of \citet{Inoue2010MNRAS} and \citet{Inoue2011}.
This method allows us to consistently calculate nebular line and continuum emission from the stellar ionizing photon production rate without performing additional photoionization calculations.

%%%%%%%%%%%%%%%%%%%%%%%%%%%%%%%%%%%%%%%%%%%%%%%%%%

\section{Energy and metal ejection from a top-heavy IMF}
\label{app:appendix_Yield}

Based on the IMFs constructed in Sec.~\ref{subsec:IMF_model}, we calculated the energy and mass ejection from star clusters under the SSP approximation using the \texttt{CELib} code \citep{Saitoh16}. 
In these calculations, only CCSN were included, and the metal yields were taken from \citet{Limongi2018ApJS}. 
The metallicities and formation redshifts of the clusters were set to 
$Z = (1\times10^{-7},\,5\times10^{-7},\,1\times10^{-6},\,5\times10^{-6},\,1\times10^{-5},\,5\times10^{-5},\,1\times10^{-4},\,5\times10^{-4},\,1\times10^{-3},\,5\times10^{-3},\,1\times10^{-2},\,5\times10^{-2},\,1\times10^{-1})$ 
and 
$z = (0,\,5,\,7.5,\,10,\,12.5,\,15,\,17.5,\,20)$, respectively. 
We calculated the cumulative SN energy and the
cumulative masses of ejected H+He and metals,
normalized by the initial SSP mass, up to an
SSP age of $10^{8}\ \mathrm{yr}$.

%Figure~\ref{fig:CELib} shows the energy,
%H+He, and metal ejection rates per unit
%stellar mass predicted for the \zZIMF\
%as functions of the formation redshift
%and metallicity of the star clusters.

%At the lowest metallicity, where the IMF becomes most top-heavy, the energy, mass, and metal ejection rates per unit stellar mass all reach their maximum values. 
%The results are used in Sec.~\ref{subsec:semiana_model} to compute the mass and metal ejection rates in the semi-analytic model. 

Figure~\ref{fig:CELib} shows these cumulative quantities as functions of the formation redshift and metallicity of the star clusters.
At the lowest metallicities, where the IMF becomes
most top-heavy, the cumulative SN energy and ejected masses of H+He and metals per unit initial stellar mass reach their maximum values.
The corresponding mass-return fractions and metal yields are used in the semi-analytic model under the instantaneous-recycling approximation (see Appendix~\ref{subsubsec:semiana_model_metal}).

\section{Additional UVLF comparisons with dust attenuation}
\label{app:UVLF_dust_comparison}

Figure~\ref{fig:UVLF_dust_calib_Trinca}
shows the UVLF at $z=7$ obtained using the
alternative dust attenuation model of
\citet{Trinca2024MNRAS}.
This comparison illustrates the calibration
of the alternative attenuation model described
in Section~\ref{subsubsec:Trinca_Dust_model}.

We also examine the effects of varying the effective dust optical depth on the UVLFs predicted by the self-consistent \zZzZ.

To quantify the effective attenuation required to reproduce the observed UVLFs, we vary the optical-depth scaling factor $R_\tau$ from 0 to 4 in steps of 0.01. At each grid point, we recalculate the UV attenuation of individual model galaxies and reconstruct the resulting UVLF. We determine the best-fitting value by minimizing
\begin{equation}
\chi^2(R_\tau)=\sum_i\left[\frac{\log_{10}\phi_{{\rm model},i}(R_\tau)-\log_{10}\phi_{{\rm obs},i}}{\sigma_{\log\phi,i}}\right]^2,
\label{eq:Rtau_chi2}
\end{equation}
where the sum includes observational detections satisfying $M_{\rm UV}<-19$ and the adopted sample-selection criteria. Asymmetric observational uncertainties are converted to logarithmic space, and the upper or lower uncertainty is selected according to the sign of the model--observation residual. Observational upper limits are included only when the predicted number density exceeds the reported limit, adopting logarithmic uncertainty of 0.3 dex. Empty model bins are assigned a floor corresponding to 0.5 counts before taking the logarithm. We fit each observational survey separately rather than combining different surveys into a single observational envelope.

Figure~\ref{fig:UVLF_all_redshifts_zZzZ_dust_comparison} shows the effect of rescaling the fiducial dust optical depth in the self-consistent \zZzZ. 
For each primary observational sample, we adopt the nearest simulation snapshot and restrict the fit to bins with $M_{\rm UV}<-19$.
The left, middle, and right panels compare the model UVLFs at $z_{\rm sim}=10.1983$, $12.0971$, and $14.3219$ with the observational UVLFs at $z_{\rm obs}=9.7$ \citep{Weibel2026ApJ}, $13.04$ \citep{Franco2025arXiv250804791F}, and $13.5$ \citep{McLeod2026arXiv260416666M}, respectively.
For reference, we also show observational data at $z=9.8$ \citep{Whitler2025ApJ}, $z=12.6$ \citep{Weibel2026ApJ}, and $z=12.8$ and $14.3$ \citep{Whitler2025ApJ} in the corresponding panels.

The first two comparisons yield $R_\tau=0.51$ at $z_{\rm obs}=9.7$ and $R_\tau=0.21$ at $z_{\rm obs}=13.04$.
The formal best-fitting value in the highest-redshift panel is $R_\tau=0.04$.
However, we do not regard this value as a robust constraint because too few bright-end bins are adequately populated by the model and the dust-free prediction remains below the observed number densities.

%%%%%%%%%%%%%%%%%%%%%%%%%%%%%%%%%%%%%%%%%%%%%%%%%%
% Don't change these lines

% Any journal BST may be used because PASJ does not provide a BST file.
\bibliographystyle{apj}
\bibliography{UVLF}

\end{document}